\documentclass[]{aa}
\usepackage{times,epsfig,amsmath} 
\usepackage{graphicx}
\usepackage{placeins}
\usepackage{float}
\usepackage{epstopdf}
\usepackage{natbib}
\bibpunct{(}{)}{;}{a}{}{,} 
\usepackage{pdflscape}

\newcommand{\cxo}{{\it Chandra}}

\newcommand{\planck}{{\it Planck}}
\newcommand{\xmm}{{\it XMM-Newton}}
\newcommand{\xmms}{{\it XMM}}

\begin{document}

\title{Hydrostatic mass profiles in X-COP galaxy clusters}

\author{S. Ettori$^{1,2}$, V. Ghirardini$^{1,3}$, D. Eckert$^4$, E. Pointecouteau$^{5}$, F. Gastaldello$^6$, M. Sereno$^{1,3}$,
M. Gaspari$^7$\thanks{{\it Einstein} and {\it Spitzer} Fellow.}, S. Ghizzardi$^6$, M. Roncarelli$^{1,3}$, M. Rossetti$^6$}
\authorrunning{S. Ettori}
\institute{
INAF, Osservatorio di Astrofisica e Scienza dello Spazio, via Pietro Gobetti 93/3, 40129 Bologna, Italy \and
INFN, Sezione di Bologna, viale Berti Pichat 6/2, I-40127 Bologna, Italy \and
Dipartimento di Fisica e Astronomia Universit\`a di Bologna, via Pietro Gobetti 93/2, 40129 Bologna, Italy  \and
MPE, Giessenbachstrasse 1, 85748 Garching bei M\"unchen, Germany \and
IRAP, Universit\'e de Toulouse, CNRS, CNES, UPS, 9 av du colonel Roche, BP44346, 31028 Toulouse cedex 4, France \and
INAF, IASF, Via E. Bassini 15, I-20133 Milano, Italy \and
Department of Astrophysical Sciences, Princeton University, 4 Ivy Lane, Princeton, NJ 08544-1001, USA
  }
\mail{stefano.ettori@inaf.it}

\abstract
{}
{We present the reconstruction of  hydrostatic mass profiles in 13 X-ray luminous galaxy clusters that have 
been mapped in their X-ray and Sunyaev-Zeldovich (SZ) signals out to $R_{200}$ for the \xmm\ Cluster Outskirts Project (X-COP).}
{Using profiles of the gas temperature, density, and pressure that have been spatially resolved out to   median values of 
$0.9 R_{500}$, $1.8 R_{500}$, and $2.3 R_{500}$, respectively, we are able to recover the hydrostatic gravitating mass
profile with several methods and using different mass models.}
{The hydrostatic masses are recovered with a relative (statistical) median error of $3$ \% at $R_{500}$ and $6$\% at $R_{200}$.
By using several different methods to solve the equation of the hydrostatic equilibrium, we evaluate 
some of the systematic uncertainties to be of the order of 5\% at both $R_{500}$ and $R_{200}$.
A Navarro-Frenk-White profile provides the best-fit in 9 cases out of 13;  the remaining 4 cases  do not show a
statistically significant tension with it. The distribution of the mass concentration follows the correlations with the total mass 
predicted from numerical simulations with a scatter of 0.18 dex, with an intrinsic scatter on the hydrostatic masses of 0.15 dex. 
We compare them with the estimates of the total gravitational mass obtained 
through X-ray scaling relations applied to $Y_X$, gas fraction, and 
$Y_{SZ}$, and from weak lensing and galaxy dynamics techniques,
and measure a substantial agreement with the results from scaling laws, from WL at both $R_{500}$ and $R_{200}$ 
(with differences below 15\%), from cluster velocity dispersions. Instead, we find  a significant tension with the caustic masses that tend to underestimate the hydrostatic masses by 40\% at $R_{200}$.
We also compare these measurements with predictions from alternative models to the cold dark matter, like
the emergent gravity and MOND scenarios, confirming that the latter underestimates  hydrostatic masses by 40\% at $R_{1000}$, 
with a decreasing tension as the radius increases, and reaches $\sim$15\%
at $R_{200}$, whereas the former reproduces $M_{500}$ within 10\%, but overestimates $M_{200}$ by about 20\%.
}
{The unprecedented accuracy of these hydrostatic mass profiles out to $R_{200}$ allows us
 to assess the level of systematic errors in the hydrostatic mass reconstruction method,  to evaluate the intrinsic scatter in the NFW $c-M$ relation, and  to robustly quantify  differences among different mass models, different mass proxies, and different gravity scenarios.
}


\maketitle 

\section{Introduction}

The use of the galaxy clusters as astrophysical laboratories and cosmological probes relies on characterizing the 
distribution of their gravitational mass  \citep[see e.g.][]{allen11,kb12}. 
In the currently favoured $\Lambda CDM$ scenario, clusters of galaxies are dominated by dark matter (80\% of the total mass), 
with a negligible contribution due to stars \citep[a small percentage; see e.g.][]{gonzalez+13}, and even less by cold/warm gas, 
and the rest (about 15\% of the total mass, i.e. $M_{\rm DM}/M_{\rm gas} \sim 4-7$) in the form of ionized  plasma emitting in X-ray and detectable 
through the Sunyaev-Zeldovich \citep[SZ, ][]{SZ} effect induced from inverse Compton scattering of the cosmic microwave background photons off the 
electrons in the hot intracluster medium (ICM).

X-ray observations can provide an estimate of the total mass in galaxy clusters under the condition that the ICM is in hydrostatic equilibrium within the
gravitational potential. This is an assumption that is satisfied if the ICM, treated as collisional fluid on timescales typical of any 
heating/cooling/dynamical process much longer than the elastic collisions time for ions and electrons, is crossed by a 
sound wave on a timescale shorter than the cluster's age \citep[e.g.][]{ettori+13}, as generally occurs in clusters that are not undergoing 
any major merger and can be considered relaxed from a dynamical point of  view.

Given this condition, the ability to resolve the mass distribution in X-rays depends on the ability to make spatially resolved measurements of the gas temperature and density.
This ability is limited severely by many systematic effects, such as the presence of gas inhomogeneities on small and large scales 
that can bias the measurements of the gas density \citep[e.g.][]{roncarelli+13}, contamination from the X-ray background that reduce the signal-to-noise ratio 
and the corresponding robustness of the spectral measurements \citep[e.g.][]{em11,reiprich+13}; adopted assumptions, for example  on  the halo
geometry \citep[see e.g.][]{buote12b,ser17} and the dynamical state \citep[see e.g.][]{nel14bias,bif16} of the objects, where mergers 
are the major source of violation of the hydrostatic equilibrium assumption \citep[e.g.][]{khatri+16}; and
inhomogeneities in the temperature distribution, which also play  a role \citep{rasia+12}.
Within $0.15\,R_{500}$, turbulence driven by the active galactic nucleus feedback \citep[e.g.][]{gaspari18} is instead responsible for the major deviations.

In recent years, the great improvement in instrumentation in detecting and characterizing the SZ signal has also allowed us to use  this piece of information to resolve
the pressure profile of the intracluster plasma.
Our \xmm\ Cluster Outskirts Project \citep[X-COP;][]{eck17xcop} has now completed the joint analysis of \xmm\ and \planck\ exposures 
of 13 nearby massive galaxy clusters, permitting us to resolve X-ray and SZ signals out to the virial radius ($\sim R_{200}$), also correcting for gas clumpiness
as resolved with the \xmms\ point spread function of about 15 arcsec ($\sim$ 17 kpc at the mean redshift of our sample of 0.06).
In this paper, which appears as a companion to the extended analysis of the X-COP sample presented in 
Ghirardini et al. (2018b, on the thermodynamical properties of the sample)
and Eckert et al. (2018, on the non-thermal pressure support), we report on the reconstruction
of the hydrostatic mass in these systems and how it compares with several independent estimates.

The paper is organized as follows. 
In Section~2 we present the observed profiles of the thermodyncamical properties.
We describe in Section~3 the different techniques adopted to recover the hydrostatic masses.
In Section~4 we compare these dark matter profiles with those recovered from scaling laws, weak lensing, and galaxy dynamics.
A comparison with predictions from the modified Newtonian dynamics and the emergent gravity scenario is discussed in Section~5.
We summarize our main findings in Section~6.
Unless mentioned otherwise, the quoted errors are statistical uncertainties at the $1 \sigma$ confidence level.

In this study, we often refer to radii, $R_{\Delta}$, and masses,  $M_{\Delta}$, which are the corresponding values estimated at the given overdensity $\Delta$ 
as $M_{\Delta} = 4/3 \, \pi \, \Delta \, \rho_{\rm c,z} R_{\Delta}^3$, where $\rho_{\rm c,z} = 3 H_z^2 / (8 \pi G)$ is the critical density of the universe 
at the observed redshift $z$ of the cluster, and 
$H_z = H_0 \, \left[\Omega_{\Lambda} +\Omega_{\rm m}(1+z)^3\right]^{0.5} = H_0 \, h_z$ is the value of the Hubble constant 
at the same redshift.
For the $\Lambda CDM$ model, we adopt the cosmological parameters
$H_0=70$ km s$^{-1}$ Mpc$^{-1}$ and $\Omega_{\rm m} = 1 - \Omega_{\Lambda}=0.3$.

\section{The X-COP sample}

The   \xmm\ Cluster Outskirts Project \citep[X-COP,][]{eck17xcop} is an \xmms\ Very Large Program (PI: Eckert) 
dedicated to the study of the X-ray emission in cluster outskirts.
It targets 12 local, massive galaxy clusters selected for their high signal-to-noise ratios in the \planck\ all-sky SZ survey
\citep[S/N> 12 in the PSZ1 sample,][]{SZcatalog} as resolved sources ($R_{500} >$ 10 arcmin)
in the redshift range $0.04 < z < 0.1$ and along the directions with a galactic absorption lower than $10^{21}$ cm$^{-2}$ to avoid
any significant suppression of the X-ray emission in the soft band where most of the spatial analysis is performed. 
These selection criteria guarantee that a joint analysis of the X-ray and SZ signals allows
the reconstruction of the ICM properties out to $R_{200}$ for all our targets. 
A complete description of the reduction and analysis of our proprietary X-ray data and of the SZ data for X-COP is provided in \citet{ghi18univ} \citep[see also][]{eckert18}.
We note here that our new \planck\ pressure profiles are extracted using exactly the same method as in \cite{planck13}, 
but on the \planck\ 2015 data release (i.e. the full intensity survey) which has a higher S/N (due to the higher exposure time) 
and improved data processing and calibration.
To the original X-COP list of 12 clusters, we added HydraA (Abell~780) due to the availability of a high-quality \xmm\ mapping \citep{degrandi+16}.
The X-ray signal extends out to $R_{200}$ and therefore it is possible to recover its hydrostatic mass with very high precision, 
even though the \planck\ signal for this cluster is contaminated by a bright central radio source.

\begin{figure}
\begin{center} 
\includegraphics[width=0.48\textwidth, keepaspectratio]{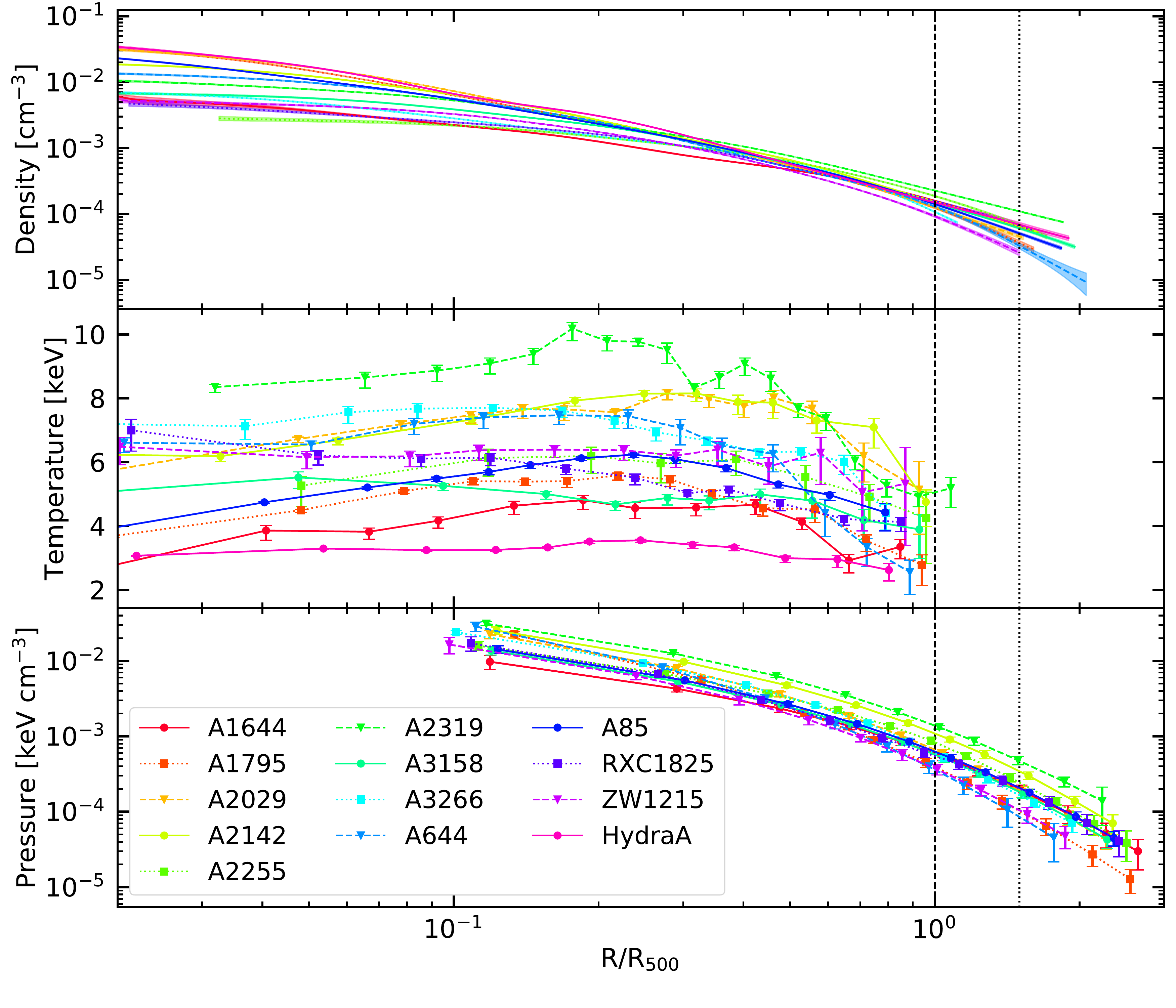}
\end{center} 
\caption{(From top to bottom) Reconstructed electron density, projected temperature, and SZ pressure profiles as functions of the radius in scale units of $R_{500}$.
Statistical error bars are overplotted.
The two vertical lines indicate $R_{500}$ and $R_{200}$.
} \label{fig:prof}
\end{figure}

Here, we summarize the properties of the quantities of interest for the reconstruction of the gravitating mass distribution.
The physical quantities directly observable are the density $n_{\rm gas}$ and temperature $T_{\rm gas}$ of the X-ray emitting gas, 
and the SZ pressure $P_{\rm gas}$ of the same plasma.
Their radial profiles are presented in Fig.~\ref{fig:prof}.
The gas density is obtained from the geometrical deprojection of the X-ray surface brightness out to a mean value of $1.8 R_{500}$.
Thanks to the observational strategy implemented in X-COP, we are able to correct the X-ray emission for the presence of gas clumps 
both by masking substructures spatially resolved with \xmm\ and by measuring the azimuthal median, instead of the azimuthal mean
\citep{zhu13,eckert+15}.
The estimates of the gas temperature are based on the modelling with an absorbed thermal component of the \xmm\ spectra extracted from 
concentric annuli around the X-ray peak in the [0.5--12] keV energy band and corrected from the local sky background components.
A typical statistical error lower than 5\% is associated with  these spectral measurements.
On average, the temperature profiles are resolved out to $0.9 R_{500}$.
The SZ electron pressure profile is obtained following the method described in Planck Collaboration  (2013) from the \planck\ PSF deconvolution and geometrical deprojection
of the azimuthally averaged integrated Comptonization parameter $y$ extracted from a re-analysis of the SZ signal mapped with \planck\ and that extends up to $\sim 2.3 R_{500}$. 

\section{Total gravitating mass from the hydrostatic equilibrium equation}
\label{sect:hee}

Under the assumption that the intracluster medium has a spherically symmetric distribution and follows the equation of state for a perfect gas,
$P_{\rm gas} = k T_{\rm gas} n_{\rm gas}$, where $k$ is the Boltzmann constant, the combination of the gas density $n_{\rm gas}$,
as the sum of the electron and proton densities $n_{\rm e} + n_{\rm p} \approx 1.83 n_{\rm e}$,
with the X-ray spectral measurements of the gas temperature and/or the SZ derived gas pressure,
allows us to evaluate the total mass within a radius $r$ through the hydrostatic equilibrium equation \citep[see e.g.][]{ettori+13}
\begin{equation}
M_{\rm tot}(<r) = - \frac{r \, P_{\rm gas}}{\mu m_{\rm u} G \, n_{\rm gas}} \frac{d \log P_{\rm gas}}{d \log r}, 
   \label{eq:mhe}
\end{equation}
where $G$ is the gravitational constant, $m_{\rm u} = 1.66 \times 10^{-24}$ g is the atomic mass unit, 
and $\mu= \rho_{\rm gas} / (m_{\rm u} n_{\rm gas}) \approx (2 X +0.75 Y +0.56 Z)^{-1} \approx 0.6$ 
is the mean molecular weight in atomic mass unit for ionized plasma, 
with $X$, $Y$, and $Z$ being the mass fraction for hydrogen, helium, and other elements, respectively 
\citep[$X+Y+Z=1$, with $X \approx 0.716$ and $Y \approx 0.278$ for a typical metallicity of 0.3 times the solar abundance from][]{ag89}.

In the present analysis, we apply both the {backward} and the {forward} method \citep[see e.g.][]{ettori+13},
as discussed and illustrated in our pilot study on A2319 \citep{ghi18}.
We use all the information available (measured pressure $P_{\rm m}$, temperature $T_{\rm m}$, and emissivity $\epsilon_{\rm m}$)
to build a joint likelihood $\mathcal{L}$,
\begin{equation}
\begin{split}
\log \mathcal{L} = &
-0.5 \left[ (P-P_{\rm{m}}) \Sigma_{tot}^{-1} (P-P_{\rm{m}})^T + n \log \left( \det \left( \Sigma_{tot} \right) \right) \right]
\\
 & -0.5  \sum_{i=1}^{n_{T}} \left[ \frac{(T_i-T_{\rm{m},i})^2}{\sigma_{T,i}^2+\sigma_{T,int}^2} + \log \left( {\sigma_{T,i}^2+\sigma_{T,int}^2} \right) \right]
\\
 &  -0.5 \left[ \sum_{j=1}^{n_{\epsilon}} \frac{(\epsilon-\epsilon_{\rm{m},j})^2}{\sigma_{\epsilon,j}^2} \right],
\end{split}
\label{eq:like}
\end{equation}
that combines in the fitting procedure 
(i) the $\chi^2$ related to the measured profiles [i.e. $\chi^2_P = (P-P_{\rm{m}}) \Sigma_{tot}^{-1} (P-P_{\rm{m}})^T$, 
$\chi^2_T = \sum_{i=1}^{n_{T}} \frac{(T_i-T_{\rm{m},i})^2}{\sigma_{T,i}^2+\sigma_{T,int}^2} $, and 
$\chi^2_{\epsilon} = \sum_{j=1}^{n_{\epsilon}} \frac{(\epsilon-\epsilon_{\rm{m},j})^2}{\sigma_{\epsilon,j}^2}$; 
see Table~\ref{tab:m_chi2}],
(ii) an intrinsic scatter $\sigma_{T,int}$ to account for any tension between X-ray and SZ measurements, and
(iii) the covariance matrix $\Sigma_{tot}$ among the data in the \planck\ pressure profile \citep[for details, see Appendix~D in][]{ghi18}. 

The emissivity $\epsilon$ is obtained from the multiscale fitting \citep[][see also Sect.~2.3 in Ghirardini et al. 2018b]{eckert+16} 
of the observed X-ray surface brightness.
In the {backward} method, a parametric mass model is assumed and combined with the gas density profile to predict 
a gas temperature profile $T$ that is then compared with the one measured $T_{\rm m}$
in the spectral analysis and the one estimated from SZ as $P /n_{\rm gas}$ 
\citep[losing the spatial resolution in the inner regions because of the modest 7 arcmin FWHM angular 
resolution of our \planck\ SZ maps, but gaining in radial extension due to the \planck\ spatial coverage;][]{planck13}
to constrain the mass model parameters. 
In the {forward} method, some functional forms are fitted to the deprojected gas temperature and pressure profiles,
as detailed in \citet{ghi18}, with instead no assumptions on the form of the gravitational potential.
We note that we neglect the three innermost \planck\ points for the analysis to avoid possible biases induced by the \planck\ beam.
The hydrostatic equilibrium equation (Eq.~\ref{eq:mhe}) is then directly applied to evaluate the radial distribution of the mass.

The profiles are fitted using an MCMC approach based on the code \textit{emcee} \citep{emcee} with 10000 steps and 
about 100 walkers, and throwing away the first 5000 points because of `burnt-in' time. 
From the resulting posterior distribution on our parameters, we estimate the reference values using the median of the distributions, and the errors 
as half the difference between the 84th and 16th percentiles.
The best-fit parameters 
In the present analysis, we investigate different mass models (Sect.~\ref{subsect:other}), and adopt 
as reference model a NFW mass model with two free parameters, the mass concentration and $R_{200}$ (see Sect.~\ref{subsect:nfw}).

\subsection{Comparison among different mass models with the {backward} method}
\label{subsect:other}

We apply the {backward} method with the following set of different mass models and estimate their maximum likelihood  
in reproducing the observed profiles of gas density, temperature and SZ pressure.

The mass profile is parametrized through the expression 
\begin{align}
M(<r) = & \,  n_0  \, r_s^3 \, f_c \, F(x) \nonumber \\
n_0 = & \, \frac{4}{3} \, \pi \, \Delta \, \rho_{\rm c,z}  = 1.14 \times 10^{14} \, h_z^2 \; {\rm M}_{\odot} {\rm Mpc}^{-3}   \nonumber \\
f_c = & \, \frac{c^3}{\log(1+c) - c/(1+c)}
\label{eq:m_mod}
\end{align}
where $\Delta = 200$, $h_z = H_z/ H_0$, and $x = r / r_{\rm s}$, with the scale radius $r_{\rm s}$ and the concentration $c$
being the two free parameters of the fit. 
The function $F(x)$ characterizes each mass model and is defined as follows:

\begin{description}
\item[NFW] $F(x) = \log(1+x) -x/(1+x)$ \citep{nfw97};
\item[EIN]  $F(x) =  a^{1-a_1} / 2^{a_1} e^{a_0} \gamma(a_1, a_0 x^{1/a})$ with $a=5$,  $a_0=2 n$, $a_1=3 n$, and $\gamma(a, y)$ being 
the incomplete gamma function equal to $\int_0^y t^{a-1} \exp(-t) dt$  \citep[from eq.~A2 in][]{ml05};
\item[ISO] $F(x) =  \log(x+\sqrt{1+x^2}) -\frac{x}{\sqrt{1+x^2}}$, which is the King approximation to the isothermal sphere \citep{king62};
\item[BUR] $F(x) =  \log(1+x^2) +2 \log(1+x) -2 \arctan(x)$ \citep{sb00};
\item[HER] $F(x) =  \frac{x^2}{(x+1)^2}$ \citep{her90}.
\end{description}

We note that our observed profiles cannot provide any robust constraint on the third parameter $a$ of the 
Einasto profile that is therefore fixed to a value of 5 as observed for massive halos \citep[e.g.][]{dutton14}.
Moreover, the parameter $c$ is defined as the `concentration' in the NFW profile, whereas it represents a way to constrain the
normalization for the other mass models. In our MCMC approach, we adopt for $c$ a uniform a-prior distribution in the linear space in the range 0.1--15.
The a-prior distributions on the scale radius (or $R_{200} = c \times r_{\rm s}$ for the NFW case) are still defined as uniform in the linear space in the following ranges: 
1--3 Mpc (NFW); 0.1--2.8 Mpc (EIN); 0.02--0.8 Mpc (ISO); 0.02--0.8 Mpc (BUR); 0.2--3 Mpc (HER).

We show all the mass profiles in Fig.~\ref{fig:m_prof}.
In Table~\ref{tab:m_mod}, we quote all the best-fit parameters, and the relative Bayesian Evidence $E$ estimated, for each mass model, as the integral of the likelihood function 
$\mathcal{L}$ (equation~\ref{eq:like}) over the a-prior distributions $P(\boldsymbol{\theta})$ of the parameters $\boldsymbol{\theta}$ 
\citep[$E =  \int \mathcal{L(\boldsymbol{\theta})} P(\boldsymbol{\theta}) d\boldsymbol{\theta}$; as implemented in e.g. {\tt MultiNest},][]{multinest}. 

\begin{figure}
\begin{center}
\includegraphics[width=0.48\textwidth, keepaspectratio]{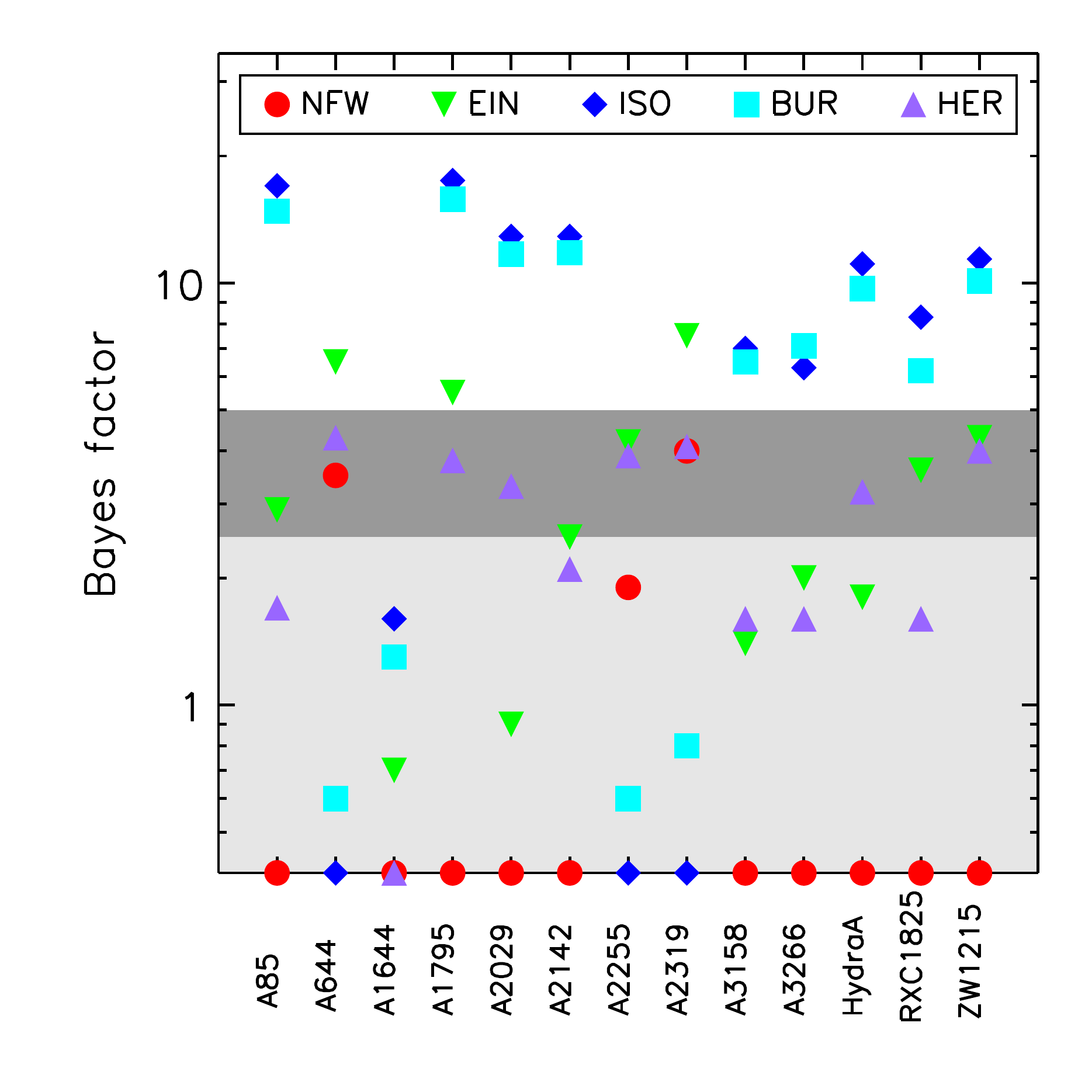}
\end{center} \vspace*{-1cm}
\caption{
Bayes factor of the mass models investigated with respect to the one with the highest evidence (see Table~\ref{tab:m_mod}).
Shaded regions identify values of the Bayes factor where the tension between the models is 
either weak ($<2.5$) or strong ($>5$) according to the Jeffreys scale \citep{jef61}.
} \label{fig:evi}
\end{figure}

In Fig.~\ref{fig:evi}, we present the Bayes factor estimated for each object as the difference between 
the logarithm of the Bayesian evidence of the mass model with the highest evidence with respect to the others.
Nine out of 13 objects prefer a NFW model fit and have data that are significantly inconsistent (Bayes factor $>5$) 
with an isothermal/Burkert mass model. 
The remaining four objects prefer different mass models (ISO for A2255 and A2319, HER for A1644, and BUR for A644), but do
not show any statistically significant (Bayes factor $<5$) tension with NFW.
Of those four objects, two (A2255 and A2319) are the ones with the highest value of the central entropy among the X-COP sample as estimated in \cite{cavagnolo+09},
strongly suggesting that they are disturbed systems that have produced a core (well modelled by an ISO profile) in the gravitational potential as a consequence of the ongoing merger.
The case of A2319 has been indeed extensively studied in \cite{ghi18}. 
A2255 shows a flat surface-brightness profile in its core and an X-ray morphology extended along the E-W axis \citep{sak_pon06}. 
Interestingly, the X-ray centroid of the cluster does not coincide with a dominant galaxy \citep{burns95}, which more strongly supports  the merger scenario. 
The cluster also exhibits  an unusual polarized radio halo which may be a radio relic seen in projection \citep{pizzo09}.
A644 is an unusual radio-quiet cool-core cluster with a temperature profile rising inward and the cD galaxy positioned approximately between the X-ray peak and centroid, 
as expected for a merger origin of the properties of the X-ray emission in the core \citep{buote05}. We also note that it shows a steep gradient 
in the outermost points of the spectral temperature profile, which also  suggests  that the object can be dynamically non-relaxed in the regions at $r >0.4 R_{500}$. 
We do not find any peculiarity in the observed profiles of A1644, confirming that, as also probed in the numerical simulations, a
NFW profile provides a good description of the spherically averaged equilibrium density profiles of CDM halos; however, deviations are  expected.

\subsection{Reference mass model: {backward} method with NFW mass model}
\label{subsect:nfw}

\begin{figure}
\begin{center}
\includegraphics[width=0.48\textwidth, keepaspectratio]{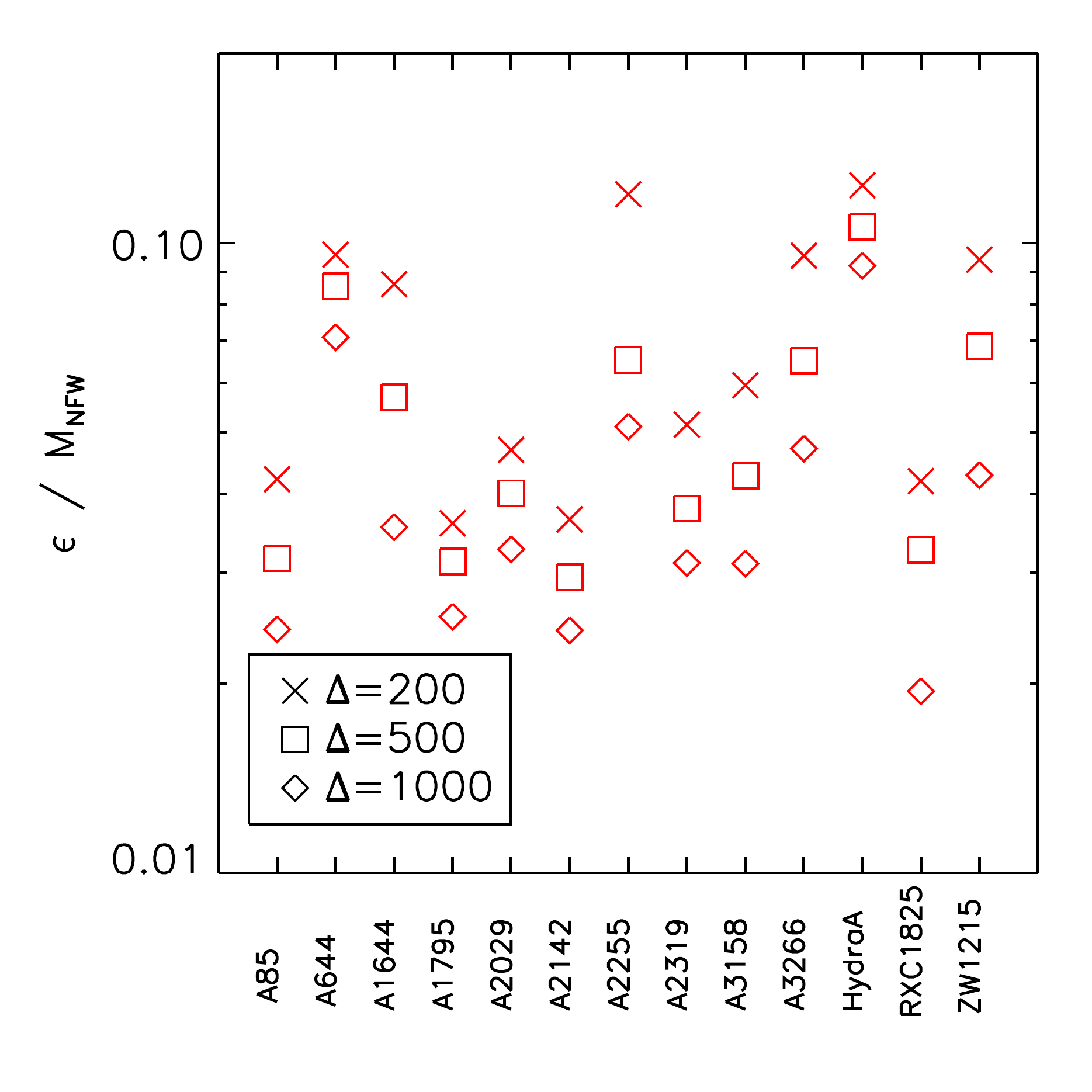}
\end{center} \vspace*{-1cm}
\caption{
Relative error (at $1 \sigma$) on the hydrostatic mass recovered with the {backward} method and a NFW model (see Table~\ref{tab:mass}).
} \label{fig:stat}
\end{figure}

We present the results of our analysis with a {backward} method and a NFW mass model in Table~\ref{tab:mass}.
We measure mean relative errors (statistical only)  lower than 8\% (median, mean, and dispersion at $\Delta=1000$, $500$, and $200$, respectively: 
$3, 4, 2$\%; $4, 5, 2$\%; $6, 7, 3$\%;  see Fig.~\ref{fig:stat}).

\begin{table*}
\begin{center}
\setlength{\tabcolsep}{3pt} 
\begin{tabular}{c c c ccc cccc} \hline \hline
Name & $z$ & $c_{200}$ & 0.5 Mpc & 1 Mpc & 1.5 Mpc & $R_{500}$ & $M_{500}$ & $R_{200}$ & $M_{200}$ \\
 & & & $10^{14} M_{\odot}$ & $10^{14} M_{\odot}$ & $10^{14} M_{\odot}$ & Mpc & $10^{14} M_{\odot}$ & Mpc & $10^{14} M_{\odot}$ \\
A85 & 0.0555 & $3.31^{+0.13}_{-0.13}$ & $1.94 \pm 0.02$ & $4.53 \pm 0.08$ & $6.81 \pm 0.18$ & $1.235 \pm 0.013$ & $5.65 \pm 0.18$ & $1.921 \pm 0.027$ & $8.50 \pm 0.36$ \\ 
A644 & 0.0704 & $5.58^{+0.65}_{-0.51}$ & $2.36 \pm 0.09$ & $4.74 \pm 0.26$ & $6.59 \pm 0.45$ & $1.230 \pm 0.035$ & $5.66 \pm 0.48$ & $1.847 \pm 0.059$ & $7.67 \pm 0.73$ \\ 
A1644 & 0.0473 & $1.46^{+0.14}_{-0.14}$ & $1.15 \pm 0.02$ & $3.24 \pm 0.10$ & $5.47 \pm 0.26$ & $1.054 \pm 0.020$ & $3.48 \pm 0.20$ & $1.778 \pm 0.051$ & $6.69 \pm 0.58$ \\ 
A1795 & 0.0622 & $4.55^{+0.16}_{-0.14}$ & $1.95 \pm 0.02$ & $4.06 \pm 0.08$ & $5.77 \pm 0.15$ & $1.153 \pm 0.012$ & $4.63 \pm 0.14$ & $1.755 \pm 0.021$ & $6.53 \pm 0.23$ \\ 
A2029 & 0.0773 & $4.26^{+0.19}_{-0.17}$ & $2.78 \pm 0.03$ & $6.25 \pm 0.13$ & $9.24 \pm 0.26$ & $1.423 \pm 0.019$ & $8.82 \pm 0.35$ & $2.173 \pm 0.034$ & $12.57 \pm 0.59$ \\ 
A2142 & 0.0909 & $3.14^{+0.10}_{-0.10}$ & $2.48 \pm 0.02$ & $6.08 \pm 0.09$ & $9.44 \pm 0.19$ & $1.424 \pm 0.014$ & $8.95 \pm 0.26$ & $2.224 \pm 0.027$ & $13.64 \pm 0.50$ \\ 
A2255 & 0.0809 & $1.37^{+0.24}_{-0.23}$ & $1.39 \pm 0.06$ & $4.08 \pm 0.11$ & $7.10 \pm 0.36$ & $1.196 \pm 0.026$ & $5.26 \pm 0.34$ & $2.033 \pm 0.081$ & $10.33 \pm 1.23$ \\ 
A2319 & 0.0557 & $4.86^{+0.51}_{-0.37}$ & $2.61 \pm 0.08$ & $5.58 \pm 0.12$ & $8.01 \pm 0.25$ & $1.346 \pm 0.017$ & $7.31 \pm 0.28$ & $2.040 \pm 0.035$ & $10.18 \pm 0.52$ \\ 
A3158 & 0.0597 & $2.88^{+0.26}_{-0.17}$ & $1.59 \pm 0.02$ & $3.76 \pm 0.09$ & $5.70 \pm 0.21$ & $1.123 \pm 0.016$ & $4.26 \pm 0.18$ & $1.766 \pm 0.035$ & $6.63 \pm 0.39$ \\ 
A3266 & 0.0589 & $2.04^{+0.25}_{-0.20}$ & $2.02 \pm 0.05$ & $5.57 \pm 0.15$ & $9.32 \pm 0.39$ & $1.430 \pm 0.031$ & $8.80 \pm 0.57$ & $2.325 \pm 0.074$ & $15.12 \pm 1.44$ \\ 
HydraA & 0.0538 & $5.51^{+0.67}_{-0.61}$ & $1.28 \pm 0.07$ & $2.40 \pm 0.20$ & $3.22 \pm 0.32$ & $0.904 \pm 0.032$ & $2.21 \pm 0.23$ & $1.360 \pm 0.056$ & $3.01 \pm 0.37$ \\ 
RXC1825 & 0.0650 & $3.35^{+0.20}_{-0.19}$ & $1.64 \pm 0.02$ & $3.69 \pm 0.07$ & $5.46 \pm 0.15$ & $1.105 \pm 0.012$ & $4.08 \pm 0.13$ & $1.719 \pm 0.024$ & $6.15 \pm 0.26$ \\ 
ZW1215 & 0.0766 & $2.11^{+0.22}_{-0.18}$ & $1.93 \pm 0.03$ & $5.22 \pm 0.15$ & $8.61 \pm 0.39$ & $1.358 \pm 0.031$ & $7.66 \pm 0.52$ & $2.200 \pm 0.069$ & $13.03 \pm 1.23$ \\ 
\hline \end{tabular}
\end{center}
\caption{Values of the total gravitating mass as estimated with the {backward} method and a NFW model at some radii of reference (0.5, 1, 1.5 Mpc) 
and at the overdensities of 500 and 200 with respect to the critical density of the universe at the cluster's redshift.
} \label{tab:mass}
\end{table*}

In the cold dark matter scenario, the structure formation is hierarchical and allows the build-up of the most massive gravitationally bound 
halos, like galaxy clusters, only at later cosmic times.
Considering that the central density of halos reflects the mean density of the Universe at the time of formation, halos with increasing mass
are expected to have lower mass concentration at given redshift \citep[e.g.][]{nfw97,bul01,dol04,dk15}.

We can investigate how our best-fit results on NFW concentration and $M_{200}$ (quoted in  Table~\ref{tab:mass}) reproduce 
the predictions from numerical simulations. We can also assess the level of the intrinsic scatter in the $c_{200}-M_{200}$ relation in this mass range.
Simulations suggest that this scatter is related to the variation in formation time and is expected to be lower in more massive halos 
that formed more recently \citep[e.g.][]{neto07}.

We model the relation with a standard power law:
\begin{equation}
\label{eq_cM_1}
c_{200} =10^\alpha \left( \frac{M_{200}}{M_\text{pivot}}\right)^\beta.
\end{equation}
The intrinsic scatter $\sigma_{c|M}$ of the concentration around a given mass, $c_{200}(M_{200})$, is taken to be lognormal \citep{duf+al08, bha13}. 

We fit the data with a linear relation in decimal logarithmic ($\log$) variables with the \textsc{R}-package \texttt{LIRA}\footnote{\texttt{LIRA} (LInear Regression in Astronomy) is available from the Comprehensive R Archive Network at \url{https://cran.r-project.org/web/packages/lira/index.html}.}. 
\texttt{LIRA} is based on a Bayesian hierarchical analysis which can deal with heteroscedastic and correlated measurements uncertainties, intrinsic scatter, scattered mass proxies, and time-evolving mass distributions \citep{ser16_lira}. 

The mass distribution of the fitted clusters has to be properly modelled to address Malmquist and Eddington biases \citep{kel07}. The Gaussian distribution can provide an adequate modelling 
\citep{se+et15_comalit_IV}. The parameters of the distribution are found within the regression procedure. 
This scheme is fully effective in modelling both selection effects at low masses and the steepness of the cosmological halo mass function at large masses.

Performing an unbiased analysis of the concentration--mass ($c$-$M$) relation requires properly addressing the uncertainties connected to the correlations and intrinsic scatter. 
Measured mass and concentration are indeed strongly anti-correlated, causing  the $c$-$M$ relation to appear steeper \citep{aug+al13,dutton14,du+fa14,ser+al15_cM}. 
By correcting for this effect, it is possible to obtain a more precise, significantly flatter relation \citep{ser+al15_cM}. 
On the other hand, the intrinsic scatter of the measurable mass with respect to the true mass can bias the estimated slope towards flatter values \citep{rasia+13,se+et15_comalit_I}. 
To correct for this effect, we measure the concentration--mass uncertainty covariance matrix for each cluster from the MCMC chains, whereas 
we model the intrinsic scatter as a free fit parameter to be found in the regression procedure.

\begin{figure*}[ht]
\centering
\includegraphics[width=0.46\hsize]{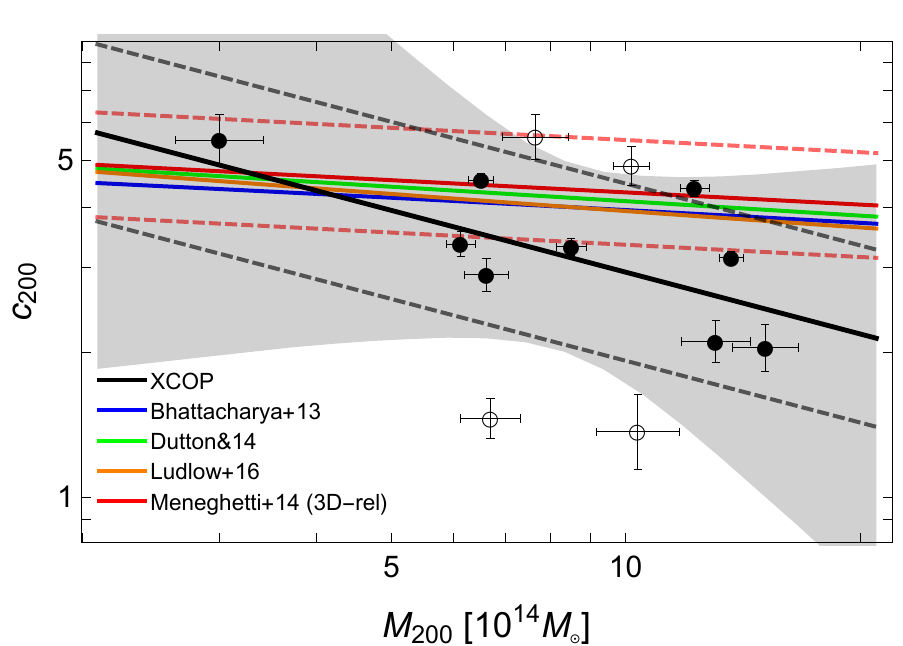} 
\includegraphics[width=0.5\hsize]{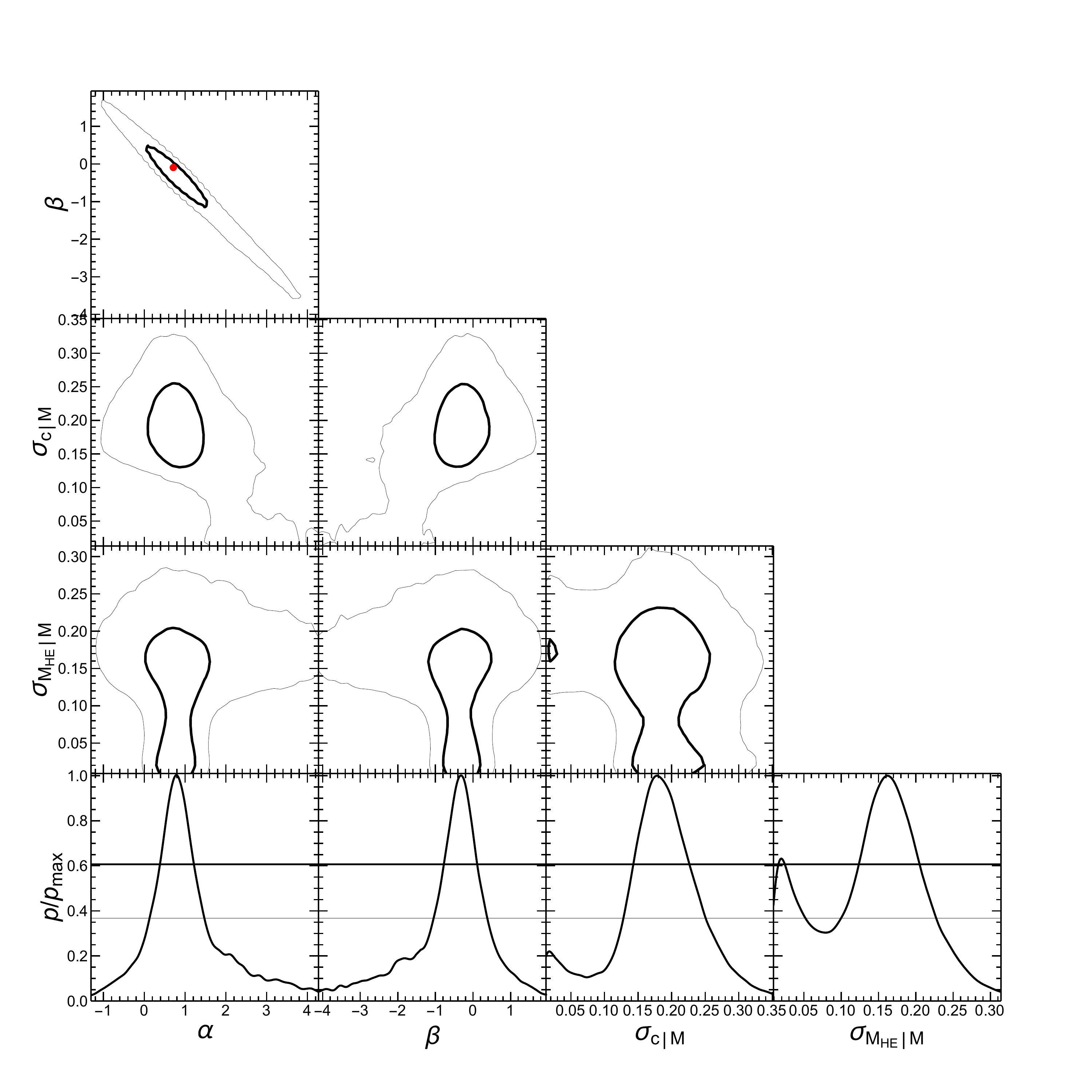}
\caption{(Left) Concentration--mass relation of the X-COP clusters. The dashed black lines show the median scaling relation (full black line) plus or minus the intrinsic scatter at redshift $z=0.06$. 
The shaded grey region encloses the $68.3\%$  confidence region around the median relation due to uncertainties on the scaling parameters. 
As reference, we plot the concentration--mass relations of \citet[][blue line]{bha13}, \citet[][green line]{dutton14}, \citet[][orange line]{lud+al16}, and \citet[solid and dashed red lines]{men+al14}.
The dashed red lines enclose the 1-$\sigma$ scatter region in the theoretical concentration--mass relation from the MUSIC-2 $N$-body/hydrodynamical simulations.
Empty circles identify the four objects  (A644, A1644, A2255, A2319; see Fig.~\ref{fig:evi}) for which NFW does not represent the best-fit mass model.
(Right) Probability distributions of the parameters of the concentration--mass relation. The thick and thin black contours include the 1-$\sigma$ and 2-$\sigma$ confidence regions in two dimensions, here defined as the regions within which the probability is larger than $\exp(-2.3/2)$ and $\exp(-6.17/2)$ of the maximum, respectively. The red disk represents the parameters found by \citet{men+al14} for the relaxed sample. The bottom row plots the marginalized 1D distributions, renormalized to the maximum probability. The thick and thin black levels denote the confidence limits in one dimension, i.e. $\exp(-1/2)$ and $\exp(-4/2)$ of the maximum. 
}
\label{fig:M200_c200}
\end{figure*}

Adopting non-informative priors \citep{ser+al17_psz2lens} we find $\alpha=0.89\pm0.90$ and $\beta=-0.42\pm0.98$, in agreement with theoretical predictions for slope and normalization
(see Fig.~\ref{fig:M200_c200}).
Statistical uncertainties are very large and we cannot discriminate between the different theoretical predictions. 

On the other hand, mass measurement uncertainties are very small and we can estimate the intrinsic scatter of the hydrostatic masses, $\sigma_{M_\text{HE}|M}=0.15\pm0.08$. 
Even though the marginalized posterior distribution of $\sigma_{M_\text{HE}|M}$ is peaked at $\sim 0.15$ (see Fig.~\ref{fig:M200_c200}), smaller values are fully consistent. 
The posterior probability that $\sigma_{M_\text{HE}|M}$ is less than 10\% (i.e.  $\sigma_{M_\text{HE}|M}<0.043$) is 15\%. 
Our estimate of hydrostatic mass scatter is in agreement with results from higher-$z$ samples \citep{se+et15_comalit_I} and larger, 
even though compatible within uncertainties, with results from numerical simulations \citep{rasia+12}.

The intrinsic scatter of the $c$-$M$ relation, $\sigma_{c|M}=0.18\pm 0.06$, is compatible with theoretical predictions \citep[$\sigma_{c|M}\sim0.15$]{bha13,men+al14}
and previous observational constraints \citep[e.g.][]{ettori+10,mantz16cm}.
The relation between mass and concentration for subsamples can differ from the general relation due to selection effects. 
Intrinsically overconcentrated clusters may be over-represented in a sample of clusters selected according to their large Einstein radii or 
to the apparent X-ray morphology \citep{men+al14,ser+al15_cM}. Intrinsic scatter of the  $c$-$M$ relation for relaxed samples is expected to be smaller 
than for the full population of mass selected halos. However, given the statistical uncertainty on the measured $\sigma_{c|M}$, 
we cannot draw any firm conclusion on the equilibrium status of the clusters.

\subsection{Comparison of different methods and systematic errors}

To evaluate some of the systematic uncertainties affecting our measurements of the hydrostatic mass, 
we estimate the mass at some fixed physical radii (500, 1000, and 1500 kpc) and at two overdensities ($R_{500}$ and $R_{200}$) 
using the {forward} method and the other mass models described in Section~\ref{subsect:other}, and
compare them to the results obtained from our model of reference ({backward} NFW).
We summarize the results of this comparison in Table~\ref{tab:syst}.

We observe that the use of the {forward} method (with or without SZ profiles) introduces a small systematic error  at any radius,
with a median difference of about 4\% at $R_{500}$, and $<$2\% (1st-3rd quartile: -5.4 / +8.1 \%) at $R_{200}$.

Any other mass model constrained with the {backward} method introduces some systematic uncertainties that
depends mainly on the shape characteristic of the model and on the fact that all the models have  only two parameters, 
implying that it is not flexibile enough to accommodate the distribution in the observed profiles.
For instance, we note that cored profiles, like ISO and BUR, present larger positive (negative) deviations in the core (outskirts)
up to about 25\%.

\begin{table*}[ht]
\begin{center}
\setlength{\tabcolsep}{2pt} 
\begin{tabular}{c c c c c c} \hline \hline
$M_i$ & \multicolumn{5}{c}{$B$ (inter-quartile range) \%} \\
  &  0.5 Mpc & 1 Mpc & 1.5 Mpc & $R_{500}$ & $R_{200}$ \\ \\
Forw & $+0.6 \, (-1.1/+3.3)$  & $-2.0 \, (-5.9/+1.4)$  & $-4.6 \, (-7.9/+1.3)$  & $-4.7 \, (-10.9/-0.5)$  & $+1.2 \, (-5.4/+8.1)$  \\
Forw (no SZ) & $-1.5 \, (-3.4/+4.1)$  & $-1.9 \, (-8.0/+0.6)$  & $-1.3 \, (-5.4/+4.5)$  & $-4.2 \, (-9.0/+1.8)$  & $+1.3 \, (-11.9/+8.0)$  \\
 \\
EIN & $-0.3 \, (-1.7/+1.3)$  & $-1.7 \, (-6.5/-0.2)$  & $-1.0 \, (-9.7/+1.5)$  & $-0.8 \, (-7.6/+1.0)$  & $-0.8 \, (-10.3/+4.3)$  \\
ISO & $+14.1 \, (+11.8/+21.0)$  & $-3.0 \, (-3.5/+5.4)$  & $-13.3 \, (-19.0/-9.7)$  & $-8.2 \, (-13.0/-5.3)$  & $-23.5 \, (-28.7/-16.5)$  \\
BUR & $+11.4 \, (+10.4/+15.1)$  & $-3.1 \, (-5.8/+4.0)$  & $-13.7 \, (-19.2/-8.3)$  & $-8.3 \, (-12.9/-5.2)$  & $-20.8 \, (-24.2/-17.9)$  \\
HER & $+1.6 \, (+0.9/+2.1)$  & $-0.7 \, (-5.6/+0.2)$  & $-5.5 \, (-11.4/-2.4)$  & $-3.7 \, (-5.3/-1.9)$  & $-9.3 \, (-13.5/-6.8)$  \\
\hline \end{tabular}
\end{center}
\caption{Systematic differences between the {forward} method (`Forw') and the other mass models described in Sect.~\ref{subsect:other}
with respect to the model of reference defined as {backward} NFW.
These differences are quoted as the median (1st-3rd quartiles, in brackets) of the quantity $B = (M_i / {\rm NFW} -1) \times 100$ \%, where
$M_i$ is listed in the first column.
} \label{tab:syst}
\end{table*}

We note that this budget of the systematic uncertainties does not include other sources of error, such as 
any non-thermal contribution to the total gas pressure \citep[e.g.][]{nel14pnt,ser17}, and
terms that account for either departures from the hydrostatic equilibrium \citep[e.g.][]{nel14bias,bif16}
or the violation of the assumed sphericity of the gas distribution \citep[e.g.][]{ser17}.
All these contributions have been shown to affect  the clusters' outskirts more significantly and tend to bias the total mass estimates higher (by 10--30\%) 
 at $r>R_{500}$, with smaller effects in the inner regions. 
In particular, by imposing the distribution of the cluster mass baryon fraction estimated in the state-of-art hydrodynamical cosmological simulations,
we evaluate in Eckert et al. (2018) a median value of about 6\% and 10\% at $R_{500}$ and $R_{200}$, respectively,
for the relative amount of non-thermal pressure support in the X-COP objects.

\section{Comparison with mass estimates from scaling laws, weak lensing, and galaxy dynamics}

\begin{figure}
\begin{center}
\includegraphics[width=0.48\textwidth, keepaspectratio]{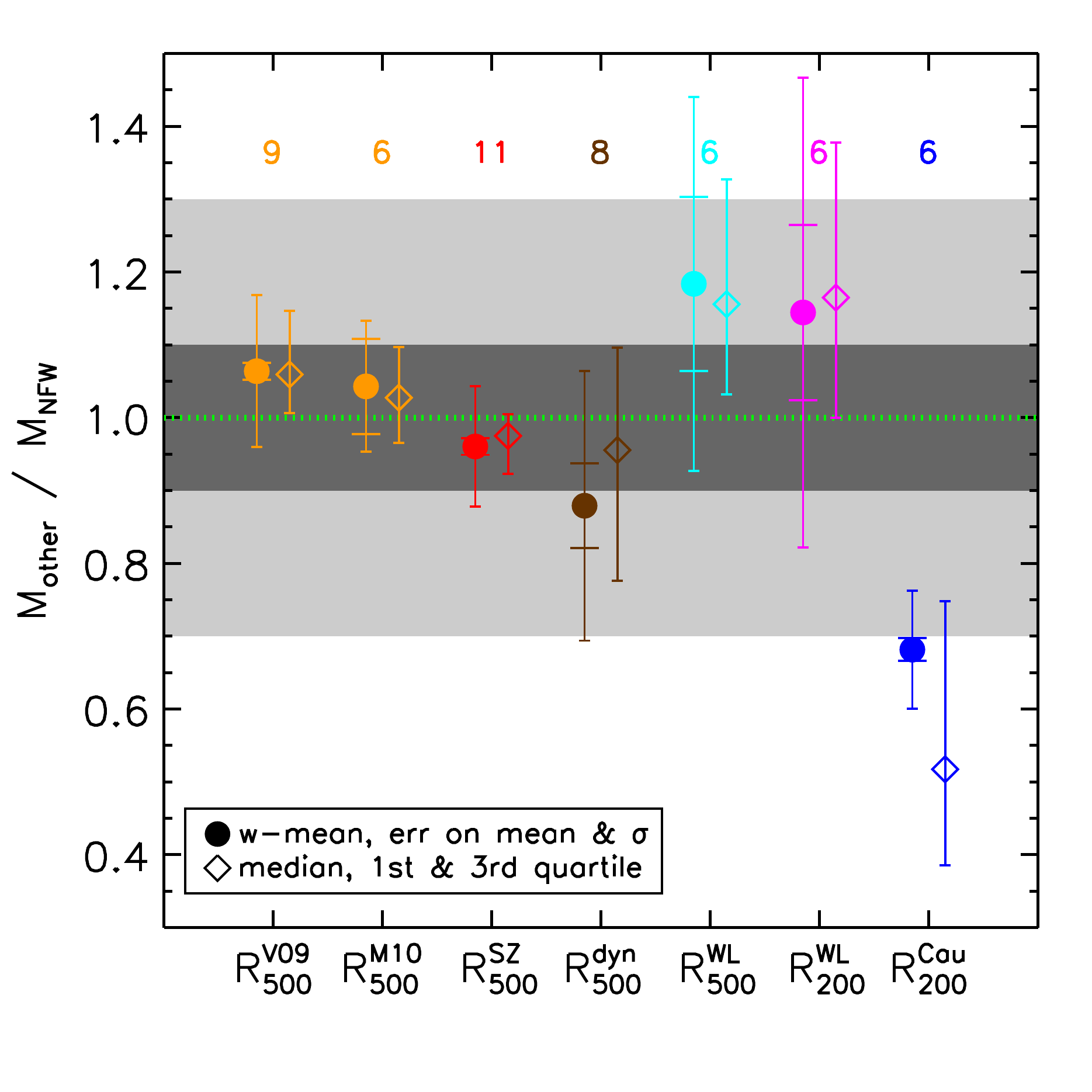}
\end{center} \vspace*{-0.5cm}
\caption{
Comparison between the {backward} NFW model and estimates of the mass from 
X-ray scaling relations ($Y_X$ from \cite{vikhlinin09} --label on X-axis: $R_{500}^{V09}$-- 
and $f_{\rm gas}$ from \cite{mantz10} --$R_{500}^{M10}$),
\planck\ PSZ2 catalog \citep[][; $R_{500}^{SZ}$]{planck15-24}, 
dynamical analysis for the HIFLUGCS sample \citep[][; $R_{500}^{dyn}$]{zhang17},
lensing (Herbonnet et al. 2018 in prep.; $R_{500}^{WL}$ and $R_{200}^{WL}$), 
caustics \citep[][; $R_{200}^{Cau}$]{rines16}.
The number of objects in common is shown.
Shaded regions indicate the $<$10\% (darkest) and $<$30\% differences.
} \label{fig:bias}
\end{figure}

We compare our estimates of the hydrostatic mass with constraints obtained from 
(i) X-ray based scaling relations applied to $Y_X = M_{\rm gas} \times T$ \citep{vikhlinin09}, 
gas mass fraction \citep[][also including a correction factor of 0.9 for \cxo\ calibration updates; see caption of their Table~4]{mantz10}, and
SZ signal \citep[the \planck\ mass proxy $M_{\rm Y_z}$ in][]{planck15-24}; 
(ii) weak lensing signals associated with the coherent distortion in the observed shape of background galaxies 
(as measured in the Multi Epoch Nearby Cluster Survey, MENeaCS; Herbonnet et al. in prep.); 
and (iii) galaxy dynamics either through the estimate of the velocity dispersion \citep{zhang17} or 
via the caustic method \citep[][; for A2029 we consider the recent measurement in \citealt{sohn18}]{rines16}, 
which calculates the mass from the escape velocity profile which is defined 
from the edges (i.e. the `caustics') of the distribution in the redshift-projected radius diagram.
The comparison is done by evaluating the hydrostatic $M_{\rm NFW}$ at the radius defined by the
other methods at a given overdensity ($\Delta=500$ for X-ray and SZ scaling laws, galaxy dynamics, and WL; 
$\Delta=200$ for WL and caustics).
We plot the medians and the error-weighted means of the ratios $M / M_{\rm NFW}$ in Fig.~\ref{fig:bias}.

Mass estimates based on X-ray scaling laws provide a very reassuring agreement:
we measure a median ratio $M / M_{\rm NFW}$ of 1.06 and 1.03 for the nine and six objects 
in common with \cite{vikhlinin09} and \cite{mantz10},  respectively. 
For the 11 objects in common with \planck-SZ catalog (Hydra-A and Zw1215 are not included there), 
we measure a median (1st-3rd quartile) of 0.98 (0.92--1.01) and an error-weighted mean of 0.96 (r.m.s. 0.08)
\footnote{Given a set of elements $x_i$ with statistical uncertainty $\sigma_i$, we adopt the following quantities 
as error-weighted mean $\bar{x}$, error on it $\epsilon$, and relative dispersion $\sigma_{\bar{x}}$: 
$\bar{x} = \sum_i w_i x_i / \sum_i w_i$; $\epsilon= (\sum_i w_i)^{-1/2}$; 
$\sigma_{\bar{x}} = ( \sum_i w_i (x_i - \bar{x})^2 / \sum_i w_i)^{1/2}$, where $w_i = 1 / \sigma_i^2$.}.

A similarly good agreement is obtained with the six WL measurements in common: 
the medians are 1.16 (error-weighted mean 1.18 $\pm$ 0.12; r.m.s. 0.26)
and 1.17 (error-weighted mean 1.14 $\pm$ 0.12; r.m.s. 0.32)
at $R_{500}$ and $R_{200}$, respectively.

Considering the eight clusters in common with \cite{zhang17}, we measure a median of 0.96, with 1st-3rd quartiles of 0.78--1.10.
On the other hand, a clear tension is measured with respect to the six mass values obtained from caustics: 
$M_{\rm Cau} / M_{\rm NFW}$ has a median of 0.52 (1st-3rd quartiles: 0.39--0.75; error-weighted mean 0.68 $\pm$ 0.02; r.m.s. 0.08), 
with ratios smaller than 0.5 in three objects (A85: 0.43; A2142: 0.39; ZW1215: 0.35), and deviations larger than 3 $\sigma$, when statistical uncertainties
on the single measurements are propagated, on three clusters (A1795: 11.5 $\sigma$; A2029: 6.5 $\sigma$; A2142: 3.9 $\sigma$).
Simulations in \cite{serra+13}, among others, show this might be the case when an insufficient number of spectroscopic member galaxies 
are adopted to constrain the caustic amplitude. 
We also note that caustic masses depend on a calibration constant that is function of the mass density, potential and galaxies' velocity anisotropy profiles.
This constant is generally calibrated with numerical simulations, and its value (between 0.5 and 0.7) has been debated in recent literature
due to its dependence on the dynamical tracers adopted in the caustic technique \citep[see e.g.][]{serra+11, gifford+13, gifford+17}. 
In the present analysis, we rely only on published results and postpone to further investigation, also on the role of this  calibration constant, 
the study of the detected tension with estimates from caustic masses for the X-COP objects.

\begin{figure*}
\begin{center}
\includegraphics[width=0.48\textwidth, keepaspectratio]{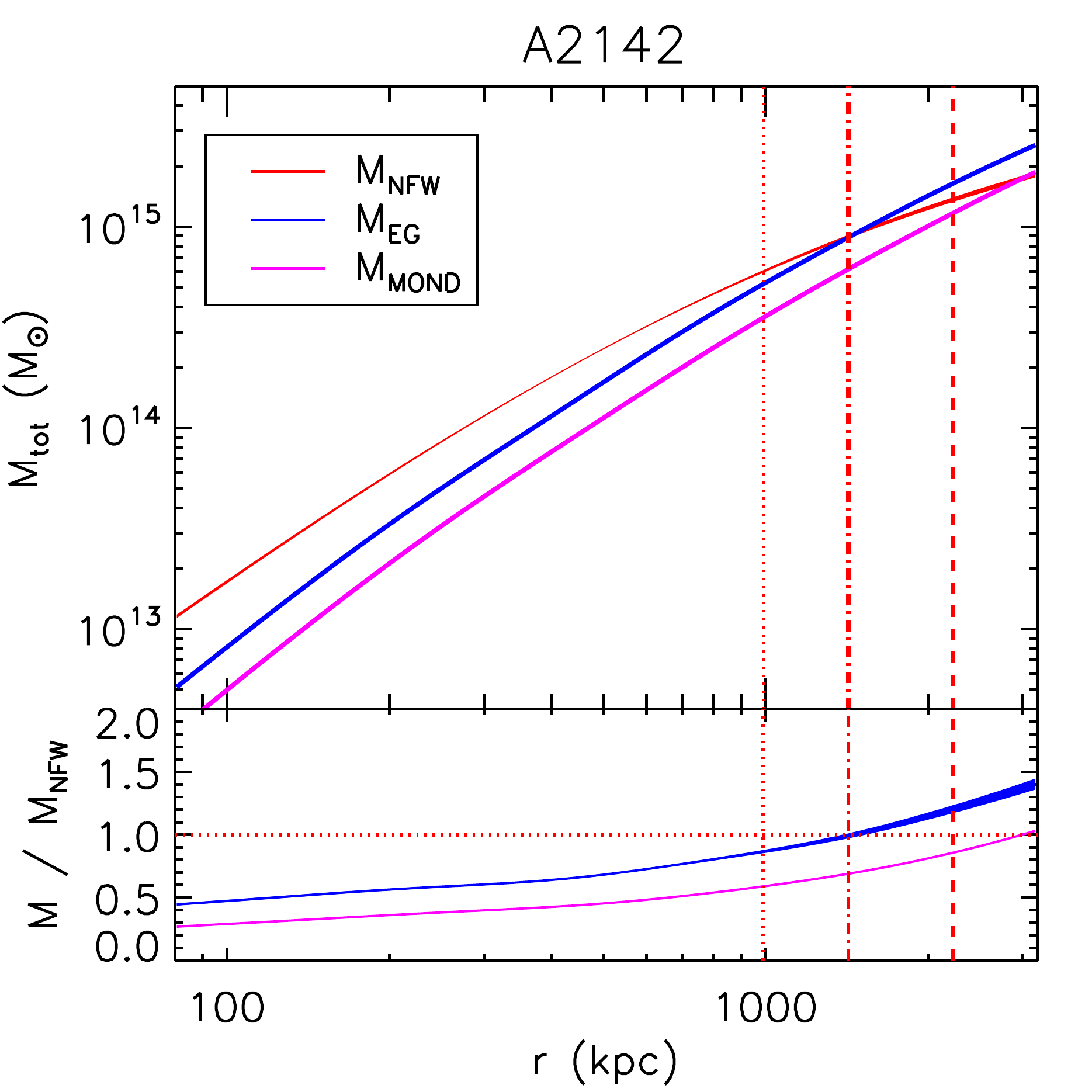}
\includegraphics[width=0.48\textwidth, keepaspectratio]{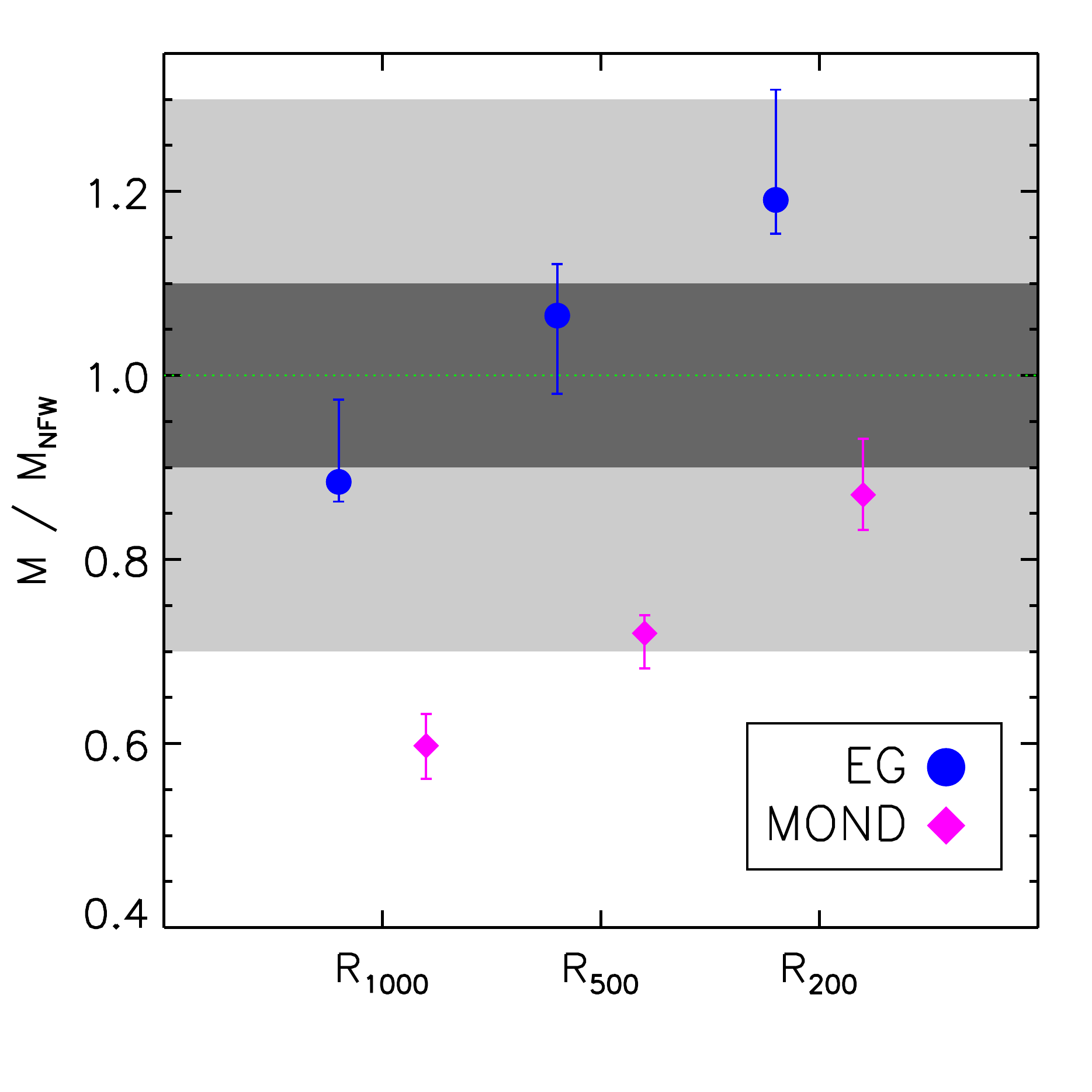}
\end{center} \vspace*{-0.5cm}
\caption{
(Left) Typical hydrostatic, EG and MOND radial mass profiles (here for A2142; plots for the remaining 12 objects in Appendix~\ref{app_eg}. 
Vertical lines indicate $R_{1000}$ (dotted line), $R_{500}$ (dash-dotted line) and $R_{200}$ (dashed line).
(Right) Medians (and 1st and 3rd quartiles) of the ratio between the hydrostatic mass and the value predicted from EG and MOND for the whole X-COP sample.
Shaded regions indicate the $<$10\% (darkest) and $<$30\% differences.
}
\label{fig:bias_eg}
\end{figure*}

\section{Comparison of predictions from the emergent gravity scenario and MOND}

The lack of a valuable candidate particle for the cold dark matter has induced part of the community to search for
alternative paradigms on how gravity works at galactic and larger scales. 
In galaxy clusters, the largest gravitationally bound structures in the universe,
visible matter can account for only a fraction of the total gravitational mass.
These systems thus  represent  a valid and robust test for models that try to  explain 
this missing mass problem. Among these rivals of the current cosmological paradigm, 
we consider here two models, one that introduces modifications in the Newtonian dynamical laws \citep[MOND, ][]{mil83} and 
another that compensates for the required extra gravitational force by an emergent gravity \citep{ver16}.
They  can both be described by similar equations, are able to describe the behaviour of the gravity on galactic scales, 
but are also known to cause trouble when applied on larger scales \citep[called the `upscaling problem' in][]{mas18}, 
where for instance MOND predicts a gravitational acceleration that is too weak, 
suggesting that it can be an incomplete theory.

In the emergent gravity scenario, dark matter can appear as a manifestation of an additional gravitational force describing the `elastic' response due to the displacement
of the entropy that can be associated with the thermal excitations carrying the positive dark energy, 
and with a strength that can be described in terms of the Hubble constant and of the baryonic mass distribution. 
For a spherically symmetric, static, and isolated astronomical system, \citet{ver16} provides a relation between the emergent dark matter and the baryonic mass 
(see his equation~7.40) that can be rearranged to isolate the dark matter component $M_{\rm DM}$.
Following \citet{ettori+17}, we can write
\begin{equation}
M_{\rm DM, EG}^2(r) =  \frac{cH_0}{6G} r^2 M_{\rm B}(r) \left(1 + 3 \delta_{\rm B} \right),
\label{eq:mdm_eg}
\end{equation}
where $M_{\rm B}(r) = \int_0^r 4\pi \rho_{\rm B} {r'^2}dr' = M_{\rm gas}(r) +M_{\rm star}(r)$ is the baryonic mass equal to the sum of the gas and stellar masses,
and $\delta_{\rm B}$ is equal to $\rho_{\rm B}(r) / \bar{\rho}_{\rm B}$ with $\bar{\rho}_{\rm B} = M_{\rm B}(r)/V(<r)$ representing the mean baryon density within the spherical volume $V(<r)$.
In our case, the gas mass has been obtained from the integral over the cluster's volume of the gas density (Fig.~\ref{fig:prof}). 
The stellar mass has been estimated using a Navarro-Frenk-White \citep[NFW,][]{nfw97} profile with a concentration of 2.9 \citep[see e.g.][]{lin04a}
and by requiring the $M_{\rm star}(<R_{500}) / M_{\rm gas}(<R_{500}) = 0.39 \, \left(M_{500}/10^{14} M_{\odot}\right)^{-0.84}$ \citep{gonzalez+13}. 
For the X-COP objects, we measure  a median $M_{\rm star}/ M_{\rm gas}$ of 0.09 (0.07--0.12 as 1st and 3rd quartile) at $R_{500}$.

As we discuss in \citet{ettori+17}, Equation~\ref{eq:mdm_eg} can be expressed as an acceleration $g_{\rm EG}$ depending on the acceleration $g_{\rm B}$ 
induced from the baryonic mass
\begin{equation}
g_{\rm EG} = g_{\rm B} \, \left( 1 + y^{-1/2} \right),
\label{eq:mond}
\end{equation}
where $y = 6 / (c H_0) \times g_{\rm B} / (1 + 3 \delta_{\rm B})$. 

Equation~\ref{eq:mond} takes a form very similar to the one implemented in MOND \citep[e.g.][]{ms16} with a characteristic acceleration $a_0 = c H_0 (1 + 3 \delta_{\rm B}) / 6$.
MOND is another theory that accounts for the mass in galaxies and galaxy clusters without a dark component and by modifying Newtonian dynamics, and requires
an acceleration $g_{\rm MOND} = g_{\rm B} \, \left( 1 + y^{-1/2} \right)$, with $y = g_{\rm B} / a_0$ and $a_0 = 10^{-8}$ cm/s$^2$.
For sake of completeness, we have also estimated a gravitating mass associated with a MOND acceleration.

The results of our comparison are shown in Fig.~\ref{fig:bias_eg} (where we present as an example the case of A2142; the other mass profiles are shown in Fig.~\ref{fig:eg}).
Although in the inner cluster regions the mismatch is indeed significant, with mass values from modified gravities that underestimate the hydrostatic quantities by a factor of few,
over the radial interval between $R_{1000}$ and $R_{200}$, the medians of the distribution of the ratios between the mass estimates 
obtained from modified accelerations and from the hydrostatic equilibrium equation are in the range $0.6-0.87$
for the MOND \citep[consistent with previous studies; e.g. ][]{pointecouteau_silk05} 
and between $0.88$ and $1.19$ for EG, with the latter that indicates a nice consistency at $R_{500}$ 
where $M_{\rm EG}  / M_{\rm Hyd} \approx 1.07$ ($0.99-1.12$ as 1st and 3rd quartiles).

\section{Conclusions}

We have investigated the hydrostatic mass profiles in the X-COP sample of 13 massive X-ray luminous galaxy clusters 
for which the gas density and temperature (from \xmm\ X-ray data) and SZ pressure profiles (from \planck) are recovered 
at very high accuracy up to about $R_{500}$ for the temperature and at $R_{200}$ for density and pressure. 

We constrain the total mass distribution by applying the hydrostatic equilibrium equation on these profiles, 
reconstructed under the assumption that the ICM follows a spherically symmetric distribution 
and using two different methods and five mass models.
By adopting as reference model a NFW mass profile constrained with the {backward} method, 
we estimate the radial mass distribution up to $R_{200}$ with a mean statistical relative error lower than 8\%.
A {forward} method, which is independent from any assumption on the shape of the gravitational potential, 
provides consistent results within 5\% both at $R_{500}$ and $R_{200}$.
Published results on the estimates of the hydrostatic mass in local massive galaxy clusters have  so far reached statistical uncertainties larger than 10\% 
\citep[e.g.][]{pointecouteau05, vikhlinin06, ettori+10}, 
confirming that the strategy implemented in the X-COP analysis is a successful way to  improve constraints on the mass.

Other sources of systematic uncertainties, like any non-thermal contribution 
to the total gas pressure that we discuss in a companion paper (Eckert et al. 2018), 
or departures from the hydrostatic equilibrium \citep[e.g.][]{nel14bias,bif16}, 
or the violation of the assumed sphericity of the gas distribution \citep[e.g.][]{ser17}
could push our mass estimates to higher values by a further 10--20\%, in particular at $r>R_{500}$.

A NFW mass profile represents the best-fit model for nine objects, where we measure a statistically significant tension with any cored mass profile.
The remaining four clusters prefer different mass models but are also consistent with a NFW.
These four systems are also the ones that deviate most in the NFW  $c_{200}-M_{200}$ plane with respect to the
theoretical predictions from numerical simulations.
Overall, we measure a scatter of 0.18 in the $c-M$ relation and of $0.15 \pm 0.08$ (with an a posteriori probability of 15 \%\ that it is below 10\%) in the hydrostatic mass measurements. 
This is in agreement with the results from the literature \citep{se+et15_comalit_I} and larger, but still 
compatible within the uncertainties, with those obtained from numerical simulations \citep{rasia+12}.

For a subsample of X-COP objects, we can quantify the average discrepancy between hydrostatic masses and estimates 
obtained from (i) scaling relations based on X-ray data and applied to the SZ signal, (ii) weak-lensing, and (iii) galaxy dynamics.
Overall, we obtain  remarkably good agreement (with an error-weighted mean and median of the ratios between hydrostatic and other masses
around 1 $\pm$ 0.2), apart from the caustic method which severely underestimates (by more than
40 \%\ on average)  
the hydrostatic values in this massive local relaxed systems.

Then, we compare these mass estimates to predictions from scenarios in which the gravitational acceleration is modified.
We note that both the traditional MOND acceleration and the value produced as a manifestation of apparent dark matter in the emergent gravity theory 
predict masses that are slightly below (by 10--40\%) the hydrostatic values in the inner 1 Mpc, with EG providing less significant tension, in particular at $R_{500}$ as estimated
from $M_{\rm Hyd}$ where we measure $M_{\rm EG}  / M_{\rm Hyd} \approx$ (0.9 --1.1).

We conclude that these estimates of the hydrostatic masses represent the best constraints ever measured, 
with a statistical error budget that is of the order of the systematic uncertainties we measure, and both within 
10 \%\ out to $R_{200}$.

Future extensions of the X-COP sample, with comparable coverage of the X-ray and SZ emission out to $R_{200}$ 
to dynamically non-relaxed systems, even at higher redshifts, will permit us to build a proper collection of hydrostatic masses that 
will provide the reference to study in detail the robustness of the assumption of the hydrostatic equilibrium.
This will be one of the main goals of the \xmm\ Multi-Year Heritage Program on galaxy clusters recently approved 
in the AO17 special call\footnote{https://www.cosmos.esa.int/web/xmm-newton/ao17} (PI: M.~Arnaud \& S.~Ettori).
It will dedicate 3 Msec of exposure time over the next 3 years to survey 118 \planck-SZ selected clusters out to $z \approx 0.6$
to map the temperature profile in $\ga$ 8 annuli up to $R_{500}$ with a relative error of 15\% 
in the in $[0.8-1.2] R_{500}$ annulus, allowing us to constrain the hydrostatic mass measurements at $R_{500}$ 
to the 15$-$20\% precision level.
A complete census of any residual kinetic energy in the gas bulk motions and turbulence as major bias in the estimates
of the hydrostatic mass requires direct measurements of the Doppler broadening and shifts of the emission lines in the ICM
over the entire cluster's volume.
{\it Hitomi} has provided the first significant results in the field, 
but limited to the inner parts of the core of the Perseus cluster \citep{hitomiPerseus16, hitomiPerseus18, zuhone18hitomi}.
The next generation of X-ray observatories equipped with high-resolution spectrometers 
like {\it XARM} \citep{ota+18} and {\it Athena} \citep{nandra+13,ettori13athena} 
will deepen our knowledge on the state of the ICM enlarging the sample and the regions of study.

\begin{acknowledgements}
We thank Ricardo Herbonnet and Henk Hoekstra for providing their weak-lensing results in advance of publication.
The research leading to these results has received funding from the European Union's Horizon 2020 Programme under the AHEAD project (grant agreement n. 654215).
S.E. and M.S. acknowledge financial contribution from the contracts NARO15 ASI-INAF I/037/12/0, ASI 2015-046-R.0, and ASI-INAF n.2017-14-H.0.
M.G. is supported by NASA through Einstein Postdoctoral Fellowship Award Number PF5-160137 issued by the Chandra X-ray Observatory Center, 
which is operated by the SAO for and on behalf of NASA under contract NAS8-03060. Support for this work was also provided by Chandra grant GO7-18121X.
X-COP data products are available for download at \url{https://www.astro.unige.ch/xcop}.
\end{acknowledgements}

\bibliographystyle{aa} 
\bibliography{mass_xcop} 

\begin{thebibliography}{85}
\expandafter\ifx\csname natexlab\endcsname\relax\def\natexlab#1{#1}\fi

\bibitem[{{Allen} {et~al.}(2011){Allen}, {Evrard}, \& {Mantz}}]{allen11}
{Allen}, S.~W., {Evrard}, A.~E., \& {Mantz}, A.~B. 2011, \araa, 49, 409

\bibitem[{{Anders} \& {Grevesse}(1989)}]{ag89}
{Anders}, E. \& {Grevesse}, N. 1989, \gca, 53, 197

\bibitem[{{Auger} {et~al.}(2013){Auger}, {Budzynski}, {Belokurov}, {Koposov},
  \& {McCarthy}}]{aug+al13}
{Auger}, M.~W., {Budzynski}, J.~M., {Belokurov}, V., {Koposov}, S.~E., \&
  {McCarthy}, I.~G. 2013, \mnras, 436, 503

\bibitem[{{Bhattacharya} {et~al.}(2013){Bhattacharya}, {Habib}, {Heitmann}, \&
  {Vikhlinin}}]{bha13}
{Bhattacharya}, S., {Habib}, S., {Heitmann}, K., \& {Vikhlinin}, A. 2013, \apj,
  766, 32

\bibitem[{{Biffi} {et~al.}(2016){Biffi}, {Borgani}, {Murante}, {Rasia},
  {Planelles}, {Granato}, {Ragone-Figueroa}, {Beck}, {Gaspari}, \&
  {Dolag}}]{bif16}
{Biffi}, V., {Borgani}, S., {Murante}, G., {et~al.} 2016, \apj, 827, 112

\bibitem[{{Bullock} {et~al.}(2001){Bullock}, {Kolatt}, {Sigad}, {Somerville},
  {Kravtsov}, {Klypin}, {Primack}, \& {Dekel}}]{bul01}
{Bullock}, J.~S., {Kolatt}, T.~S., {Sigad}, Y., {et~al.} 2001, \mnras, 321, 559

\bibitem[{{Buote} \& {Humphrey}(2012)}]{buote12b}
{Buote}, D.~A. \& {Humphrey}, P.~J. 2012, \mnras, 421, 1399

\bibitem[{{Buote} {et~al.}(2005){Buote}, {Humphrey}, \& {Stocke}}]{buote05}
{Buote}, D.~A., {Humphrey}, P.~J., \& {Stocke}, J.~T. 2005, \apj, 630, 750

\bibitem[{{Burns} {et~al.}(1995){Burns}, {Roettiger}, {Pinkney}, {Perley},
  {Owen}, \& {Voges}}]{burns95}
{Burns}, J.~O., {Roettiger}, K., {Pinkney}, J., {et~al.} 1995, \apj, 446, 583

\bibitem[{{Cavagnolo} {et~al.}(2009){Cavagnolo}, {Donahue}, {Voit}, \&
  {Sun}}]{cavagnolo+09}
{Cavagnolo}, K.~W., {Donahue}, M., {Voit}, G.~M., \& {Sun}, M. 2009, \apjs,
  182, 12

\bibitem[{{De Grandi} {et~al.}(2016){De Grandi}, {Eckert}, {Molendi},
  {Girardi}, {Roediger}, {Gaspari}, {Gastaldello}, {Ghizzardi}, {Nonino}, \&
  {Rossetti}}]{degrandi+16}
{De Grandi}, S., {Eckert}, D., {Molendi}, S., {et~al.} 2016, \aap, 592, A154

\bibitem[{{Diemer} \& {Kravtsov}(2015)}]{dk15}
{Diemer}, B. \& {Kravtsov}, A.~V. 2015, \apj, 799, 108

\bibitem[{{Dolag} {et~al.}(2004){Dolag}, {Bartelmann}, {Perrotta},
  {Baccigalupi}, {Moscardini}, {Meneghetti}, \& {Tormen}}]{dol04}
{Dolag}, K., {Bartelmann}, M., {Perrotta}, F., {et~al.} 2004, \aap, 416, 853

\bibitem[{{Du} \& {Fan}(2014)}]{du+fa14}
{Du}, W. \& {Fan}, Z. 2014, \apj, 785, 57

\bibitem[{{Duffy} {et~al.}(2008){Duffy}, {Schaye}, {Kay}, \& {Dalla
  Vecchia}}]{duf+al08}
{Duffy}, A.~R., {Schaye}, J., {Kay}, S.~T., \& {Dalla Vecchia}, C. 2008,
  \mnras, 390, L64

\bibitem[{{Dutton} \& {Macci{\`o}}(2014)}]{dutton14}
{Dutton}, A.~A. \& {Macci{\`o}}, A.~V. 2014, \mnras, 441, 3359

\bibitem[{{Eckert} {et~al.}(2016){Eckert}, {Ettori}, {Coupon}, {Gastaldello},
  {Pierre}, {Melin}, {Le Brun}, {McCarthy}, {Adami}, {Chiappetti}, {Faccioli},
  {Giles}, {Lavoie}, {Lef{\`e}vre}, {Lieu}, {Mantz}, {Maughan}, {McGee},
  {Pacaud}, {Paltani}, {Sadibekova}, {Smith}, \& {Ziparo}}]{eckert+16}
{Eckert}, D., {Ettori}, S., {Coupon}, J., {et~al.} 2016, \aap, 592, A12

\bibitem[{{Eckert} {et~al.}(2017){Eckert}, {Ettori}, {Pointecouteau},
  {Molendi}, {Paltani}, \& {Tchernin}}]{eck17xcop}
{Eckert}, D., {Ettori}, S., {Pointecouteau}, E., {et~al.} 2017, Astronomische
  Nachrichten, 338, 293

\bibitem[{{Eckert} {et~al.}(2018){Eckert}, {Ghirardini}, {Ettori}, {Rasia},
  {Biffi}, {Pointecouteau}, {Rossetti}, {Molendi}, {Vazza}, {Gastaldello},
  {Gaspari}, {De Grandi}, {Ghizzardi}, {Bourdin}, {Tchernin}, \&
  {Roncarelli}}]{eckert18}
{Eckert}, D., {Ghirardini}, V., {Ettori}, S., {et~al.} 2018, ArXiv e-prints
  [\eprint[arXiv]{1805.00034}]

\bibitem[{{Eckert} {et~al.}(2015){Eckert}, {Roncarelli}, {Ettori}, {Molendi},
  {Vazza}, {Gastaldello}, \& {Rossetti}}]{eckert+15}
{Eckert}, D., {Roncarelli}, M., {Ettori}, S., {et~al.} 2015, \mnras, 447, 2198

\bibitem[{{Ettori} {et~al.}(2013{\natexlab{a}}){Ettori}, {Donnarumma},
  {Pointecouteau}, {Reiprich}, {Giodini}, {Lovisari}, \& {Schmidt}}]{ettori+13}
{Ettori}, S., {Donnarumma}, A., {Pointecouteau}, E., {et~al.}
  2013{\natexlab{a}}, \ssr, 177, 119

\bibitem[{{Ettori} {et~al.}(2010){Ettori}, {Gastaldello}, {Leccardi},
  {Molendi}, {Rossetti}, {Buote}, \& {Meneghetti}}]{ettori+10}
{Ettori}, S., {Gastaldello}, F., {Leccardi}, A., {et~al.} 2010, \aap, 524, A68

\bibitem[{{Ettori} {et~al.}(2017){Ettori}, {Ghirardini}, {Eckert}, {Dubath}, \&
  {Pointecouteau}}]{ettori+17}
{Ettori}, S., {Ghirardini}, V., {Eckert}, D., {Dubath}, F., \& {Pointecouteau},
  E. 2017, \mnras, 470, L29

\bibitem[{{Ettori} \& {Molendi}(2011)}]{em11}
{Ettori}, S. \& {Molendi}, S. 2011, Memorie della Societa Astronomica Italiana
  Supplementi, 17, 47

\bibitem[{{Ettori} {et~al.}(2013{\natexlab{b}}){Ettori}, {Pratt}, {de Plaa},
  {Eckert}, {Nevalainen}, {Battistelli}, {Borgani}, {Croston}, {Finoguenov},
  {Kaastra}, {Gaspari}, {Gastaldello}, {Gitti}, {Molendi}, {Pointecouteau},
  {Ponman}, {Reiprich}, {Roncarelli}, {Rossetti}, {Sanders}, {Sun},
  {Trinchieri}, {Vazza}, {Arnaud}, {B{\"o}ringher}, {Brighenti}, {Dahle}, {De
  Grandi}, {Mohr}, {Moretti}, \& {Schindler}}]{ettori13athena}
{Ettori}, S., {Pratt}, G.~W., {de Plaa}, J., {et~al.} 2013{\natexlab{b}}, ArXiv
  e-prints [\eprint[arXiv]{1306.2322}]

\bibitem[{{Feroz} {et~al.}(2009){Feroz}, {Hobson}, \& {Bridges}}]{multinest}
{Feroz}, F., {Hobson}, M.~P., \& {Bridges}, M. 2009, \mnras, 398, 1601

\bibitem[{{Foreman-Mackey} {et~al.}(2013){Foreman-Mackey}, {Hogg}, {Lang}, \&
  {Goodman}}]{emcee}
{Foreman-Mackey}, D., {Hogg}, D.~W., {Lang}, D., \& {Goodman}, J. 2013, \pasp,
  125, 306

\bibitem[{{Gaspari} {et~al.}(2018){Gaspari}, {McDonald}, {Hamer}, {Brighenti},
  {Temi}, {Gendron-Marsolais}, {Hlavacek-Larrondo}, {Edge}, {Werner}, {Tozzi},
  {Sun}, {Stone}, {Tremblay}, {Hogan}, {Eckert}, {Ettori}, {Yu}, {Biffi}, \&
  {Planelles}}]{gaspari18}
{Gaspari}, M., {McDonald}, M., {Hamer}, S.~L., {et~al.} 2018, \apj, 854, 167

\bibitem[{{Ghirardini} {et~al.}(2018{\natexlab{a}}){Ghirardini}, {Ettori},
  {Eckert}, {Molendi}, {Gastaldello}, {Pointecouteau}, {Hurier}, \&
  {Bourdin}}]{ghi18}
{Ghirardini}, V., {Ettori}, S., {Eckert}, D., {et~al.} 2018{\natexlab{a}},
  \aap, 614, A7

\bibitem[{{Ghirardini} {et~al.}(2018{\natexlab{b}}){Ghirardini}, {Eckert},
  {Ettori}, {Pointecouteau}, {Molendi}, {Gaspari}, {Rossetti}, {De Grandi},
  {Roncarelli}, {Bourdin}, {Mazzotta}, {Rasia}, \& {Vazza}}]{ghi18univ}
{Ghirardini}, V., {Eckert}, D., {Ettori}, S., {et~al.} 2018{\natexlab{b}},
  ArXiv e-prints [\eprint[arXiv]{1805.00042}]

\bibitem[{{Gifford} {et~al.}(2017){Gifford}, {Kern}, \& {Miller}}]{gifford+17}
{Gifford}, D., {Kern}, N., \& {Miller}, C.~J. 2017, \apj, 834, 204

\bibitem[{{Gifford} {et~al.}(2013){Gifford}, {Miller}, \& {Kern}}]{gifford+13}
{Gifford}, D., {Miller}, C., \& {Kern}, N. 2013, \apj, 773, 116

\bibitem[{{Gonzalez} {et~al.}(2013){Gonzalez}, {Sivanandam}, {Zabludoff}, \&
  {Zaritsky}}]{gonzalez+13}
{Gonzalez}, A.~H., {Sivanandam}, S., {Zabludoff}, A.~I., \& {Zaritsky}, D.
  2013, \apj, 778, 14

\bibitem[{{Hernquist}(1990)}]{her90}
{Hernquist}, L. 1990, \apj, 356, 359

\bibitem[{{Hitomi Collaboration} {et~al.}(2016){Hitomi Collaboration},
  {Aharonian}, {Akamatsu}, {Akimoto}, {Allen}, {Anabuki}, {Angelini}, {Arnaud},
  {Audard}, {Awaki}, {Axelsson}, {Bamba}, {Bautz}, {Blandford}, {Brenneman},
  {Brown}, {Bulbul}, {Cackett}, {Chernyakova}, {Chiao}, {Coppi}, {Costantini},
  {de Plaa}, {den Herder}, {Done}, {Dotani}, {Ebisawa}, {Eckart}, {Enoto},
  {Ezoe}, {Fabian}, {Ferrigno}, {Foster}, {Fujimoto}, {Fukazawa}, {Furuzawa},
  {Galeazzi}, {Gallo}, {Gandhi}, {Giustini}, {Goldwurm}, {Gu}, {Guainazzi},
  {Haba}, {Hagino}, {Hamaguchi}, {Harrus}, {Hatsukade}, {Hayashi}, {Hayashi},
  {Hayashida}, {Hiraga}, {Hornschemeier}, {Hoshino}, {Hughes}, {Iizuka},
  {Inoue}, {Inoue}, {Ishibashi}, {Ishida}, {Ishikawa}, {Ishisaki}, {Itoh},
  {Iyomoto}, {Kaastra}, {Kallman}, {Kamae}, {Kara}, {Kataoka}, {Katsuda},
  {Katsuta}, {Kawaharada}, {Kawai}, {Kelley}, {Khangulyan}, {Kilbourne},
  {King}, {Kitaguchi}, {Kitamoto}, {Kitayama}, {Kohmura}, {Kokubun}, {Koyama},
  {Koyama}, {Kretschmar}, {Krimm}, {Kubota}, {Kunieda}, {Laurent}, {Lebrun},
  {Lee}, {Leutenegger}, {Limousin}, {Loewenstein}, {Long}, {Lumb}, {Madejski},
  {Maeda}, {Maier}, {Makishima}, {Markevitch}, {Matsumoto}, {Matsushita},
  {McCammon}, {McNamara}, {Mehdipour}, {Miller}, {Miller}, {Mineshige},
  {Mitsuda}, {Mitsuishi}, {Miyazawa}, {Mizuno}, {Mori}, {Mori}, {Moseley},
  {Mukai}, {Murakami}, {Murakami}, {Mushotzky}, {Nagino}, {Nakagawa},
  {Nakajima}, {Nakamori}, {Nakano}, {Nakashima}, {Nakazawa}, {Nobukawa},
  {Noda}, {Nomachi}, {O'Dell}, {Odaka}, {Ohashi}, {Ohno}, {Okajima}, {Ota},
  {Ozaki}, {Paerels}, {Paltani}, {Parmar}, {Petre}, {Pinto}, {Pohl}, {Porter},
  {Pottschmidt}, {Ramsey}, {Reynolds}, {Russell}, {Safi-Harb}, {Saito},
  {Sakai}, {Sameshima}, {Sato}, {Sato}, {Sato}, {Sawada}, {Schartel},
  {Serlemitsos}, {Seta}, {Shidatsu}, {Simionescu}, {Smith}, {Soong}, {Stawarz},
  {Sugawara}, {Sugita}, {Szymkowiak}, {Tajima}, {Takahashi}, {Takahashi},
  {Takeda}, {Takei}, {Tamagawa}, {Tamura}, {Tamura}, {Tanaka}, {Tanaka},
  {Tanaka}, {Tashiro}, {Tawara}, {Terada}, {Terashima}, {Tombesi}, {Tomida},
  {Tsuboi}, {Tsujimoto}, {Tsunemi}, {Tsuru}, {Uchida}, {Uchiyama}, {Uchiyama},
  {Ueda}, {Ueda}, {Ueno}, {Uno}, {Urry}, {Ursino}, {de Vries}, {Watanabe},
  {Werner}, {Wik}, {Wilkins}, {Williams}, {Yamada}, {Yamaguchi}, {Yamaoka},
  {Yamasaki}, {Yamauchi}, {Yamauchi}, {Yaqoob}, {Yatsu}, {Yonetoku}, {Yoshida},
  {Yuasa}, {Zhuravleva}, \& {Zoghbi}}]{hitomiPerseus16}
{Hitomi Collaboration}, {Aharonian}, F., {Akamatsu}, H., {et~al.} 2016, \nat,
  535, 117

\bibitem[{{Hitomi Collaboration} {et~al.}(2018){Hitomi Collaboration},
  {Aharonian}, {Akamatsu}, {Akimoto}, {Allen}, {Angelini}, {Audard}, {Awaki},
  {Axelsson}, {Bamba}, {Bautz}, {Blandford}, {Brenneman}, {Brown}, {Bulbul},
  {Cackett}, {Canning}, {Chernyakova}, {Chiao}, {Coppi}, {Costantini}, {de
  Plaa}, {de Vries}, {den Herder}, {Done}, {Dotani}, {Ebisawa}, {Eckart},
  {Enoto}, {Ezoe}, {Fabian}, {Ferrigno}, {Foster}, {Fujimoto}, {Fukazawa},
  {Furuzawa}, {Galeazzi}, {Gallo}, {Gandhi}, {Giustini}, {Goldwurm}, {Gu},
  {Guainazzi}, {Haba}, {Hagino}, {Hamaguchi}, {Harrus}, {Hatsukade}, {Hayashi},
  {Hayashi}, {Hayashi}, {Hayashida}, {Hiraga}, {Hornschemeier}, {Hoshino},
  {Hughes}, {Ichinohe}, {Iizuka}, {Inoue}, {Inoue}, {Inoue}, {Ishida},
  {Ishikawa}, {Ishisaki}, {Iwai}, {Kaastra}, {Kallman}, {Kamae}, {Kataoka},
  {Katsuda}, {Kawai}, {Kelley}, {Kilbourne}, {Kitaguchi}, {Kitamoto},
  {Kitayama}, {Kohmura}, {Kokubun}, {Koyama}, {Koyama}, {Kretschmar}, {Krimm},
  {Kubota}, {Kunieda}, {Laurent}, {Lee}, {Leutenegger}, {Limousin},
  {Loewenstein}, {Long}, {Lumb}, {Madejski}, {Maeda}, {Maier}, {Makishima},
  {Markevitch}, {Matsumoto}, {Matsushita}, {McCammon}, {McNamara}, {Mehdipour},
  {Miller}, {Miller}, {Mineshige}, {Mitsuda}, {Mitsuishi}, {Miyazawa},
  {Mizuno}, {Mori}, {Mori}, {Mukai}, {Murakami}, {Mushotzky}, {Nakagawa},
  {Nakajima}, {Nakamori}, {Nakashima}, {Nakazawa}, {Nobukawa}, {Nobukawa},
  {Noda}, {Odaka}, {Ohashi}, {Ohno}, {Okajima}, {Ota}, {Ozaki}, {Paerels},
  {Paltani}, {Petre}, {Pinto}, {Porter}, {Pottschmidt}, {Reynolds},
  {Safi-Harb}, {Saito}, {Sakai}, {Sasaki}, {Sato}, {Sato}, {Sato}, {Sawada},
  {Schartel}, {Serlemtsos}, {Seta}, {Shidatsu}, {Simionescu}, {Smith}, {Soong},
  {Stawarz}, {Sugawara}, {Sugita}, {Szymkowiak}, {Tajima}, {Takahashi},
  {Takahashi}, {Takeda}, {Takei}, {Tamagawa}, {Tamura}, {Tanaka}, {Tanaka},
  {Tanaka}, {Tanaka}, {Tashiro}, {Tawara}, {Terada}, {Terashima}, {Tombesi},
  {Tomida}, {Tsuboi}, {Tsujimoto}, {Tsunemi}, {Tsuru}, {Uchida}, {Uchiyama},
  {Uchiyama}, {Ueda}, {Ueda}, {Uno}, {Urry}, {Ursino}, {Wang}, {Watanabe},
  {Werner}, {Wilkins}, {Williams}, {Yamada}, {Yamaguchi}, {Yamaoka},
  {Yamasaki}, {Yamauchi}, {Yamauchi}, {Yaqoob}, {Yatsu}, {Yonetoku},
  {Zhuravleva}, \& {Zoghbi}}]{hitomiPerseus18}
{Hitomi Collaboration}, {Aharonian}, F., {Akamatsu}, H., {et~al.} 2018, \pasj,
  70, 9

\bibitem[{{Jeffreys}(1961)}]{jef61}
{Jeffreys}, H. 1961, {The Theory of Probability} (Oxford)

\bibitem[{{Kelly}(2007)}]{kel07}
{Kelly}, B.~C. 2007, \apj, 665, 1489

\bibitem[{{Khatri} \& {Gaspari}(2016)}]{khatri+16}
{Khatri}, R. \& {Gaspari}, M. 2016, \mnras, 463, 655

\bibitem[{{King}(1962)}]{king62}
{King}, I. 1962, \aj, 67, 471

\bibitem[{{Kravtsov} \& {Borgani}(2012)}]{kb12}
{Kravtsov}, A.~V. \& {Borgani}, S. 2012, \araa, 50, 353

\bibitem[{{Lin} {et~al.}(2004){Lin}, {Mohr}, \& {Stanford}}]{lin04a}
{Lin}, Y.-T., {Mohr}, J.~J., \& {Stanford}, S.~A. 2004, \apj, 610, 745

\bibitem[{{Ludlow} {et~al.}(2016){Ludlow}, {Bose}, {Angulo}, {Wang},
  {Hellwing}, {Navarro}, {Cole}, \& {Frenk}}]{lud+al16}
{Ludlow}, A.~D., {Bose}, S., {Angulo}, R.~E., {et~al.} 2016, \mnras, 460, 1214

\bibitem[{{Mamon} \& {{\L}okas}(2005)}]{ml05}
{Mamon}, G.~A. \& {{\L}okas}, E.~L. 2005, \mnras, 362, 95

\bibitem[{{Mantz} {et~al.}(2010){Mantz}, {Allen}, {Ebeling}, {Rapetti}, \&
  {Drlica-Wagner}}]{mantz10}
{Mantz}, A., {Allen}, S.~W., {Ebeling}, H., {Rapetti}, D., \& {Drlica-Wagner},
  A. 2010, \mnras, 406, 1773

\bibitem[{{Mantz} {et~al.}(2016){Mantz}, {Allen}, \& {Morris}}]{mantz16cm}
{Mantz}, A.~B., {Allen}, S.~W., \& {Morris}, R.~G. 2016, \mnras, 462, 681

\bibitem[{{Massimi}(2018)}]{mas18}
{Massimi}, M. 2018, ArXiv e-prints [\eprint[arXiv]{1804.07704}]

\bibitem[{{Meneghetti} {et~al.}(2014){Meneghetti}, {Rasia}, {Vega}, {Merten},
  {Postman}, {Yepes}, {Sembolini}, {Donahue}, {Ettori}, {Umetsu}, {Balestra},
  {Bartelmann}, {Ben{\'{\i}}tez}, {Biviano}, {Bouwens}, {Bradley},
  {Broadhurst}, {Coe}, {Czakon}, {De Petris}, {Ford}, {Giocoli},
  {Gottl{\"o}ber}, {Grillo}, {Infante}, {Jouvel}, {Kelson}, {Koekemoer},
  {Lahav}, {Lemze}, {Medezinski}, {Melchior}, {Mercurio}, {Molino},
  {Moscardini}, {Monna}, {Moustakas}, {Moustakas}, {Nonino}, {Rhodes},
  {Rosati}, {Sayers}, {Seitz}, {Zheng}, \& {Zitrin}}]{men+al14}
{Meneghetti}, M., {Rasia}, E., {Vega}, J., {et~al.} 2014, \apj, 797, 34

\bibitem[{{Milgrom}(1983)}]{mil83}
{Milgrom}, M. 1983, \apj, 270, 365

\bibitem[{{Milgrom} \& {Sanders}(2016)}]{ms16}
{Milgrom}, M. \& {Sanders}, R.~H. 2016, ArXiv e-prints
  [\eprint[arXiv]{1612.09582}]

\bibitem[{{Nandra} {et~al.}(2013){Nandra}, {Barret}, {Barcons}, {Fabian}, {den
  Herder}, {Piro}, {Watson}, {Adami}, {Aird}, {Afonso}, \& et~al.}]{nandra+13}
{Nandra}, K., {Barret}, D., {Barcons}, X., {et~al.} 2013, ArXiv e-prints
  [\eprint[arXiv]{1306.2307}]

\bibitem[{{Navarro} {et~al.}(1997){Navarro}, {Frenk}, \& {White}}]{nfw97}
{Navarro}, J.~F., {Frenk}, C.~S., \& {White}, S.~D.~M. 1997, \apj, 490, 493

\bibitem[{{Nelson} {et~al.}(2014{\natexlab{a}}){Nelson}, {Lau}, \&
  {Nagai}}]{nel14pnt}
{Nelson}, K., {Lau}, E.~T., \& {Nagai}, D. 2014{\natexlab{a}}, \apj, 792, 25

\bibitem[{{Nelson} {et~al.}(2014{\natexlab{b}}){Nelson}, {Lau}, {Nagai},
  {Rudd}, \& {Yu}}]{nel14bias}
{Nelson}, K., {Lau}, E.~T., {Nagai}, D., {Rudd}, D.~H., \& {Yu}, L.
  2014{\natexlab{b}}, \apj, 782, 107

\bibitem[{{Neto} {et~al.}(2007){Neto}, {Gao}, {Bett}, {Cole}, {Navarro},
  {Frenk}, {White}, {Springel}, \& {Jenkins}}]{neto07}
{Neto}, A.~F., {Gao}, L., {Bett}, P., {et~al.} 2007, \mnras, 381, 1450

\bibitem[{{Ota} {et~al.}(2018){Ota}, {Nagai}, \& {Lau}}]{ota+18}
{Ota}, N., {Nagai}, D., \& {Lau}, E.~T. 2018, \pasj, 70, 51

\bibitem[{{Pizzo} \& {de Bruyn}(2009)}]{pizzo09}
{Pizzo}, R.~F. \& {de Bruyn}, A.~G. 2009, \aap, 507, 639

\bibitem[{{Planck Collaboration} {et~al.}(2014){Planck Collaboration}, {Ade},
  {Aghanim}, {Armitage-Caplan}, {Arnaud}, {Ashdown}, {Atrio-Barandela},
  {Aumont}, {Aussel}, {Baccigalupi}, \& et~al.}]{SZcatalog}
{Planck Collaboration}, {Ade}, P.~A.~R., {Aghanim}, N., {et~al.} 2014, \aap,
  571, A29

\bibitem[{{Planck Collaboration} {et~al.}(2013){Planck Collaboration}, {Ade},
  {Aghanim}, {Arnaud}, {Ashdown}, {Atrio-Barandela}, {Aumont}, {Baccigalupi},
  {Balbi}, {Banday}, \& et~al.}]{planck13}
{Planck Collaboration}, {Ade}, P.~A.~R., {Aghanim}, N., {et~al.} 2013, \aap,
  550, A131

\bibitem[{{Planck Collaboration} {et~al.}(2016){Planck Collaboration}, {Ade},
  {Aghanim}, {Arnaud}, {Ashdown}, {Aumont}, {Baccigalupi}, {Banday},
  {Barreiro}, {Bartlett}, \& et~al.}]{planck15-24}
{Planck Collaboration}, {Ade}, P.~A.~R., {Aghanim}, N., {et~al.} 2016, \aap,
  594, A24

\bibitem[{{Pointecouteau} {et~al.}(2005){Pointecouteau}, {Arnaud}, \&
  {Pratt}}]{pointecouteau05}
{Pointecouteau}, E., {Arnaud}, M., \& {Pratt}, G.~W. 2005, \aap, 435, 1

\bibitem[{{Pointecouteau} \& {Silk}(2005)}]{pointecouteau_silk05}
{Pointecouteau}, E. \& {Silk}, J. 2005, \mnras, 364, 654

\bibitem[{{Rasia} {et~al.}(2013){Rasia}, {Borgani}, {Ettori}, {Mazzotta}, \&
  {Meneghetti}}]{rasia+13}
{Rasia}, E., {Borgani}, S., {Ettori}, S., {Mazzotta}, P., \& {Meneghetti}, M.
  2013, \apj, 776, 39

\bibitem[{{Rasia} {et~al.}(2012){Rasia}, {Meneghetti}, {Martino}, {Borgani},
  {Bonafede}, {Dolag}, {Ettori}, {Fabjan}, {Giocoli}, {Mazzotta}, {Merten},
  {Radovich}, \& {Tornatore}}]{rasia+12}
{Rasia}, E., {Meneghetti}, M., {Martino}, R., {et~al.} 2012, New Journal of
  Physics, 14, 055018

\bibitem[{{Reiprich} {et~al.}(2013){Reiprich}, {Basu}, {Ettori}, {Israel},
  {Lovisari}, {Molendi}, {Pointecouteau}, \& {Roncarelli}}]{reiprich+13}
{Reiprich}, T.~H., {Basu}, K., {Ettori}, S., {et~al.} 2013, \ssr, 177, 195

\bibitem[{{Rines} {et~al.}(2016){Rines}, {Geller}, {Diaferio}, \&
  {Hwang}}]{rines16}
{Rines}, K.~J., {Geller}, M.~J., {Diaferio}, A., \& {Hwang}, H.~S. 2016, \apj,
  819, 63

\bibitem[{{Roncarelli} {et~al.}(2013){Roncarelli}, {Ettori}, {Borgani},
  {Dolag}, {Fabjan}, \& {Moscardini}}]{roncarelli+13}
{Roncarelli}, M., {Ettori}, S., {Borgani}, S., {et~al.} 2013, \mnras, 432, 3030

\bibitem[{{Sakelliou} \& {Ponman}(2006)}]{sak_pon06}
{Sakelliou}, I. \& {Ponman}, T.~J. 2006, \mnras, 367, 1409

\bibitem[{{Salucci} \& {Burkert}(2000)}]{sb00}
{Salucci}, P. \& {Burkert}, A. 2000, \apjl, 537, L9

\bibitem[{{Sereno}(2016)}]{ser16_lira}
{Sereno}, M. 2016, \mnras, 455, 2149

\bibitem[{{Sereno} {et~al.}(2017{\natexlab{a}}){Sereno}, {Covone}, {Izzo},
  {Ettori}, {Coupon}, \& {Lieu}}]{ser+al17_psz2lens}
{Sereno}, M., {Covone}, G., {Izzo}, L., {et~al.} 2017{\natexlab{a}}, \mnras,
  472, 1946

\bibitem[{{Sereno} \& {Ettori}(2015{\natexlab{a}})}]{se+et15_comalit_IV}
{Sereno}, M. \& {Ettori}, S. 2015{\natexlab{a}}, \mnras, 450, 3675

\bibitem[{{Sereno} \& {Ettori}(2015{\natexlab{b}})}]{se+et15_comalit_I}
{Sereno}, M. \& {Ettori}, S. 2015{\natexlab{b}}, \mnras, 450, 3633

\bibitem[{{Sereno} {et~al.}(2017{\natexlab{b}}){Sereno}, {Ettori},
  {Meneghetti}, {Sayers}, {Umetsu}, {Merten}, {Chiu}, \& {Zitrin}}]{ser17}
{Sereno}, M., {Ettori}, S., {Meneghetti}, M., {et~al.} 2017{\natexlab{b}},
  \mnras, 467, 3801

\bibitem[{{Sereno} {et~al.}(2015){Sereno}, {Giocoli}, {Ettori}, \&
  {Moscardini}}]{ser+al15_cM}
{Sereno}, M., {Giocoli}, C., {Ettori}, S., \& {Moscardini}, L. 2015, \mnras,
  449, 2024

\bibitem[{{Serra} \& {Diaferio}(2013)}]{serra+13}
{Serra}, A.~L. \& {Diaferio}, A. 2013, \apj, 768, 116

\bibitem[{{Serra} {et~al.}(2011){Serra}, {Diaferio}, {Murante}, \&
  {Borgani}}]{serra+11}
{Serra}, A.~L., {Diaferio}, A., {Murante}, G., \& {Borgani}, S. 2011, \mnras,
  412, 800

\bibitem[{{Sohn} {et~al.}(2018){Sohn}, {Geller}, {Walker}, {Dell'Antonio},
  {Diaferio}, \& {Rines}}]{sohn18}
{Sohn}, J., {Geller}, M.~J., {Walker}, S.~A., {et~al.} 2018, ArXiv e-prints
  [\eprint[arXiv]{1808.00488}]

\bibitem[{{Sunyaev} \& {Zeldovich}(1972)}]{SZ}
{Sunyaev}, R.~A. \& {Zeldovich}, Y.~B. 1972, Comments on Astrophysics and Space
  Physics, 4, 173

\bibitem[{{Verlinde}(2016)}]{ver16}
{Verlinde}, E.~P. 2016, ArXiv e-prints [\eprint[arXiv]{1611.02269}]

\bibitem[{{Vikhlinin} {et~al.}(2009){Vikhlinin}, {Burenin}, {Ebeling},
  {Forman}, {Hornstrup}, {Jones}, {Kravtsov}, {Murray}, {Nagai}, {Quintana}, \&
  {Voevodkin}}]{vikhlinin09}
{Vikhlinin}, A., {Burenin}, R.~A., {Ebeling}, H., {et~al.} 2009, \apj, 692,
  1033

\bibitem[{{Vikhlinin} {et~al.}(2006){Vikhlinin}, {Kravtsov}, {Forman}, {Jones},
  {Markevitch}, {Murray}, \& {Van Speybroeck}}]{vikhlinin06}
{Vikhlinin}, A., {Kravtsov}, A., {Forman}, W., {et~al.} 2006, \apj, 640, 691

\bibitem[{{Zhang} {et~al.}(2017){Zhang}, {Reiprich}, {Schneider}, {Clerc},
  {Merloni}, {Schwope}, {Borm}, {Andernach}, {Caretta}, \& {Wu}}]{zhang17}
{Zhang}, Y.-Y., {Reiprich}, T.~H., {Schneider}, P., {et~al.} 2017, \aap, 599,
  A138

\bibitem[{{Zhuravleva} {et~al.}(2013){Zhuravleva}, {Churazov}, {Kravtsov},
  {Lau}, {Nagai}, \& {Sunyaev}}]{zhu13}
{Zhuravleva}, I., {Churazov}, E., {Kravtsov}, A., {et~al.} 2013, \mnras, 428,
  3274

\bibitem[{{ZuHone} {et~al.}(2018){ZuHone}, {Miller}, {Bulbul}, \&
  {Zhuravleva}}]{zuhone18hitomi}
{ZuHone}, J.~A., {Miller}, E.~D., {Bulbul}, E., \& {Zhuravleva}, I. 2018, \apj,
  853, 180

\end{thebibliography}

\begin{appendix}

\onecolumn 

\section{Parameters and profiles of the best-fit mass models}

\begin{table*}[ht]
\hbox{
\setlength{\tabcolsep}{6pt} 
\begin{tabular}{c cccc}
\hline \hline
 Name & \multicolumn{4}{c}{NFW} \\ \\
  & kpc & $c$ & $\sigma_{T, int}$ & $\ln E$ \\
A85 & $1921^{+30}_{-24}$ & $3.31^{+0.13}_{-0.13}$ & $0.019^{+0.012}_{-0.005}$ & 7.5 \\
A644 & $1847^{+61}_{-58}$ & $5.58^{+0.65}_{-0.51}$ & $0.063^{+0.017}_{-0.016}$ & -9.7 \\
A1644 & $1778^{+55}_{-48}$ & $1.46^{+0.14}_{-0.14}$ & $0.001^{+0.000}_{-0.000}$ & -2.5 \\
A1795 & $1755^{+22}_{-21}$ & $4.55^{+0.16}_{-0.14}$ & $0.009^{+0.012}_{-0.007}$ & 9.0 \\
A2029 & $2173^{+35}_{-33}$ & $4.26^{+0.19}_{-0.17}$ & $0.023^{+0.006}_{-0.005}$ & -2.5 \\
A2142 & $2224^{+29}_{-25}$ & $3.14^{+0.10}_{-0.10}$ & $0.001^{+0.000}_{-0.000}$ & -0.6 \\
A2255 & $2033^{+88}_{-74}$ & $1.37^{+0.24}_{-0.23}$ & $0.002^{+0.006}_{-0.001}$ & -4.2 \\
A2319 & $2040^{+34}_{-30}$ & $4.86^{+0.51}_{-0.37}$ & $0.055^{+0.009}_{-0.008}$ & -12.4 \\
A3158 & $1766^{+34}_{-37}$ & $2.88^{+0.26}_{-0.17}$ & $0.002^{+0.015}_{-0.001}$ & 1.2 \\
A3266 & $2325^{+74}_{-75}$ & $2.04^{+0.25}_{-0.20}$ & $0.036^{+0.008}_{-0.009}$ & -7.1 \\
HydraA & $1360^{+58}_{-55}$ & $5.51^{+0.67}_{-0.61}$ & $0.079^{+0.020}_{-0.016}$ & 698.1 \\
RXC1825 & $1719^{+24}_{-25}$ & $3.35^{+0.20}_{-0.19}$ & $0.001^{+0.001}_{-0.000}$ & 1.7 \\
ZW1215 & $2200^{+74}_{-64}$ & $2.11^{+0.22}_{-0.18}$ & $0.003^{+0.006}_{-0.002}$ & 5.4 \\
\hline 
\end{tabular}

\setlength{\tabcolsep}{6pt} 
\begin{tabular}{c cccc}
\hline \hline
 Name & \multicolumn{4}{c}{EIN} \\ \\
  & kpc & $c$ & $\sigma_{T, int}$ & $\ln E$ \\
A85 & $636^{+40}_{-73}$ & $1.49^{+0.18}_{-0.08}$ & $0.026^{+0.021}_{-0.008}$ & 4.6 \\
A644 & $325^{+60}_{-53}$ & $2.99^{+0.52}_{-0.42}$ & $0.062^{+0.017}_{-0.012}$ & -12.7 \\
A1644 & $1119^{+107}_{-207}$ & $0.66^{+0.15}_{-0.06}$ & $0.005^{+0.047}_{-0.003}$ & -3.1 \\
A1795 & $480^{+23}_{-21}$ & $1.92^{+0.08}_{-0.07}$ & $0.009^{+0.007}_{-0.006}$ & 3.5 \\
A2029 & $571^{+29}_{-38}$ & $1.95^{+0.13}_{-0.08}$ & $0.020^{+0.010}_{-0.008}$ & -3.4 \\
A2142 & $866^{+42}_{-39}$ & $1.27^{+0.05}_{-0.05}$ & $0.001^{+0.000}_{-0.000}$ & -3.1 \\
A2255 & $959^{+97}_{-124}$ & $0.88^{+0.14}_{-0.09}$ & $0.002^{+0.009}_{-0.001}$ & -6.5 \\
A2319 & $403^{+46}_{-47}$ & $2.64^{+0.34}_{-0.26}$ & $0.067^{+0.017}_{-0.013}$ & -15.9 \\
A3158 & $530^{+34}_{-52}$ & $1.57^{+0.16}_{-0.09}$ & $0.004^{+0.028}_{-0.003}$ & -0.2 \\
A3266 & $765^{+61}_{-91}$ & $1.34^{+0.18}_{-0.10}$ & $0.057^{+0.015}_{-0.013}$ & -9.1 \\
HydraA & $299^{+33}_{-29}$ & $2.45^{+0.21}_{-0.21}$ & $0.046^{+0.009}_{-0.006}$ & 696.3 \\
RXC1825 & $495^{+26}_{-66}$ & $1.71^{+0.25}_{-0.08}$ & $0.002^{+0.011}_{-0.001}$ & -1.9 \\
ZW1215 & $1341^{+145}_{-161}$ & $0.75^{+0.10}_{-0.07}$ & $0.004^{+0.008}_{-0.002}$ & 1.1 \\
\hline 
\end{tabular}
}
\hbox{
\setlength{\tabcolsep}{6pt} 
\begin{tabular}{c cccc}
\hline \hline
 Name & \multicolumn{4}{c}{ISO} \\ \\
  & kpc & $c$ & $\sigma_{T, int}$ & $\ln E$ \\
A85 & $190^{+9}_{-9}$ & $7.71^{+0.29}_{-0.28}$ & $0.057^{+0.011}_{-0.008}$ & -9.5 \\
A644 & $168^{+9}_{-12}$ & $9.03^{+0.59}_{-0.32}$ & $0.029^{+0.028}_{-0.027}$ & -6.2 \\
A1644 & $298^{+19}_{-13}$ & $4.50^{+0.16}_{-0.19}$ & $0.001^{+0.000}_{-0.000}$ & -4.0 \\
A1795 & $144^{+8}_{-6}$ & $9.45^{+0.32}_{-0.39}$ & $0.055^{+0.012}_{-0.008}$ & -8.5 \\
A2029 & $195^{+13}_{-10}$ & $8.81^{+0.37}_{-0.39}$ & $0.078^{+0.015}_{-0.012}$ & -15.4 \\
A2142 & $240^{+13}_{-11}$ & $7.35^{+0.27}_{-0.26}$ & $0.060^{+0.011}_{-0.008}$ & -13.5 \\
A2255 & $444^{+42}_{-41}$ & $3.62^{+0.28}_{-0.22}$ & $0.001^{+0.001}_{-0.000}$ & -2.3 \\
A2319 & $241^{+14}_{-9}$ & $7.22^{+0.23}_{-0.32}$ & $0.041^{+0.018}_{-0.009}$ & -8.4 \\
A3158 & $223^{+15}_{-13}$ & $6.19^{+0.33}_{-0.29}$ & $0.030^{+0.009}_{-0.011}$ & -5.8 \\
A3266 & $312^{+30}_{-21}$ & $5.61^{+0.31}_{-0.30}$ & $0.076^{+0.015}_{-0.013}$ & -13.4 \\
HydraA & $221^{+63}_{-31}$ & $6.22^{+0.45}_{-0.42}$ & $0.324^{+0.029}_{-0.028}$ & 687.0 \\
RXC1825 & $233^{+10}_{-9}$ & $5.98^{+0.19}_{-0.18}$ & $0.001^{+0.001}_{-0.000}$ & -6.6 \\
ZW1215 & $313^{+18}_{-16}$ & $5.25^{+0.19}_{-0.17}$ & $0.001^{+0.000}_{-0.000}$ & -6.0 \\
\hline 
\end{tabular}

\setlength{\tabcolsep}{6pt} 
\begin{tabular}{c cccc}
\hline \hline
 Name & \multicolumn{4}{c}{BUR} \\ \\
  & kpc & $c$ & $\sigma_{T, int}$ & $\ln E$ \\
A85 & $188^{+9}_{-11}$ & $4.87^{+0.21}_{-0.19}$ & $0.060^{+0.013}_{-0.009}$ & -7.3 \\
A644 & $172^{+6}_{-7}$ & $5.64^{+0.16}_{-0.15}$ & $0.001^{+0.000}_{-0.000}$ & -6.8 \\
A1644 & $294^{+17}_{-16}$ & $2.74^{+0.13}_{-0.11}$ & $0.001^{+0.000}_{-0.000}$ & -3.7 \\
A1795 & $146^{+7}_{-6}$ & $5.86^{+0.22}_{-0.22}$ & $0.047^{+0.012}_{-0.009}$ & -6.8 \\
A2029 & $186^{+9}_{-10}$ & $5.73^{+0.25}_{-0.21}$ & $0.064^{+0.010}_{-0.009}$ & -14.2 \\
A2142 & $237^{+11}_{-11}$ & $4.59^{+0.19}_{-0.16}$ & $0.060^{+0.014}_{-0.010}$ & -12.4 \\
A2255 & $409^{+28}_{-28}$ & $2.31^{+0.14}_{-0.12}$ & $0.002^{+0.002}_{-0.001}$ & -2.9 \\
A2319 & $229^{+11}_{-13}$ & $4.68^{+0.23}_{-0.18}$ & $0.042^{+0.013}_{-0.012}$ & -9.2 \\
A3158 & $217^{+12}_{-13}$ & $3.90^{+0.21}_{-0.17}$ & $0.026^{+0.013}_{-0.022}$ & -5.3 \\
A3266 & $313^{+30}_{-26}$ & $3.44^{+0.26}_{-0.21}$ & $0.074^{+0.015}_{-0.013}$ & -14.2 \\
HydraA & $96^{+15}_{-11}$ & $7.01^{+0.53}_{-0.57}$ & $0.097^{+0.019}_{-0.013}$ & 688.4 \\
RXC1825 & $220^{+10}_{-10}$ & $3.86^{+0.15}_{-0.14}$ & $0.001^{+0.000}_{-0.000}$ & -4.5 \\
ZW1215 & $303^{+20}_{-19}$ & $3.30^{+0.15}_{-0.14}$ & $0.002^{+0.002}_{-0.001}$ & -4.7 \\
\hline 
\end{tabular}
}
\hbox{
\setlength{\tabcolsep}{6pt} 
\begin{tabular}{c cccc}
\hline \hline
 Name & \multicolumn{4}{c}{HER} \\ \\
  & kpc & $c$ & $\sigma_{T, int}$ & $\ln E$ \\
A85 & $914^{+55}_{-109}$ & $1.92^{+0.26}_{-0.09}$ & $0.027^{+0.019}_{-0.010}$ & 5.8 \\
A644 & $478^{+73}_{-74}$ & $3.73^{+0.67}_{-0.45}$ & $0.090^{+0.027}_{-0.022}$ & -10.5 \\
A1644 & $2026^{+230}_{-188}$ & $0.72^{+0.07}_{-0.06}$ & $0.001^{+0.001}_{-0.000}$ & -2.4 \\
A1795 & $721^{+31}_{-37}$ & $2.37^{+0.12}_{-0.09}$ & $0.017^{+0.016}_{-0.007}$ & 5.2 \\
A2029 & $925^{+69}_{-42}$ & $2.27^{+0.10}_{-0.13}$ & $0.028^{+0.008}_{-0.007}$ & -5.8 \\
A2142 & $1262^{+46}_{-49}$ & $1.63^{+0.06}_{-0.05}$ & $0.006^{+0.011}_{-0.004}$ & -2.7 \\
A2255 & $2235^{+203}_{-289}$ & $0.73^{+0.11}_{-0.05}$ & $0.001^{+0.001}_{-0.000}$ & -6.2 \\
A2319 & $674^{+55}_{-77}$ & $2.99^{+0.41}_{-0.22}$ & $0.068^{+0.014}_{-0.010}$ & -12.5 \\
A3158 & $1203^{+137}_{-80}$ & $1.36^{+0.09}_{-0.12}$ & $0.001^{+0.001}_{-0.000}$ & -0.4 \\
A3266 & $1952^{+285}_{-189}$ & $1.03^{+0.10}_{-0.12}$ & $0.036^{+0.008}_{-0.008}$ & -8.7 \\
HydraA & $473^{+64}_{-43}$ & $2.86^{+0.24}_{-0.26}$ & $0.055^{+0.010}_{-0.007}$ & 694.9 \\
RXC1825 & $913^{+49}_{-69}$ & $1.76^{+0.14}_{-0.09}$ & $0.002^{+0.007}_{-0.001}$ & 0.1 \\
ZW1215 & $1325^{+86}_{-113}$ & $1.38^{+0.12}_{-0.07}$ & $0.004^{+0.020}_{-0.002}$ & 1.4 \\
\hline 
\end{tabular}
}
\caption{For each mass model we have investigated (see Sect.~\ref{subsect:other}), we quote the following best-fit parameters:
scale radius $r_{\rm s}$ ($R_{200}$ in the case of NFW)  and the concentration (or normalization) as defined in equation~\ref{eq:m_mod}; 
the intrinsic scatter $\sigma_{T,int}$ of equation~\ref{eq:like}; and the logarithmic value of the evidence $E$.}
\label{tab:m_mod}
\end{table*}

\begin{table}[b]
\centering
\setlength{\tabcolsep}{3pt} 
\begin{tabular}{c c c c c} \hline \hline
 Name & $\chi^2_{\epsilon}$ (DOF) & $\chi^2_{T}$ (DOF) & $\chi^2_{P}$ (DOF) & $\chi^2_{\rm TOT}$ (DOF) \\
A85 & 66.9 (70) & 86.1 (13) & 10.2 (7) & 163.2 (90) \\
A644 & 43.5 (68) & 48.0 (12) & 85.1 (6) & 171.6 (86) \\
A1644 & 72.0 (65) & 53.5 (12) & 2.2 (8) & 127.7 (85) \\
A1795 & 45.7 (65) & 31.3 (13) & 8.2 (7) & 85.2 (85) \\
A2029 & 31.2 (65) & 120.3 (14) & 11.3 (7) & 162.8 (86) \\
A2142 & 77.5 (65) & 23.5 (12) & 2.0 (7) & 103.0 (84) \\
A2255 & 16.1 (58) & 3.3 (8) & 6.5 (8) & 25.9 (74) \\
A2319 & 69.5 (69) & 168.9 (19) & 5.0 (7) & 143.4 (95) \\
A3158 & 41.9 (68) & 20.4 (11) & 2.4 (7) & 64.7 (86) \\
A3266 & 90.3 (69) & 122.7 (13) & 23.8 (7) & 236.8 (89) \\
HydraA & 99.0 (67) & 2235.3 (12) &  -- (--) & 2334.3 (79) \\
RXC1825 & 51.7 (57) & 22.9 (11) & 5.0 (8) & 79.6 (76) \\
ZW1215 & 32.1 (61) & 9.4 (12) & 13.4 (7) & 54.9 (80) \\
\hline 
\end{tabular}
\caption{Values of  $\chi^2$ (and relative degrees of freedom DOF) of the three components (emissivity $\epsilon$, temperature $T$, and pressure $P$) 
of the likelihood presented in equation~\ref{eq:like} and of the total $\chi^2$ for the best-fit mass model. }
\label{tab:m_chi2}
\end{table}

\FloatBarrier

\begin{figure*}[hbt]
\centering
\hbox{
\includegraphics[width=0.245\textwidth, keepaspectratio]{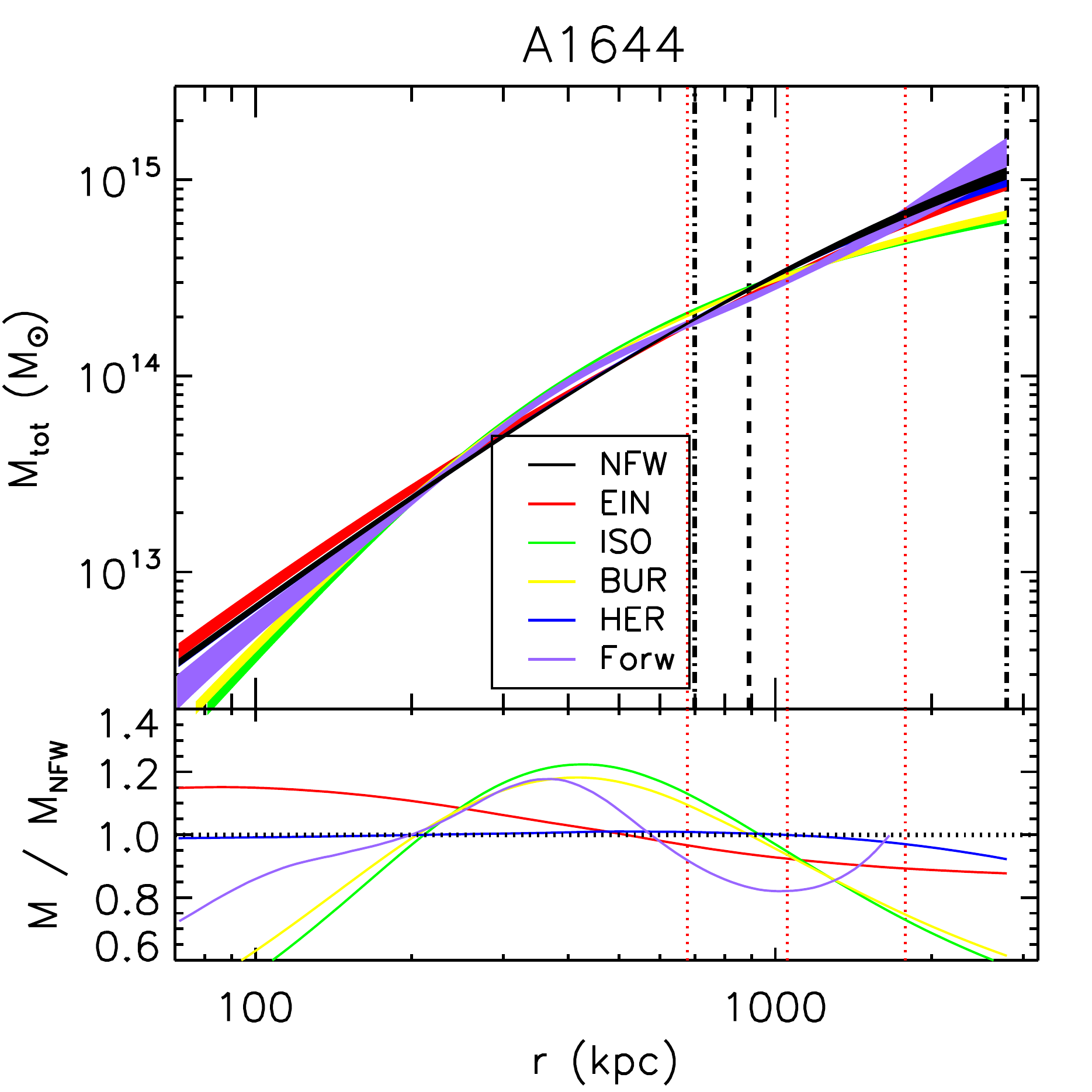}
\includegraphics[width=0.245\textwidth, keepaspectratio]{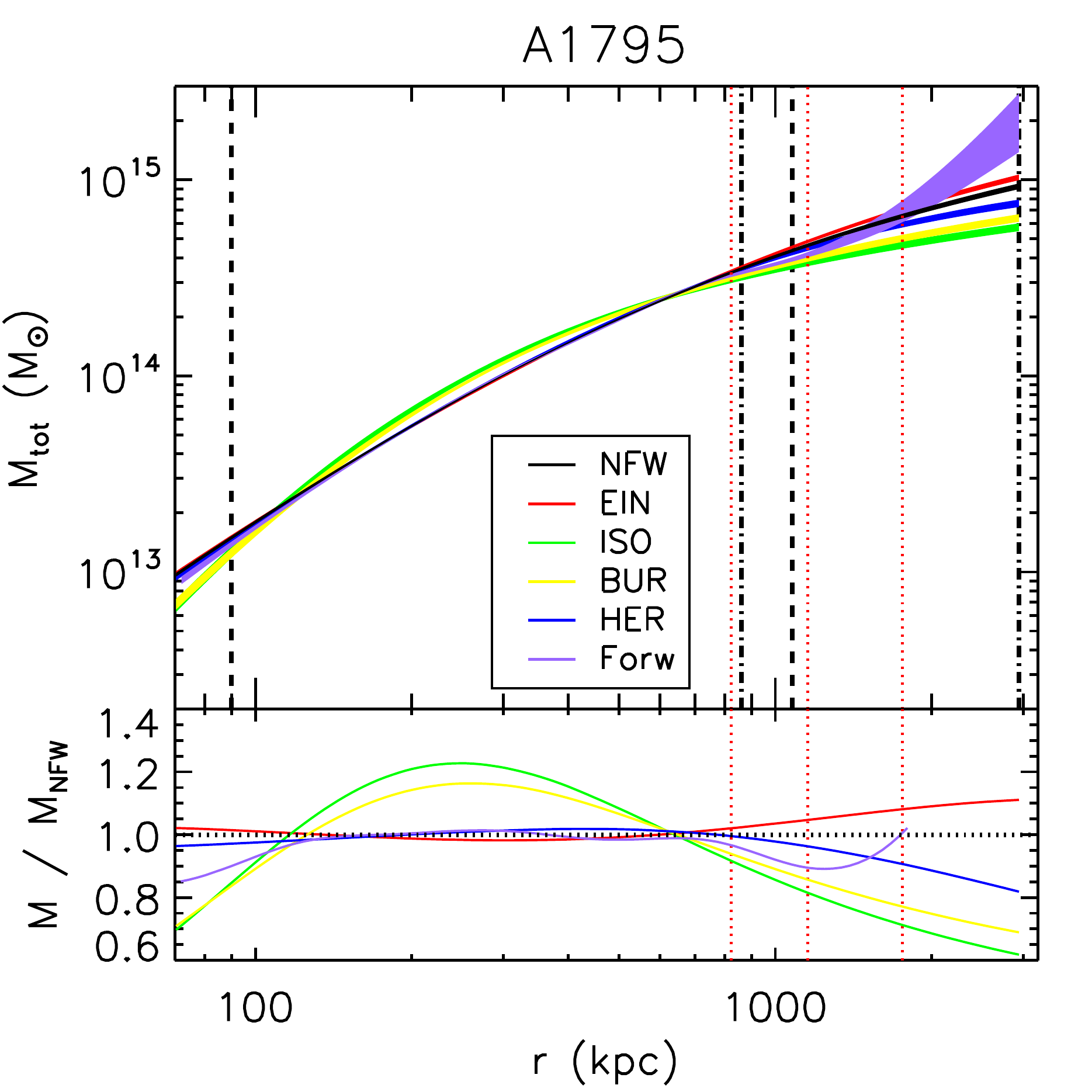}
\includegraphics[width=0.245\textwidth, keepaspectratio]{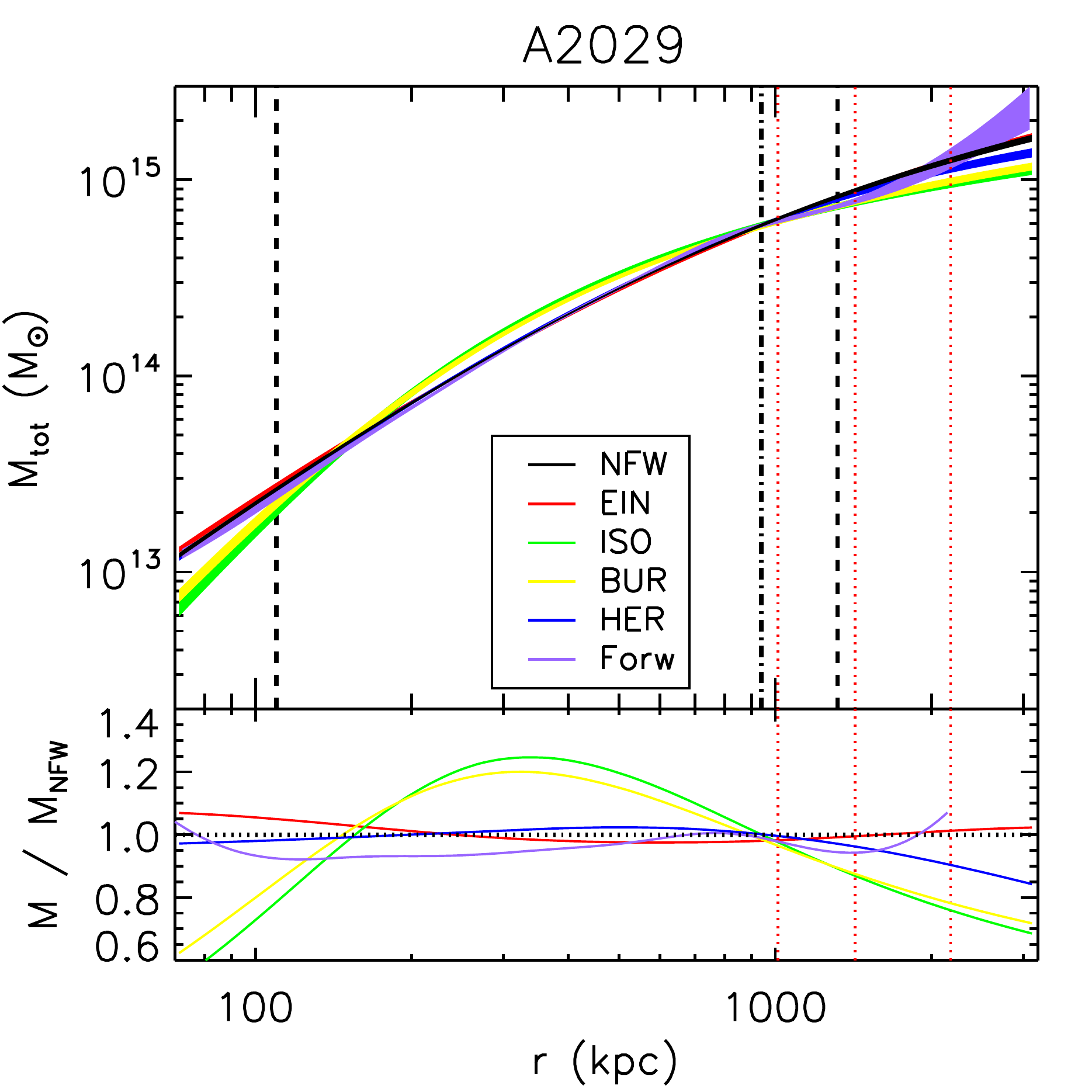}
\includegraphics[width=0.245\textwidth, keepaspectratio]{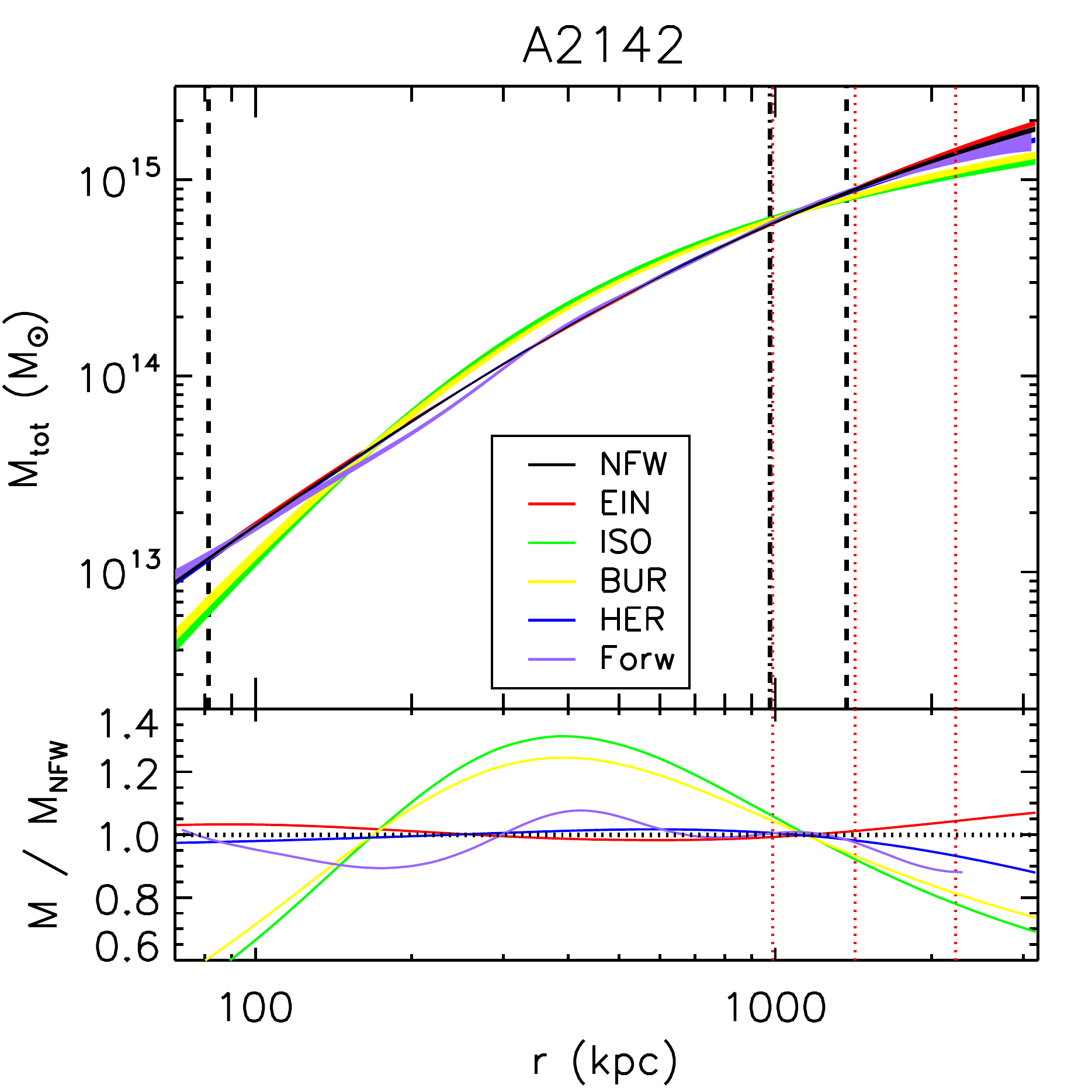}
} \hbox{
\includegraphics[width=0.245\textwidth, keepaspectratio]{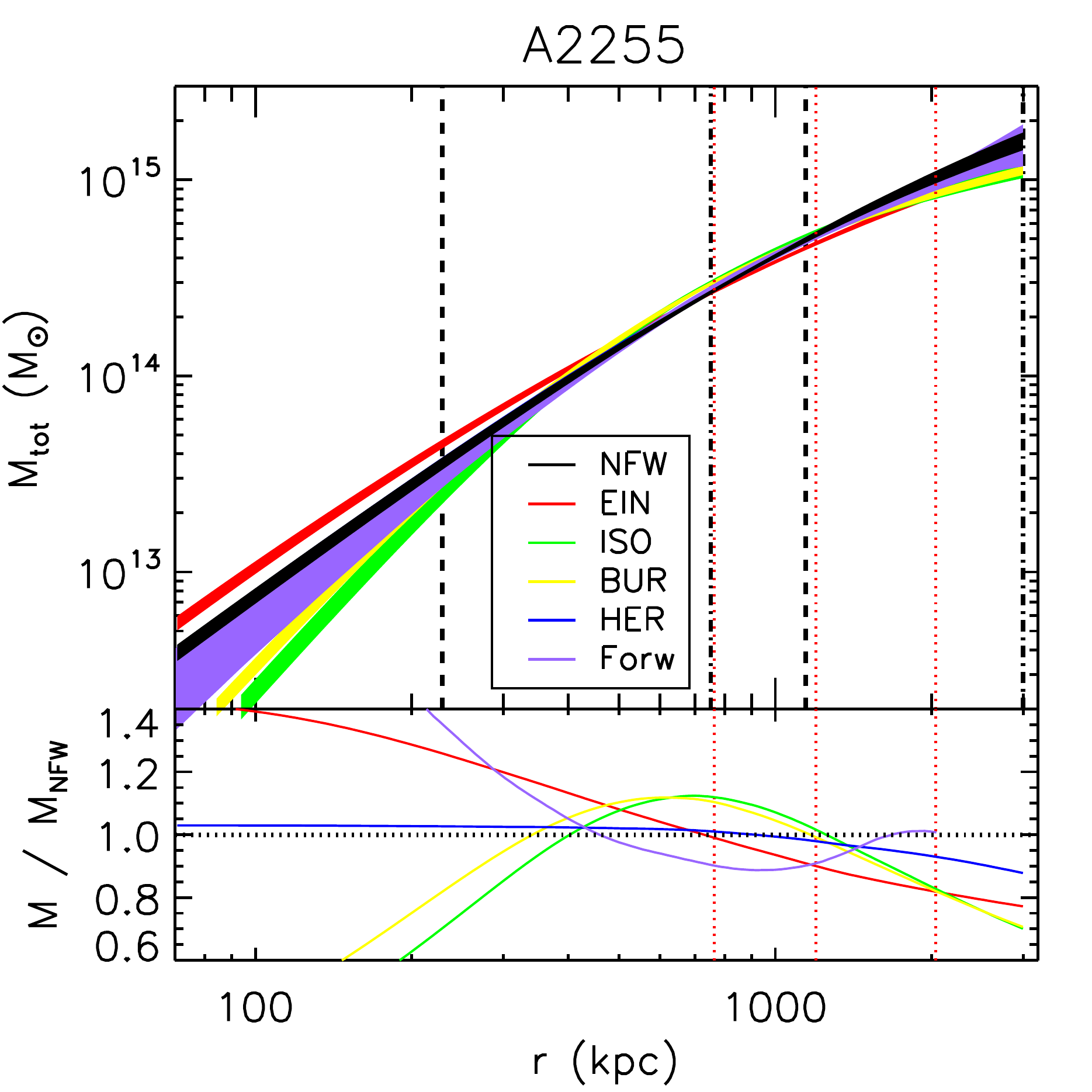}
\includegraphics[width=0.245\textwidth, keepaspectratio]{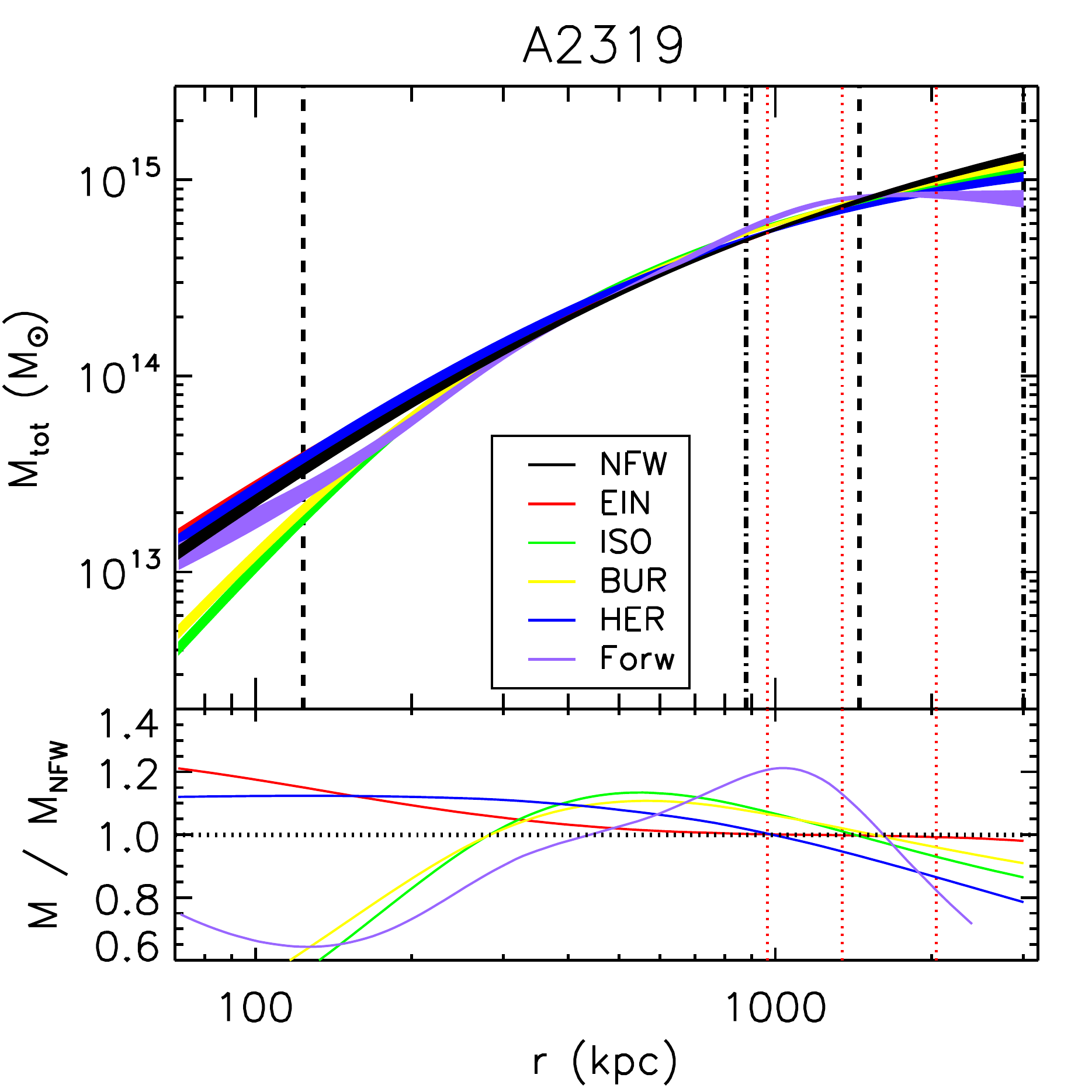}
\includegraphics[width=0.245\textwidth, keepaspectratio]{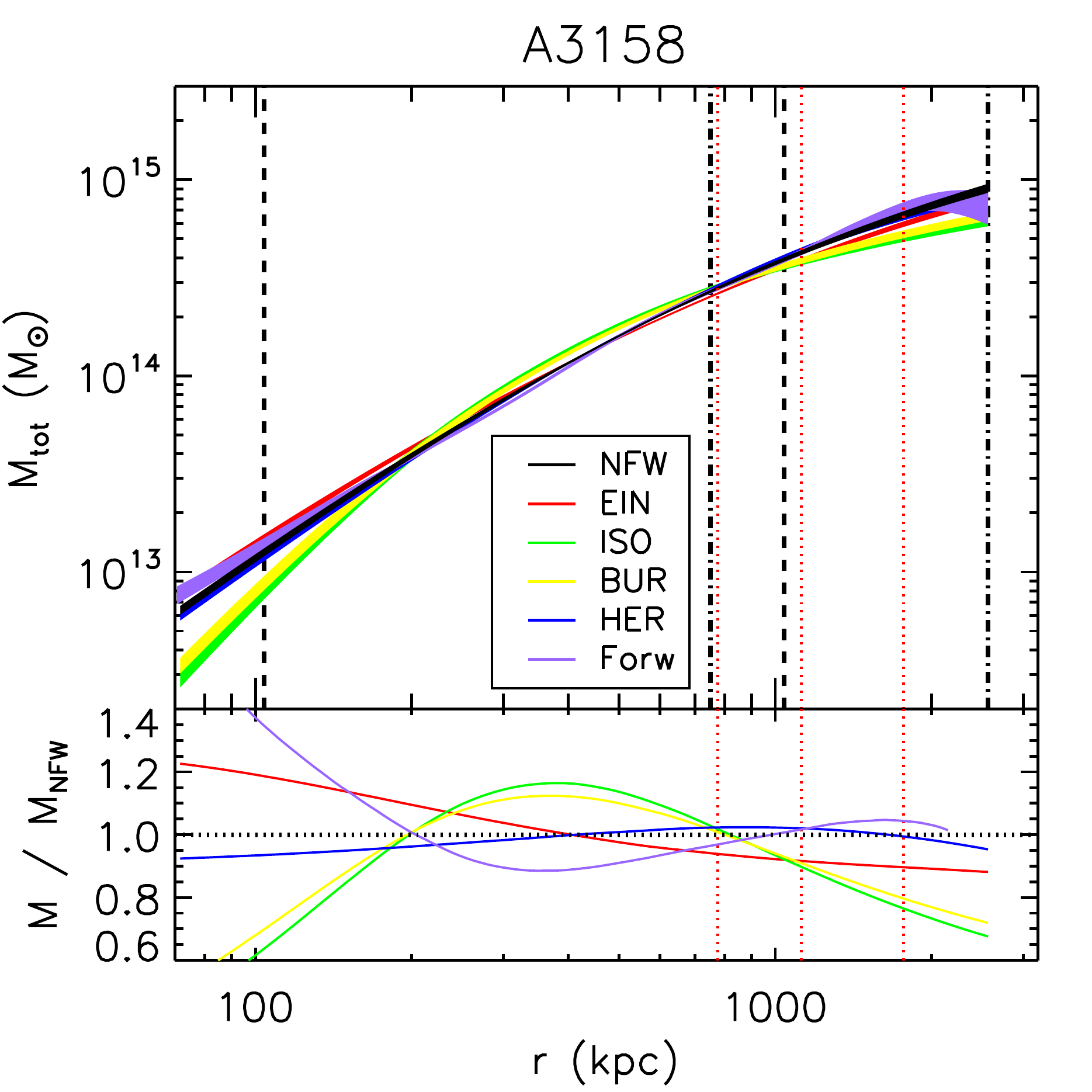}
\includegraphics[width=0.245\textwidth, keepaspectratio]{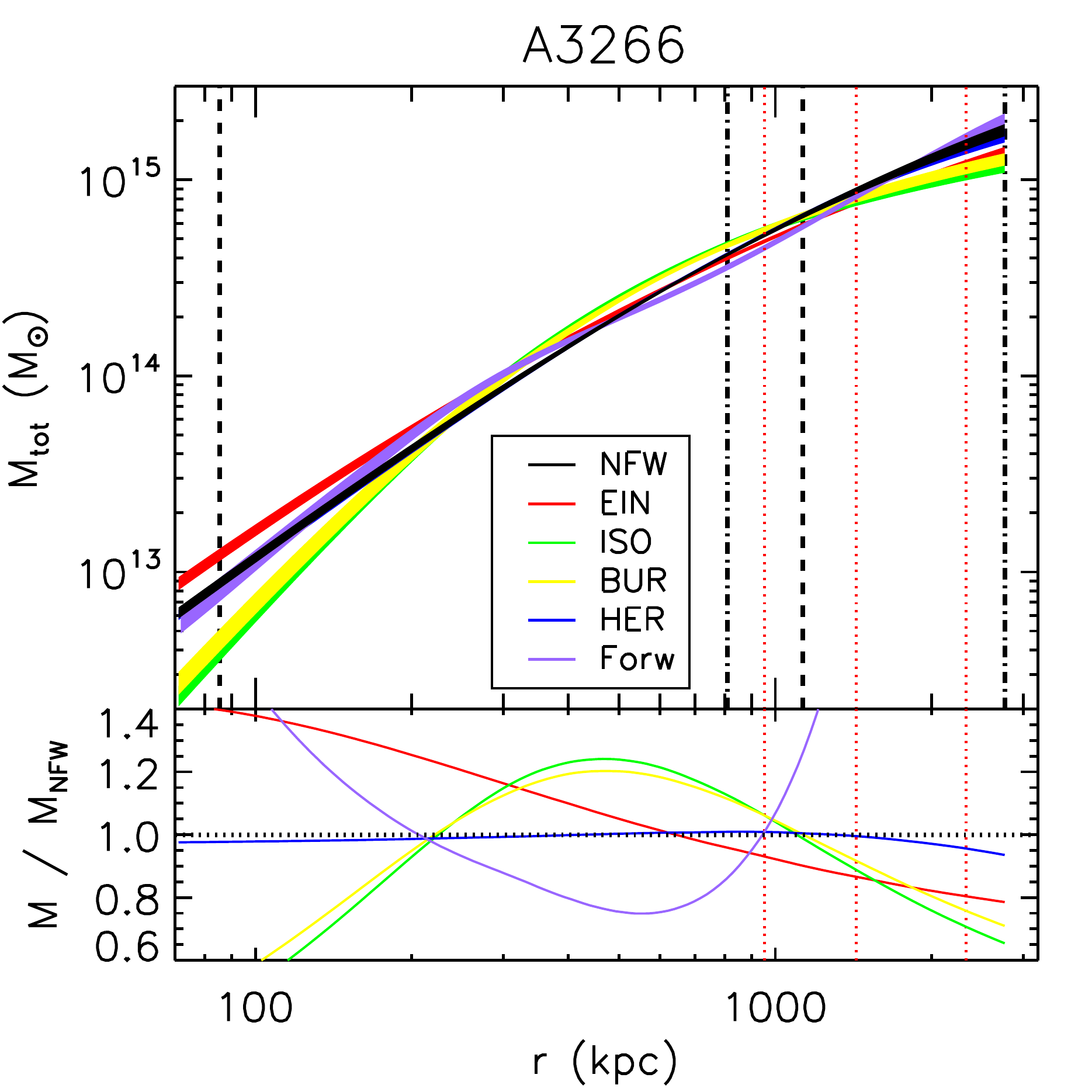}
} \hbox{
\includegraphics[width=0.245\textwidth, keepaspectratio]{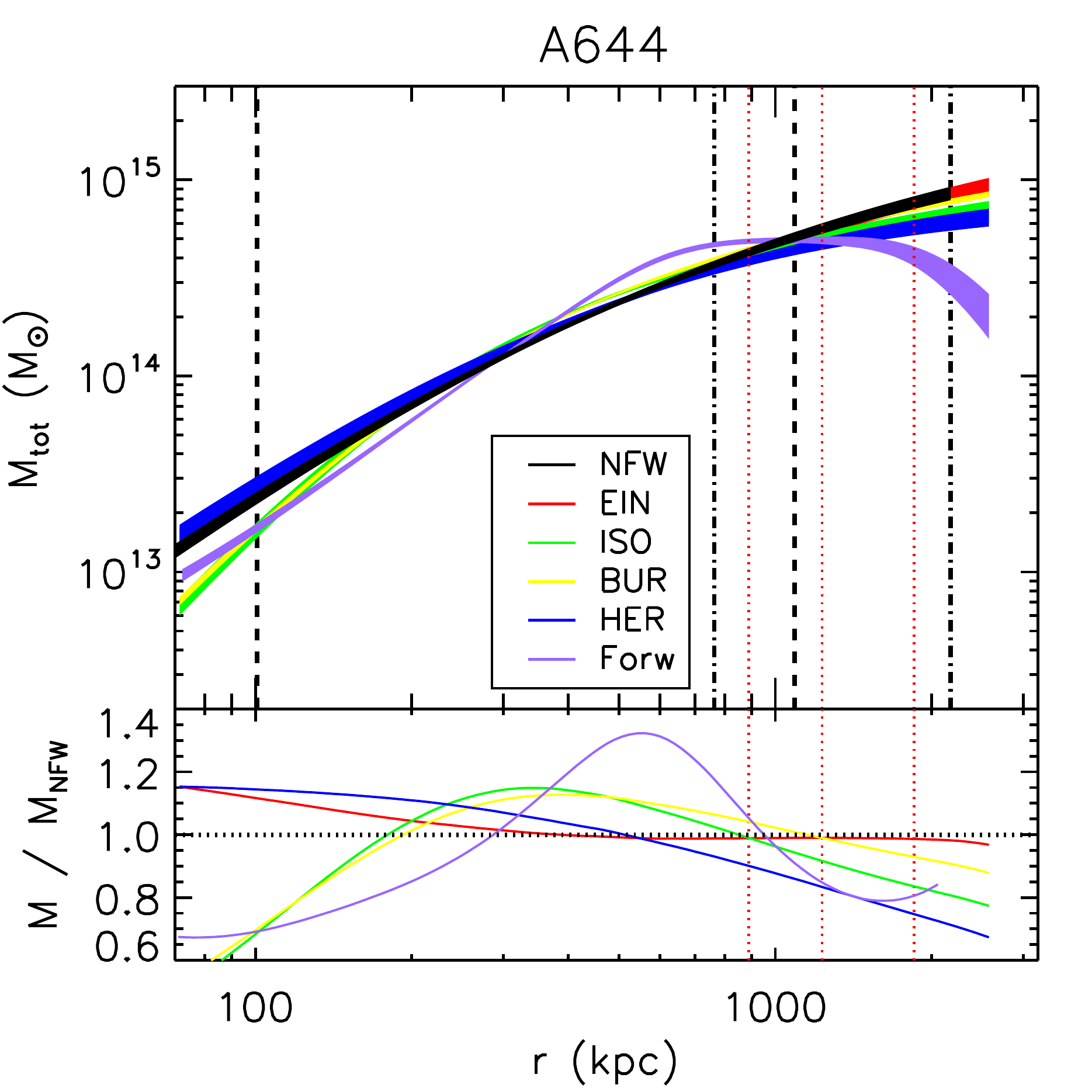}
\includegraphics[width=0.245\textwidth, keepaspectratio]{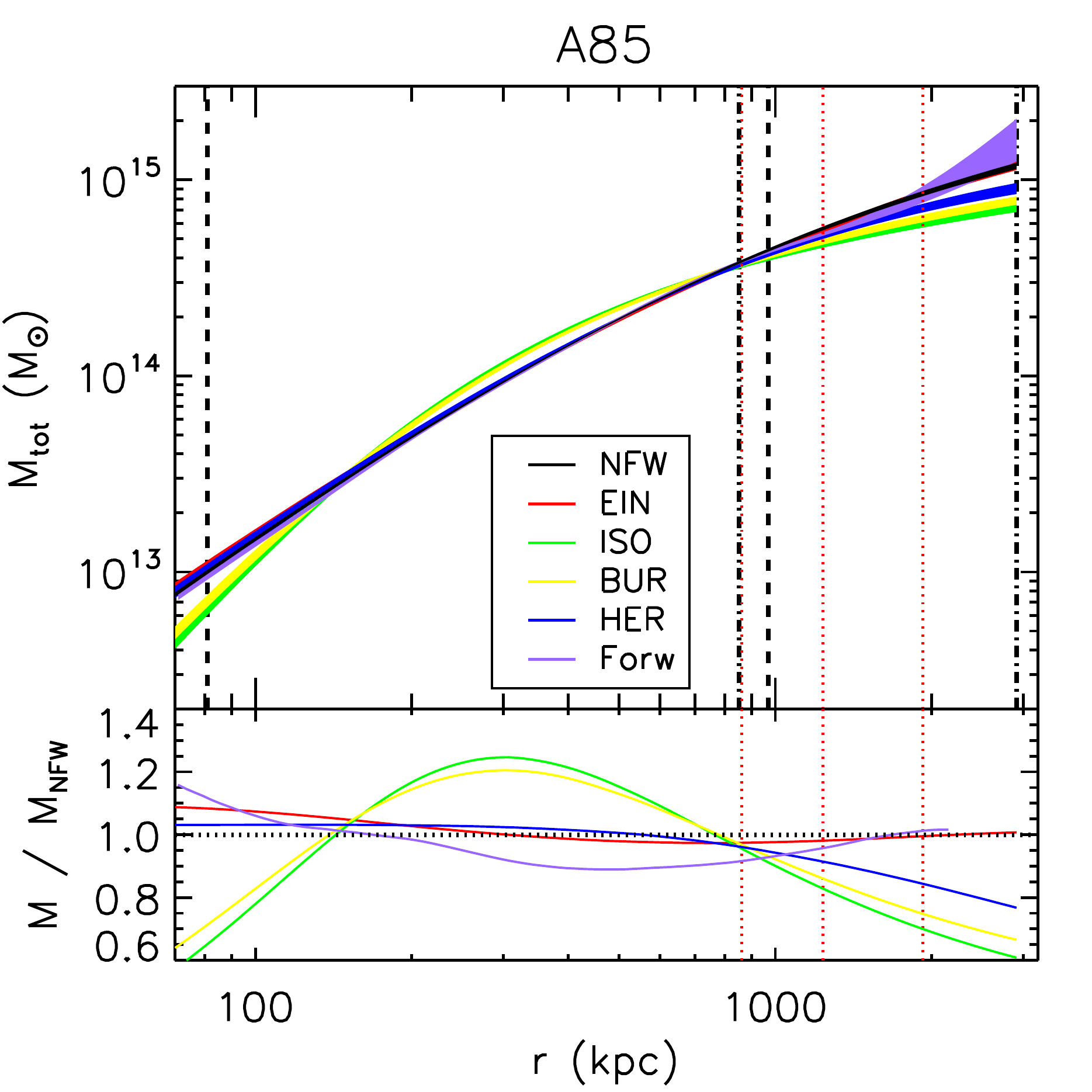}
\includegraphics[width=0.245\textwidth, keepaspectratio]{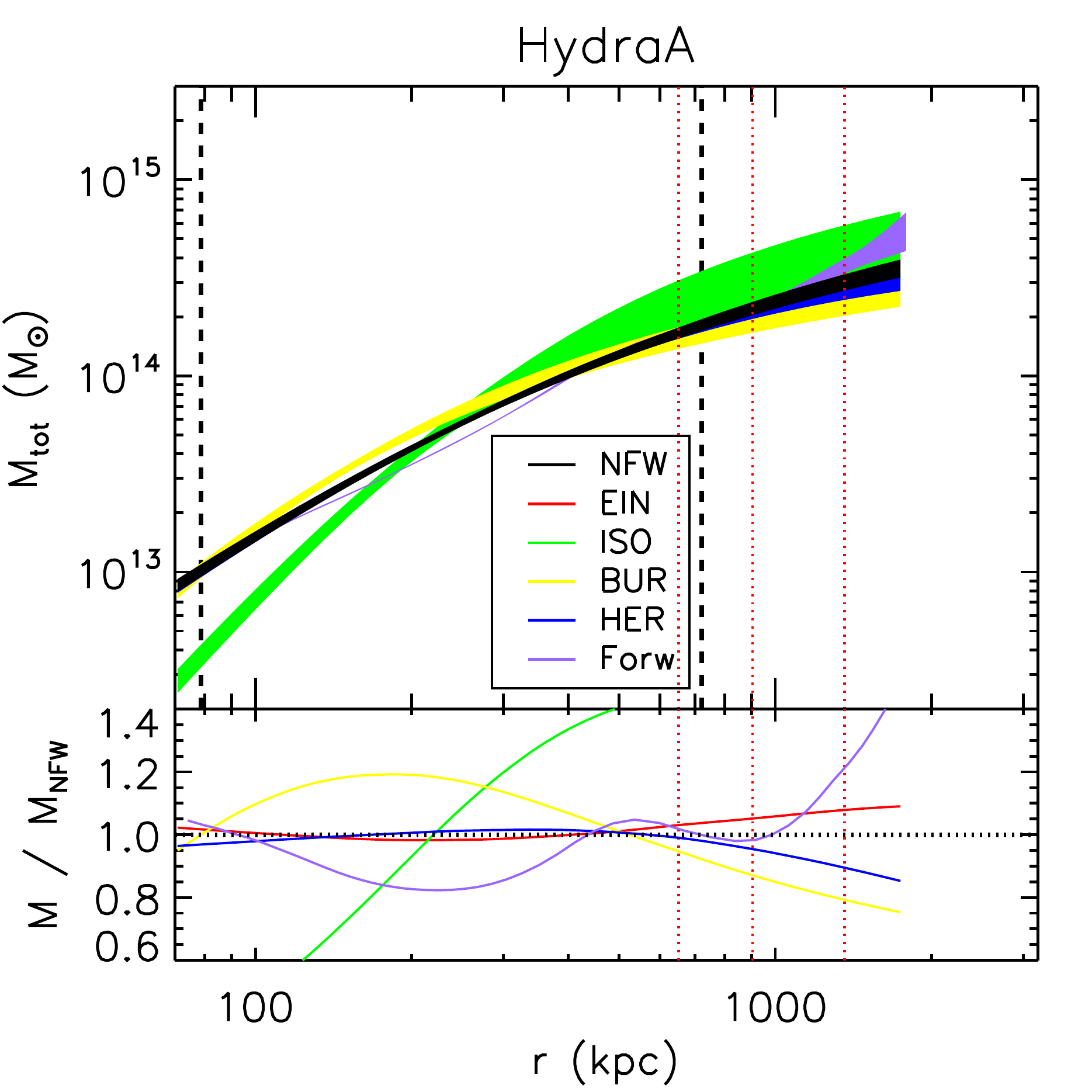}
\includegraphics[width=0.245\textwidth, keepaspectratio]{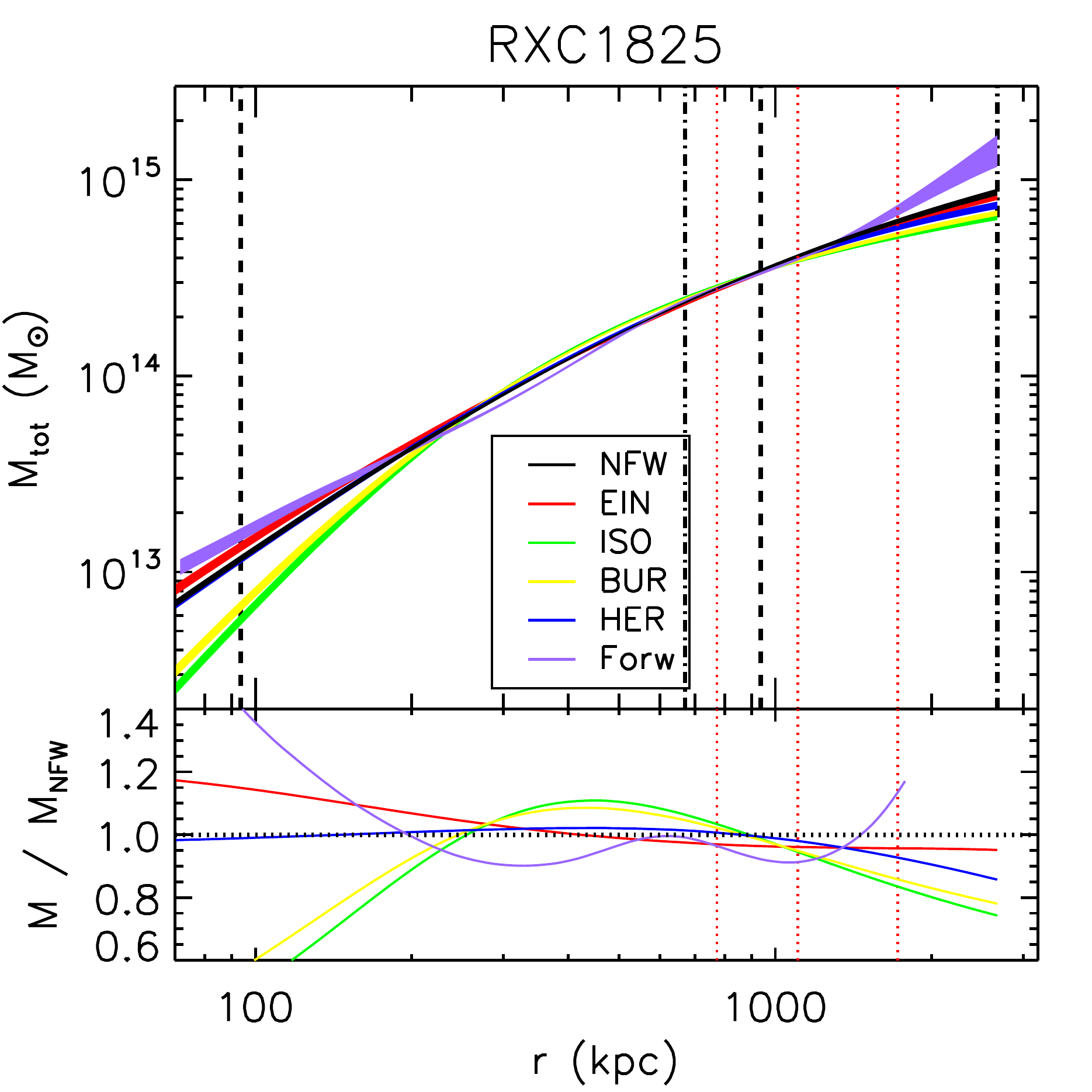}
} \hbox{
\includegraphics[width=0.245\textwidth, keepaspectratio]{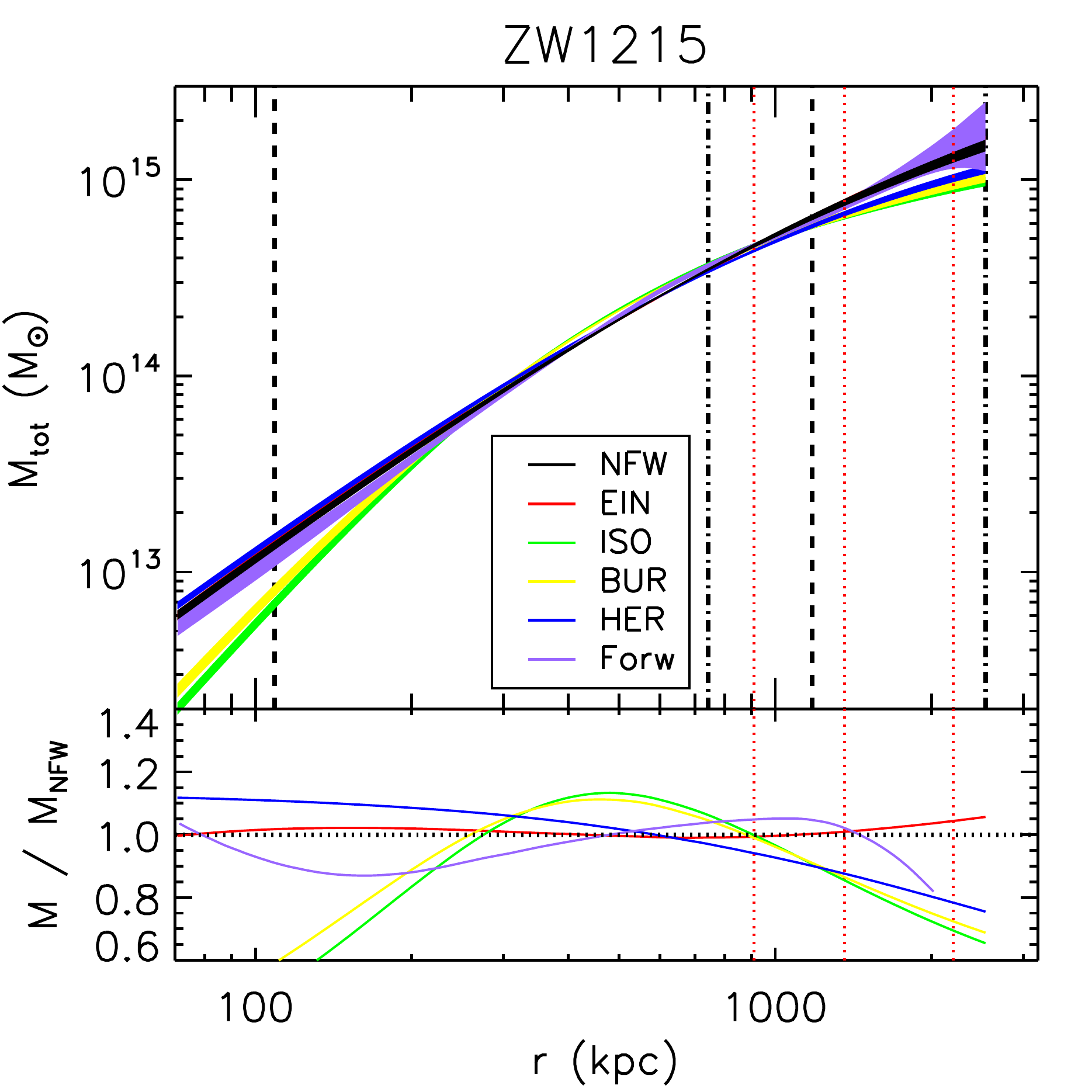}
}
\caption{Hydrostatic mass radial profiles obtained from the {forward} method and 
the mass models applied to the {backward} method and described in Sect.~\ref{subsect:other}.
Vertical lines indicate: (dotted) $R_{1000}$, $R_{500}$, and $R_{200}$; 
(dashed) third radial bin (a minimum number of three independent bins are needed to constrain a mass model with two free parameters) 
and upper limit in measurements of the spectroscopic temperature; 
(dot-dashed) radial range covered from SZ pressure profile after the exclusion of the three inner points and used in the reconstruction of the mass profile.
(Bottom panels) Ratios of the different mass profiles to the NFW profile adopted as reference model.
} \label{fig:m_prof}
\end{figure*}

\clearpage
\section{EG and MOND mass profiles}
\label{app_eg}

We present here the mass profiles for each X-COP object as obtained from (i) the {backward} NFW method, (ii) emergent gravity, and (iii) Modified Newtonian Dynamics.

\begin{figure*}[hbt]
\centering
\hbox{
\includegraphics[width=0.245\textwidth, keepaspectratio]{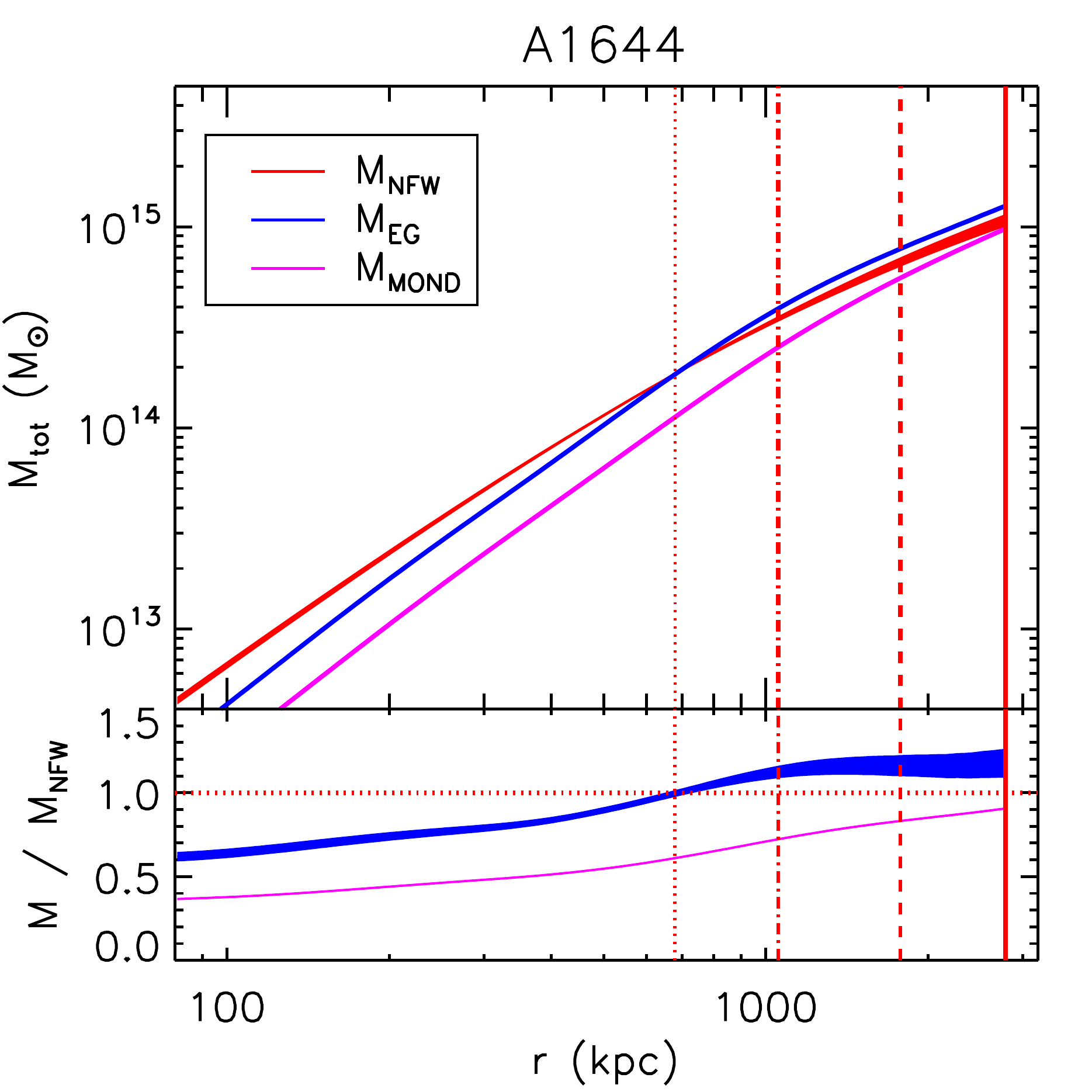}
\includegraphics[width=0.245\textwidth, keepaspectratio]{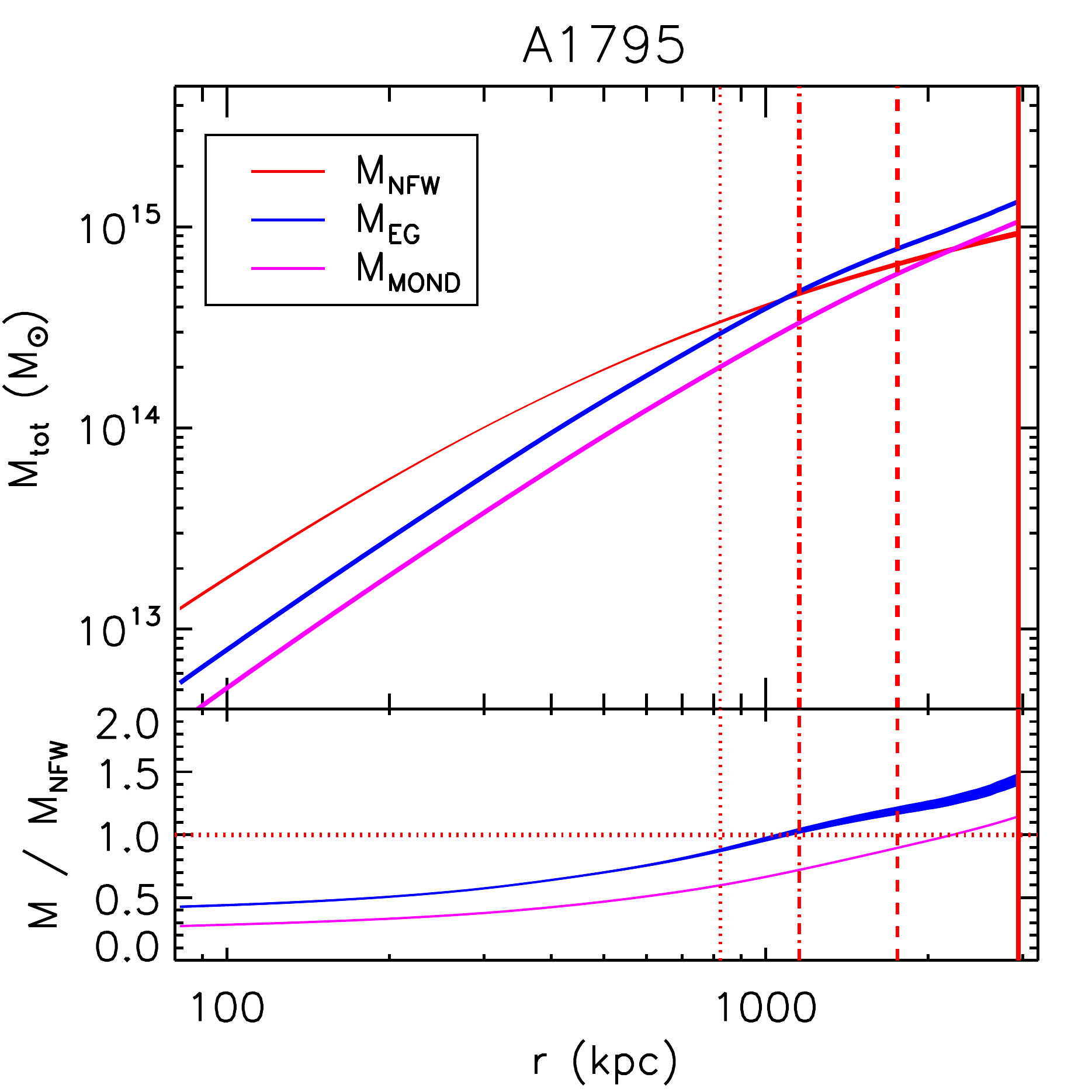}
\includegraphics[width=0.245\textwidth, keepaspectratio]{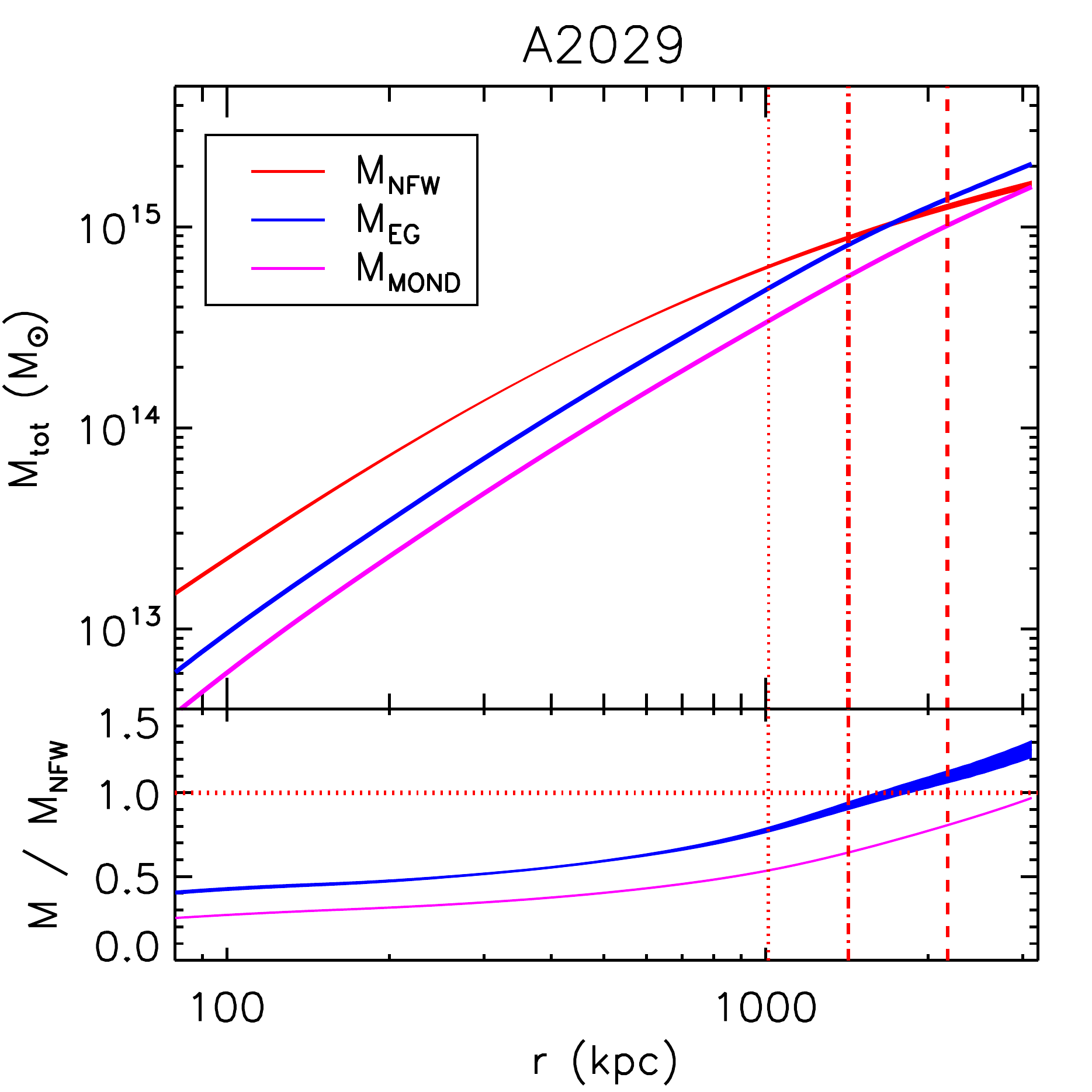}
\includegraphics[width=0.245\textwidth, keepaspectratio]{figs/A2142_eg.pdf}
} \hbox{
\includegraphics[width=0.245\textwidth, keepaspectratio]{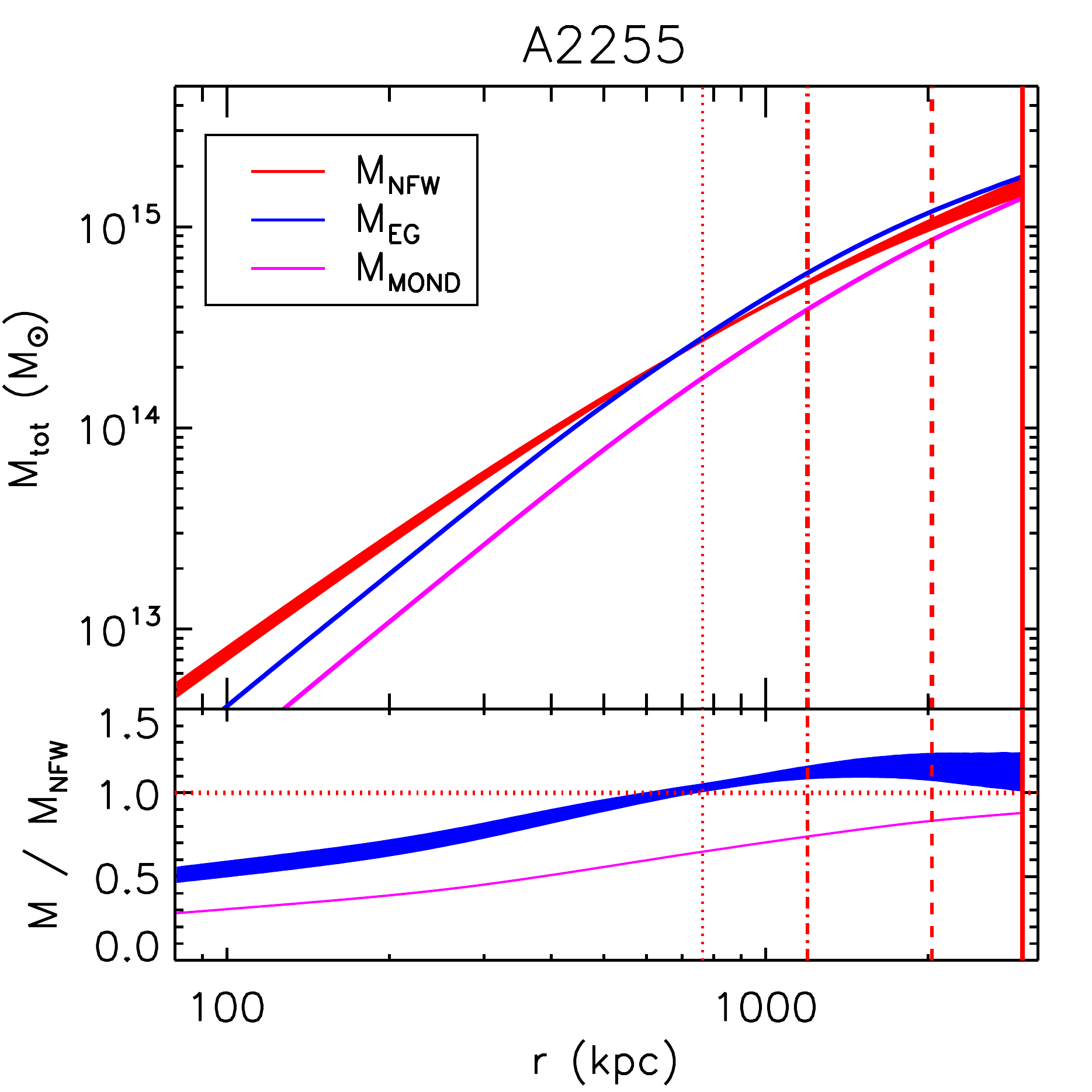}
\includegraphics[width=0.245\textwidth, keepaspectratio]{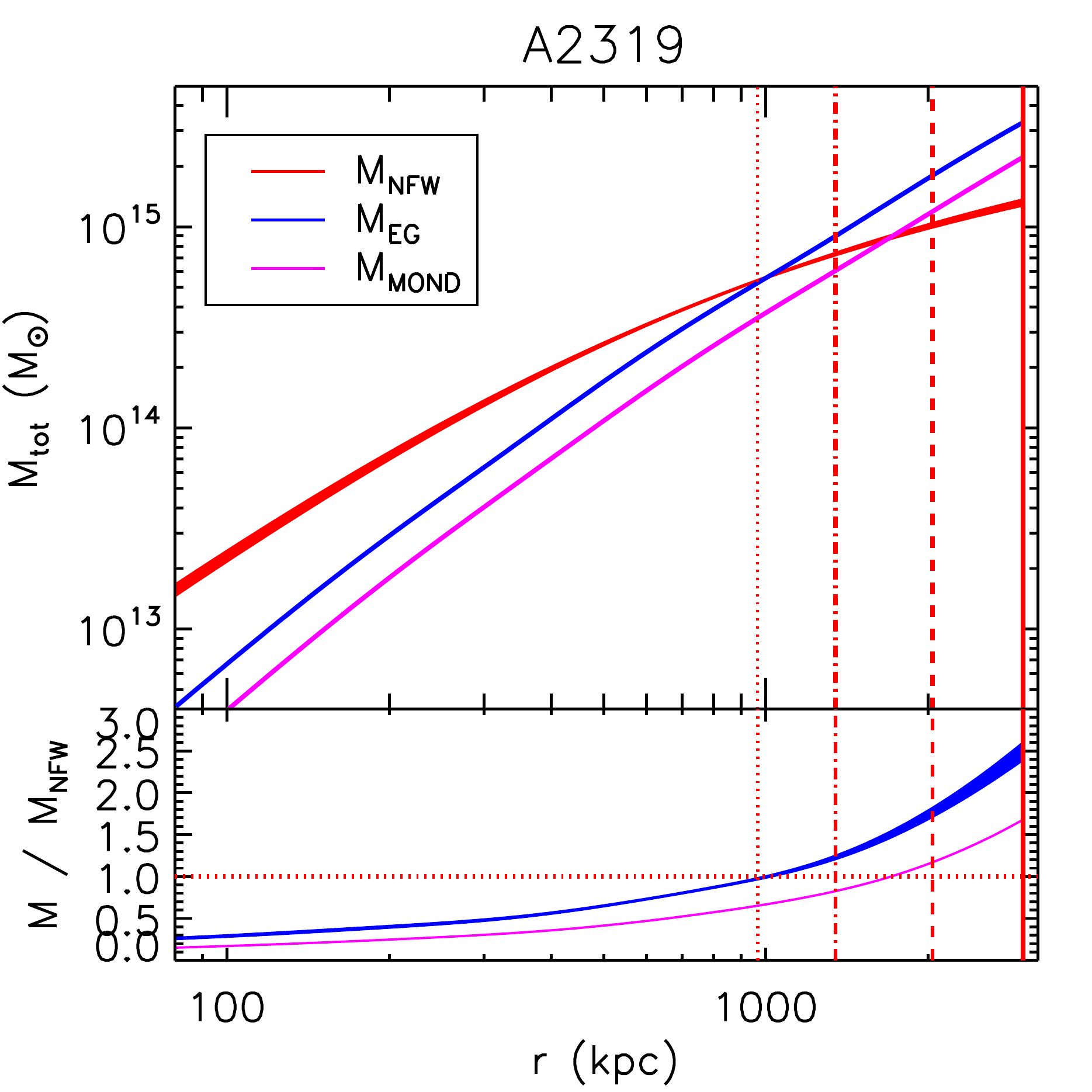}
\includegraphics[width=0.245\textwidth, keepaspectratio]{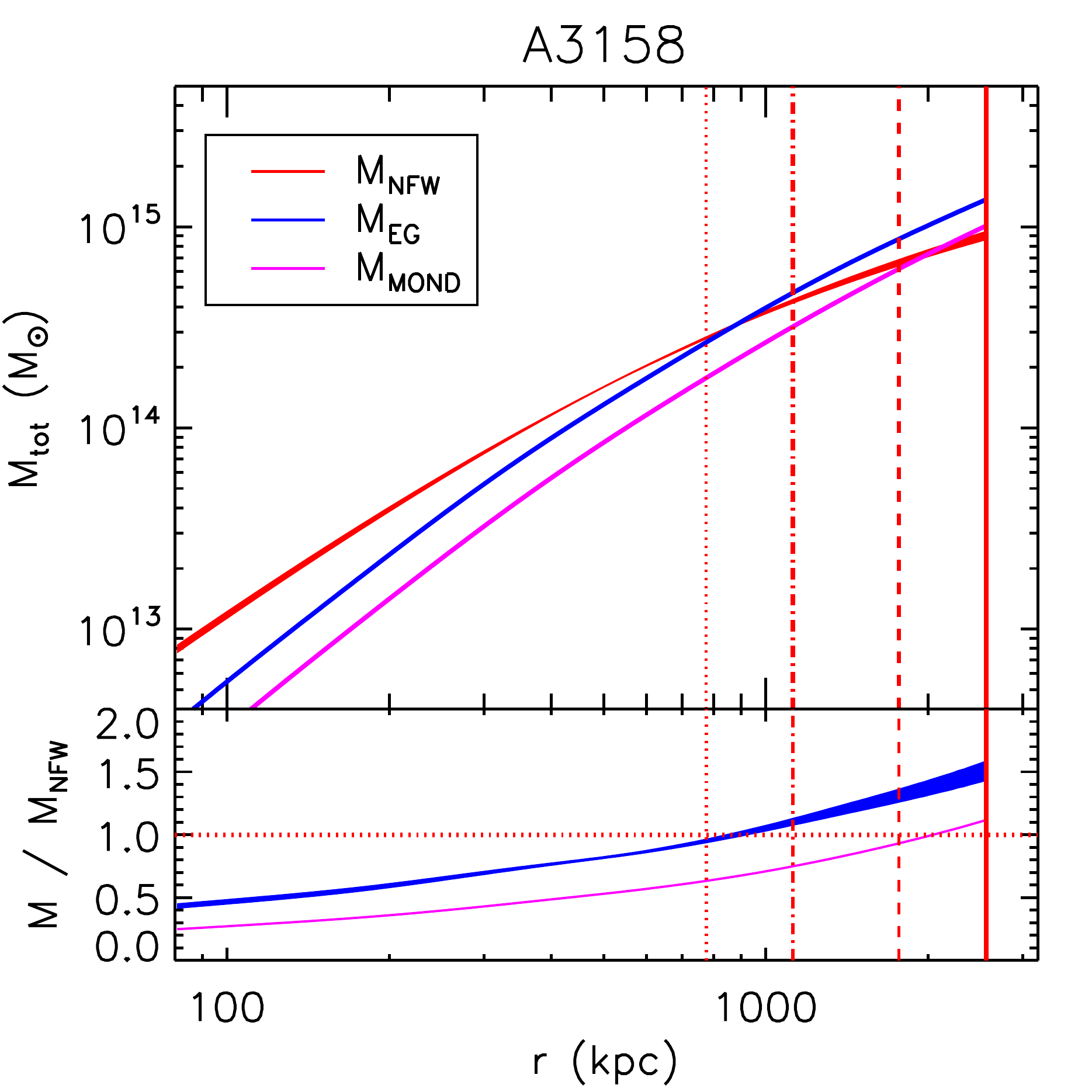}
\includegraphics[width=0.245\textwidth, keepaspectratio]{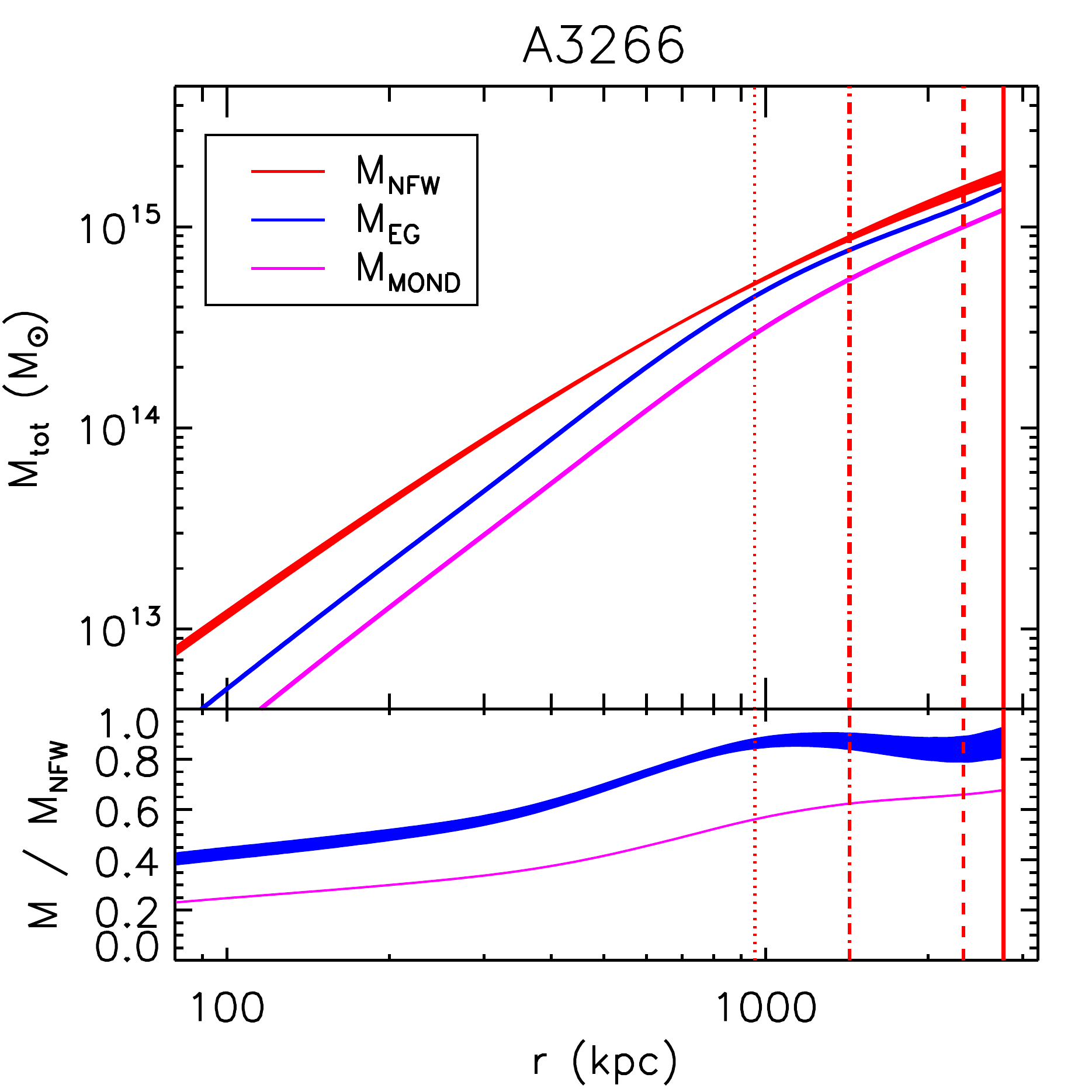}
} \hbox{
\includegraphics[width=0.245\textwidth, keepaspectratio]{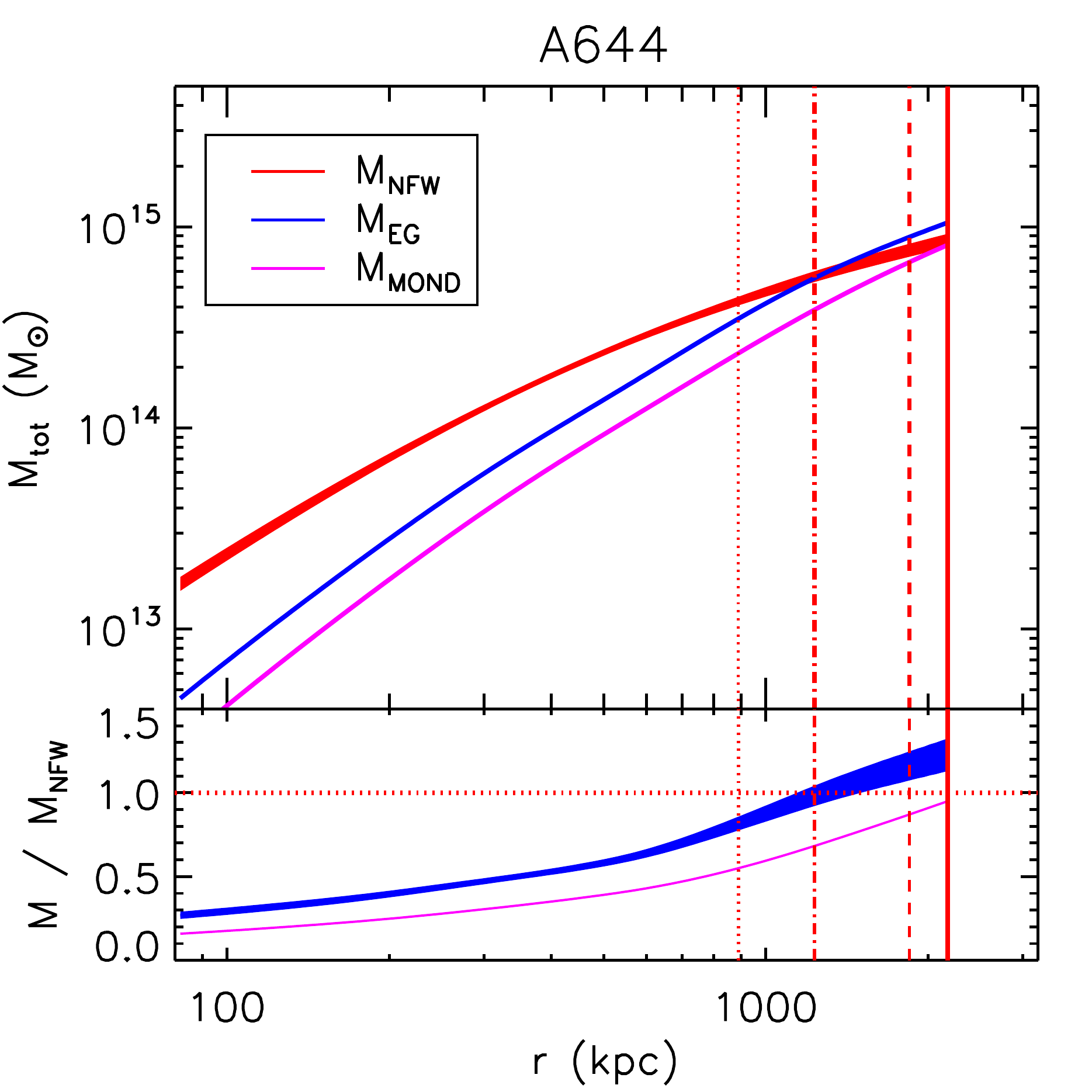}
\includegraphics[width=0.245\textwidth, keepaspectratio]{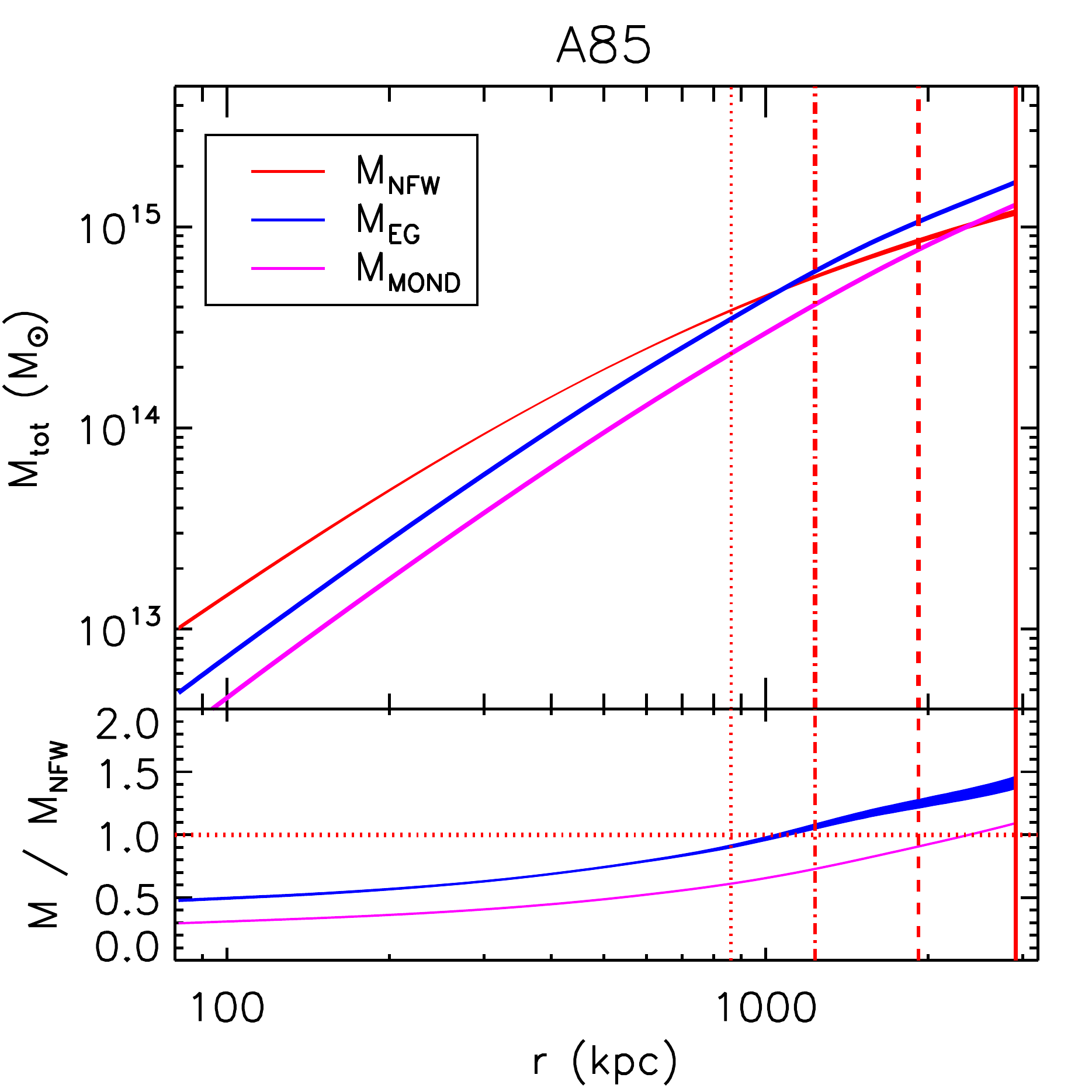}
\includegraphics[width=0.245\textwidth, keepaspectratio]{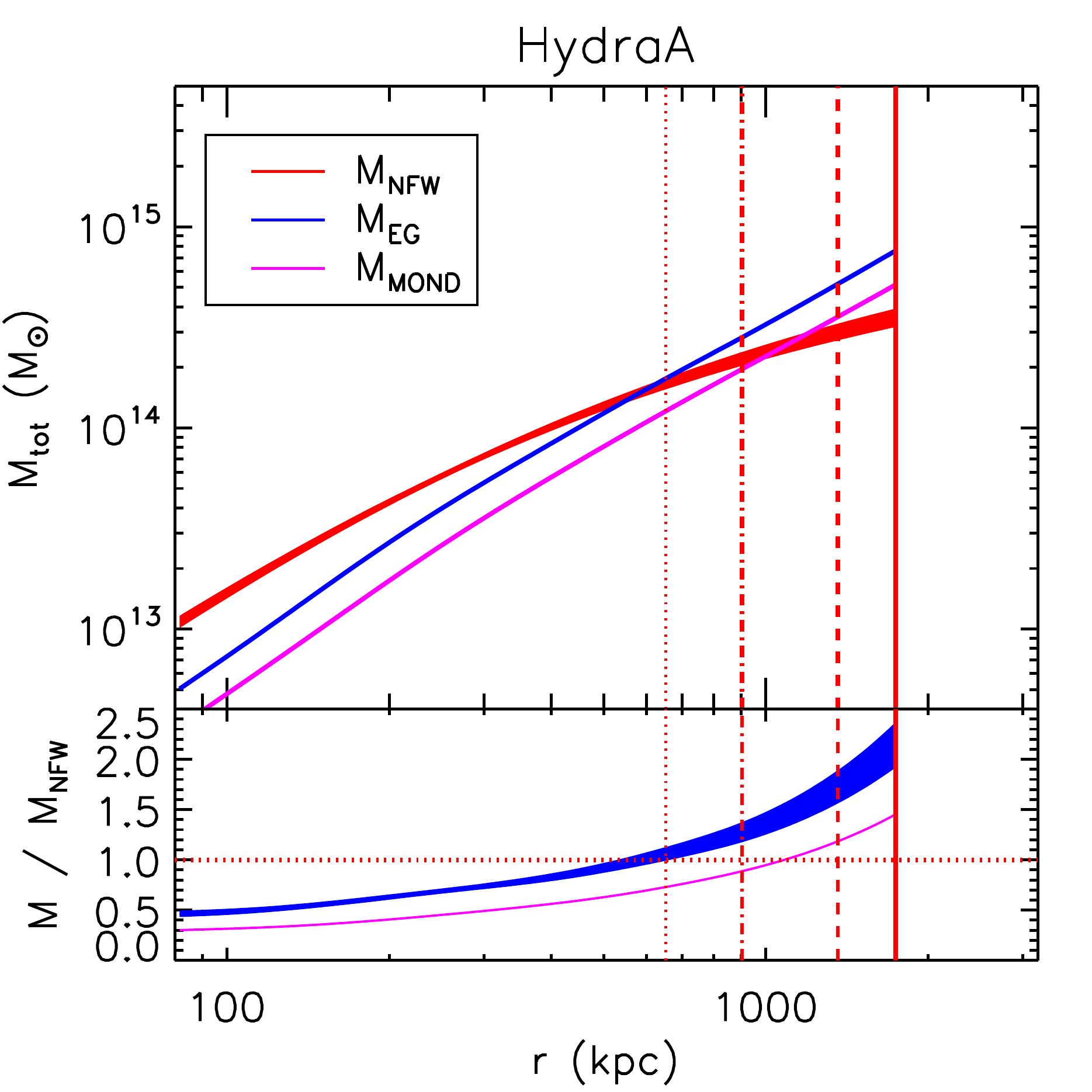}
\includegraphics[width=0.245\textwidth, keepaspectratio]{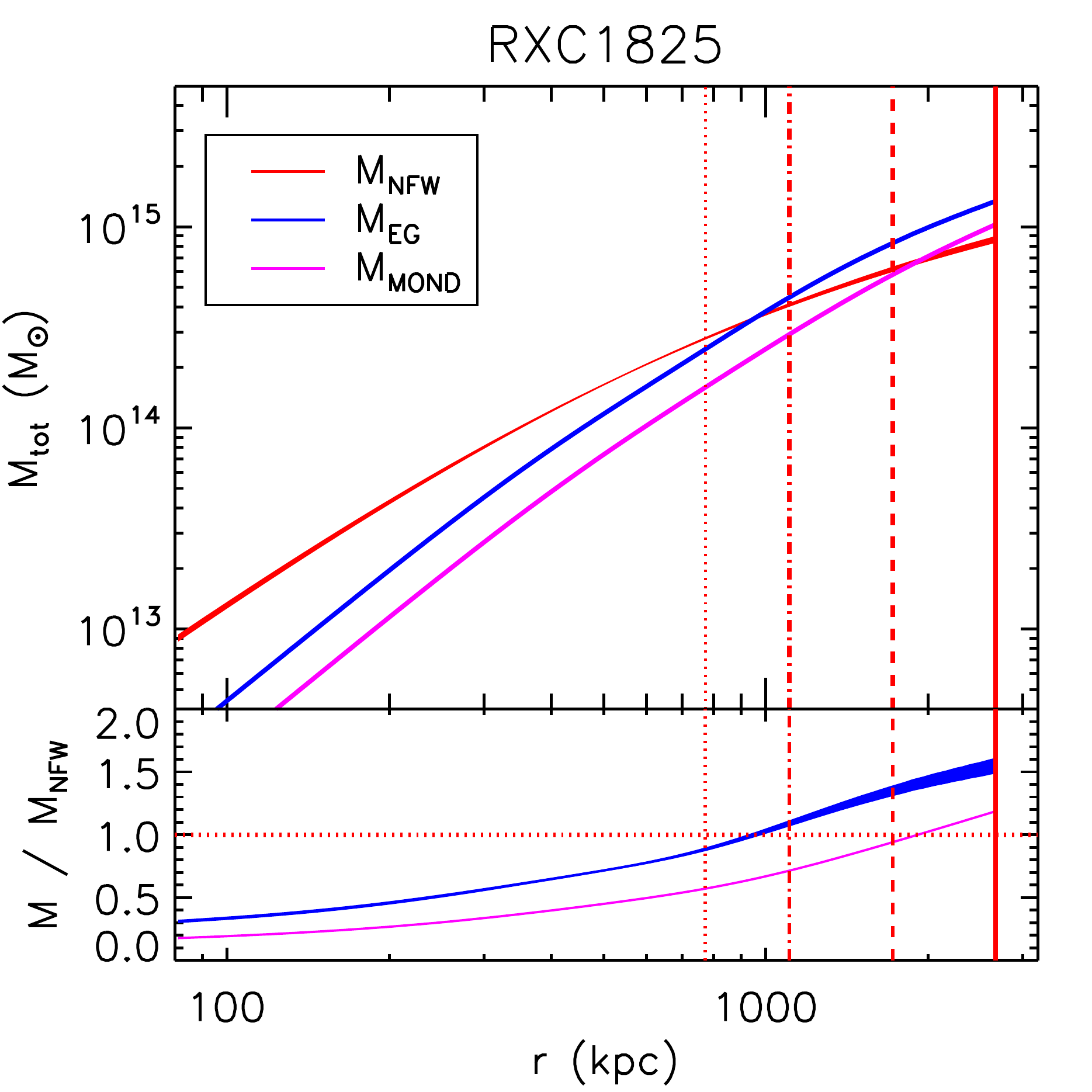}
} \hbox{
\includegraphics[width=0.245\textwidth, keepaspectratio]{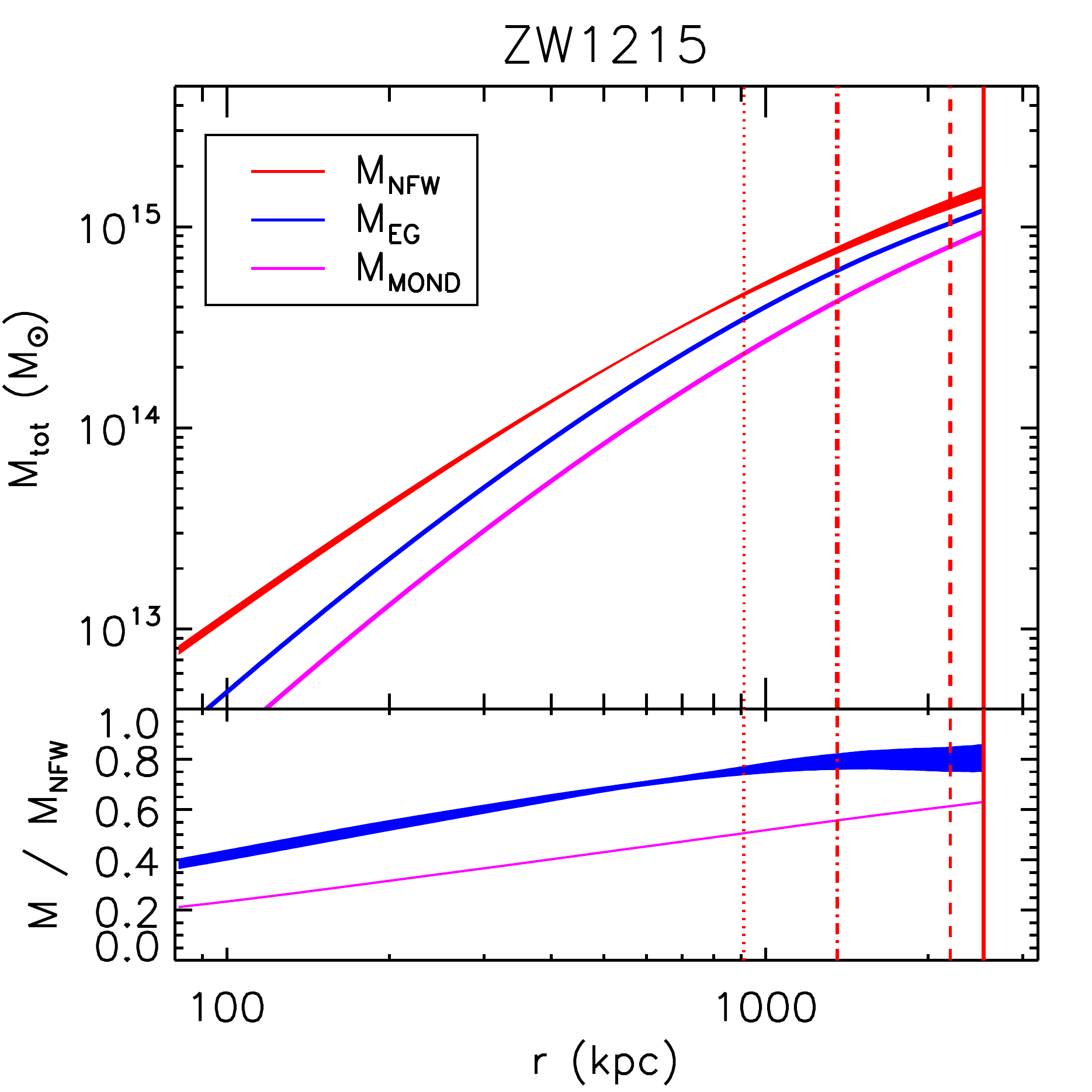}
}
\caption{Hydrostatic, EG, and MOND mass radial profiles and relative ratios.
Vertical lines indicate $R_{1000}$ (dotted line), $R_{500}$ (dash-dotted line), $R_{200}$ (dashed line),
and the outermost radial bin in the gas density profile (solid  line).
} \label{fig:eg}
\end{figure*}

\end{appendix}

\end{document}